\documentclass[apsrev4-1,twocolumn,footinbib,superscriptaddress,floatfix,bibliography]{revtex4-1}
\usepackage{times}
\usepackage{float}
\usepackage{multirow}
\usepackage[dvipsnames]{xcolor}
\usepackage{amsmath}
\usepackage{amsthm}
\usepackage{amssymb}
\usepackage{amsbsy}
\usepackage{enumitem}
\usepackage{wasysym}
\usepackage[english]{babel}
\usepackage[T1]{fontenc}
\usepackage[utf8]{inputenc} 
\usepackage{graphicx}
\usepackage[colorlinks,bookmarks=false,citecolor=blue,linkcolor=red,urlcolor=blue]{hyperref}
\usepackage{ulem}
\usepackage{pstricks}
\usepackage{rotating}			       % creates rotated page (landscape) 
\usepackage{tabularx,hhline}	
\usepackage[caption=false]{subfig}		
\newcolumntype{P}[1]{>{\centering\arraybackslash}p{#1}}

\graphicspath{{./figs/}} 		% Specifies the directory where pictures are stored
\newcommand{\CCO}{Ca$_{10}$Cr$_7$O$_{28}$}

\colorlet{mylinkcolor}{violet}
\colorlet{mycitecolor}{YellowOrange}
\colorlet{myurlcolor}{Aquamarine}

\newcommand{\nic}{\textcolor{OliveGreen}}

\begin{document}
%%%%%%%%%%%%%%%%%%%%%%%%%%%%%%%%%%%%%

%\title{How many spin liquids are there in Ca$_{10}$Cr$_7$O$_{28}$?}
%\title{Origin of spin-liquid behavior in bilayer breathing-kagome magnet Ca$_{10}$Cr$_7$O$_{28}$}

%%%%%%%%%%%%%%%%%%%%%%%%%%%%%%%%%%%%%
\title{Theory of Ca$_{10}$Cr$_7$O$_{28}$ as a bilayer breathing--kagome magnet:\\
Classical thermodynamics and semi--classical dynamics}
%%%%%%%%%%%%%%%%%%%%%%%%%%%%%%%%%%%%%

%%%%%%%%%%%%%%%%%%%%%%%%%%%%%%%%%%%%%
\author{Rico Pohle}
%%%%%%%%%%%%%%%%%%%%%%%%%%%%%%%%%%%%%
\email{rico.pohle@aoni.waseda.jp}
\affiliation{Theory of Quantum Matter Unit, Okinawa Institute of Science and Technology Graduate University, Onna-son, Okinawa 904-0412, Japan}
\affiliation{Department of Applied Physics, University of Tokyo, Hongo,  Bunkyo-ku,
Tokyo, 113-8656, Japan}
\affiliation{Department of Applied Physics, Waseda University, Okubo, Shinjuku-ku, 
Tokyo 169-8555, Japan}

%%%%%%%%%%%%%%%%%%%%%%%%%%%%%%%%%%%%%
\author{Han Yan}
%%%%%%%%%%%%%%%%%%%%%%%%%%%%%%%%%%%%%
\email{han.yan@oist.jp}
\affiliation{Theory of Quantum Matter Unit, Okinawa Institute of Science and Technology Graduate University, Onna-son, Okinawa 904-0412, Japan} 
%%%%%%%%%%%%%%%%%%%%%%%%%%%%%%%%%%%%%
\author{Nic Shannon}
%%%%%%%%%%%%%%%%%%%%%%%%%%%%%%%%%%%%%
\email{nic.shannon@oist.jp}
\affiliation{Theory of Quantum Matter Unit, Okinawa Institute of Science and Technology Graduate University, Onna-son, Okinawa 904-0412, Japan}

%%%%%%%%%%%%%%%%%%%%%%%%%%%%%%%%%%%%%
\date{\today}
%%%%%%%%%%%%%%%%%%%%%%%%%%%%%%%%%%%%%

%%%%%%%%%%%%%%%%%%%%%%%%%%%%%%%%%%%%%
\begin{abstract}
%%%%%%%%%%%%%%%%%%%%%%%%%%%%%%%%%%%%%

\CCO\ is a novel spin--$1/2$ magnet exhibiting %with 
spin liquid behaviour which 
sets it apart from any previously studied model or material.
However, understanding \CCO\ presents a significant 
challenge, because the low symmetry of the crystal structure 
leads to very complex interactions, with up to seven inequivalent 
coupling parameters 
%six inequivalent magnetic sites 
in the unit cell.
Here we explore the origin of the spin-liquid behaviour in \CCO, 
starting from the simplest microscopic model consistent with experiment 
--- a Heisenberg model on a single bilayer of the breathing--kagome (BBK) lattice.
We use a combination of classical Monte Carlo (MC) simulation and 
(semi--)classical Molecular Dynamics (MD) simulation to explore the 
thermodynamic and dynamic properties of this model, and compare these
with experimental results for \CCO.
We uncover qualitatively different behaviours on different timescales, 
and argue that the ground state of \CCO\ is born out of 
a slowly-fluctuating ``spiral spin liquid'', while faster fluctuations 
echo the U(1) spin liquid found in the kagome antiferromagnet.
We also identify key differences between longitudinal and transverse 
spin excitations in applied magnetic field, and argue that these are a 
distinguishing feature of the spin liquid in the BBK model.

%%%%%%%%%%%%%%%%%%%%%%%%%%%%%%%%%%%%%
\end{abstract}
%%%%%%%%%%%%%%%%%%%%%%%%%%%%%%%%%%%%%


%%%%%%%%%%%%%%%%%%%%%%%%%%%%%%%%%%%%%%%%%%
\maketitle
%%%%%%%%%%%%%%%%%%%%%%%%%%%%%%%%%%%%%%%%%%

%%%%%%%%%%%%%%%%%%%%%%%%%%%%%%%%%%%%%
\section{Introduction}
%%%%%%%%%%%%%%%%%%%%%%%%%%%%%%%%%%%%%

The search for quantum spin liquids (QSL), exotic phases which host new forms 
of magnetic excitations, has become one of the central themes of modern 
condensed matter physics \cite{anderson73-MatResBull8,balents10-Nature464,Savary_2016,knolle19}.  
Fortunately, after a long ``drought'' \cite{lee08-Science321}, recent years have seen 
an explosion in the number of materials under study, with examples including 
quasi-2D organics \cite{shimizu03-PRL91,zhou17}, 
thin films of $^3$He \cite{masutomi04-PRL92}, 
spin--1/2 magnets with a kagome lattice \cite{lee08-Science321,han12-Nature492}, 
``Kitaev'' magnets with strongly anisotropic exchange 
\cite{kitaev06-AnnPhys321,jackeli09-PRL102,banerjee16-NatMater15,baek17-PRL119,winter17,hermanns17},  
and quantum analogues of spin ice 
\cite{hermele04-PRB69,banerjee08-PRL100,benton12-PRB86,sibille16,Benton2018-PRL121}. 
Another new arrival on this scene is the quasi--2D magnet \CCO, 
a system which appears to have qualitatively different properties from 
any previously--studied spin liquid 
\cite{Balz2016,Balz2017-PRB95,Sonnenschein2019}.

%%%%%%%%%%%%%%%%%%%%%%%%%%%%%%%%%%%%%

\begin{figure}[t]
	\centering
  	\includegraphics[width=0.26\textwidth]{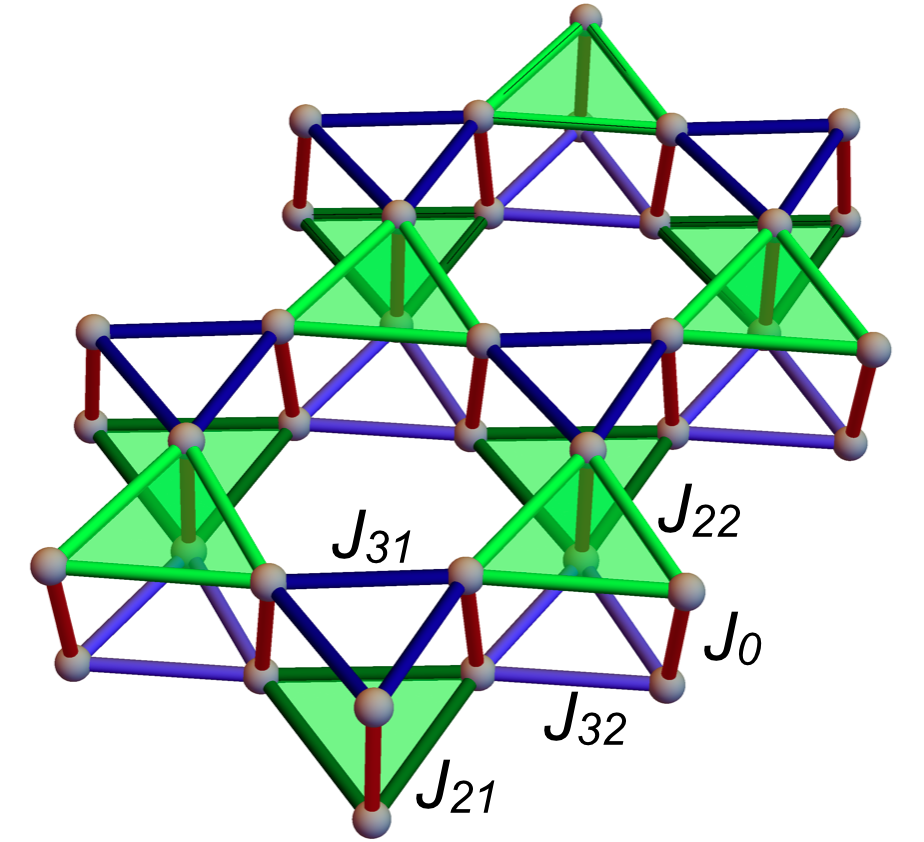}	
	\caption{ 
	Bilayer breathing--Kagome (BBK) lattice realised by \mbox{spin--1/2} Cr$^{5+}$ 
	ions in \CCO.
	Interactions for the minimal \mbox{spin--1/2} BBK model [Eq.~(\ref{eq:H.BBK})] 
	are labelled following the conventions of Balz~{\it et al.}
	\cite{Balz2016, Balz2017-PRB95}.
	Experimental estimates of these parameters can be found 
	in Table~\ref{tab:experimental.parameters}.
	}
	\label{fig:BBK.model}
\end{figure}

\begin{figure}[t]
	\captionsetup[subfigure]{labelformat=empty, farskip=2pt,captionskip=1pt}
	\centering	
	\subfloat[(a) $E =$ 0.25 meV  \label{fig:Exp_025meV}]{
  		\includegraphics[width=0.5\columnwidth]{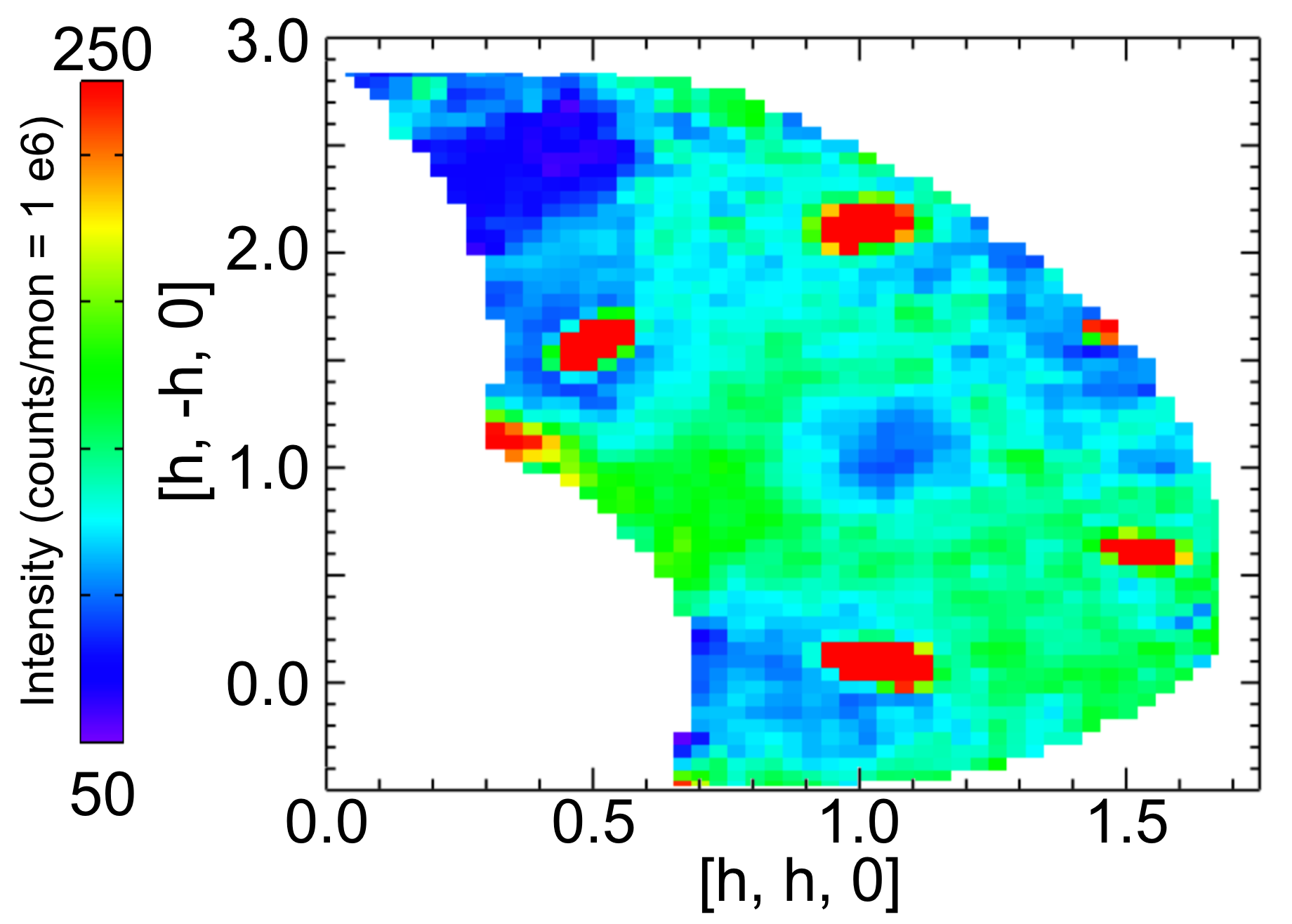}
		}
	\subfloat[(b) $E =$ 0.26 meV 	\label{fig:MD_023meV}]{ 
  		\includegraphics[width=0.46\columnwidth]{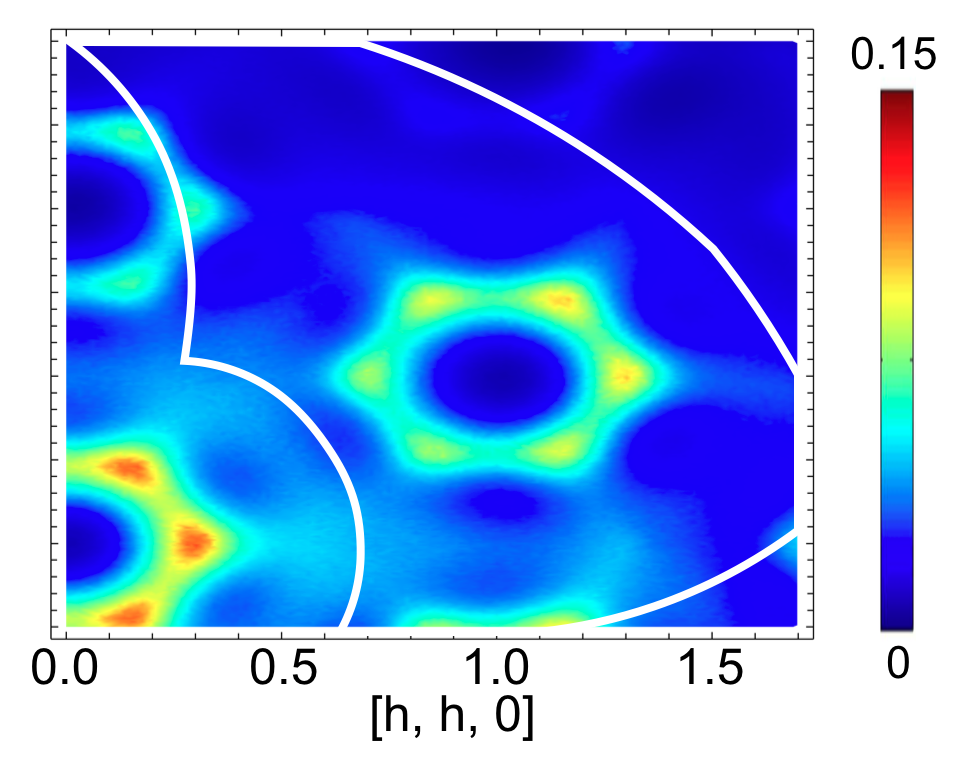}
		}
	\\[1ex]
	
	\subfloat[(c) $E =$ 0.65 meV \label{fig:Exp_065meV}]{
  		\includegraphics[width=0.5\columnwidth]{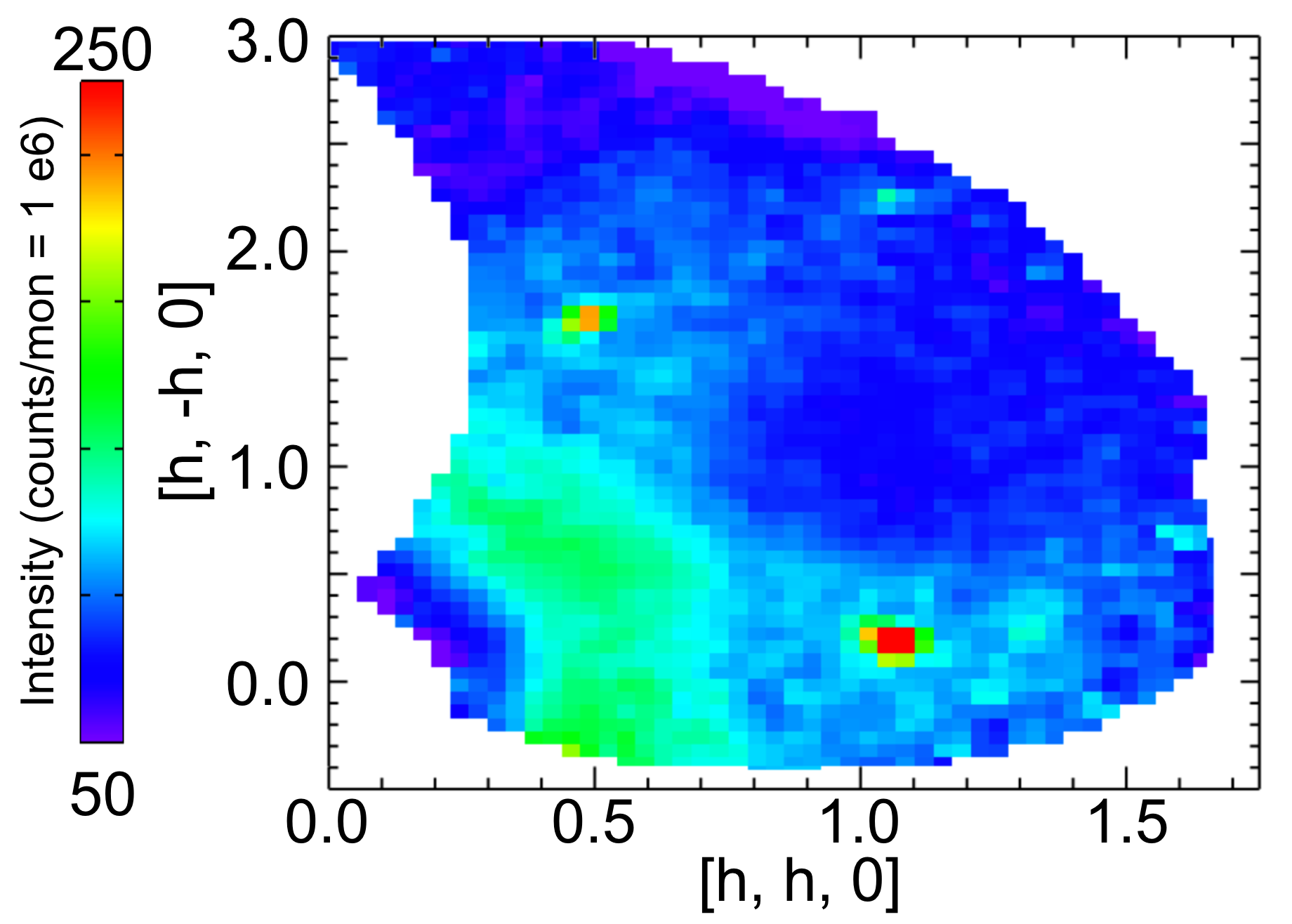}
		}
	\subfloat[(d) $E =$ 0.46 meV  \label{fig:MD_52meV}]{ 
  		\includegraphics[width=0.46\columnwidth]{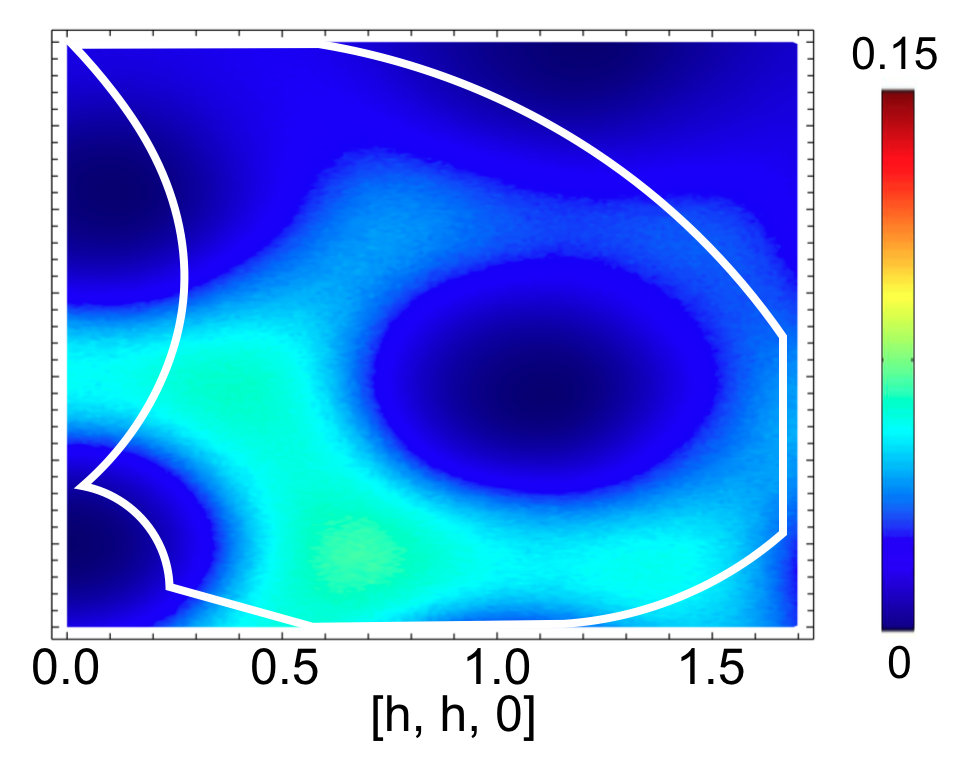}
		}
	\\[1ex]

	\subfloat[(e) $E =$ 0.9 meV \label{fig:Exp_090meV}]{
  		\includegraphics[width=0.5\columnwidth]{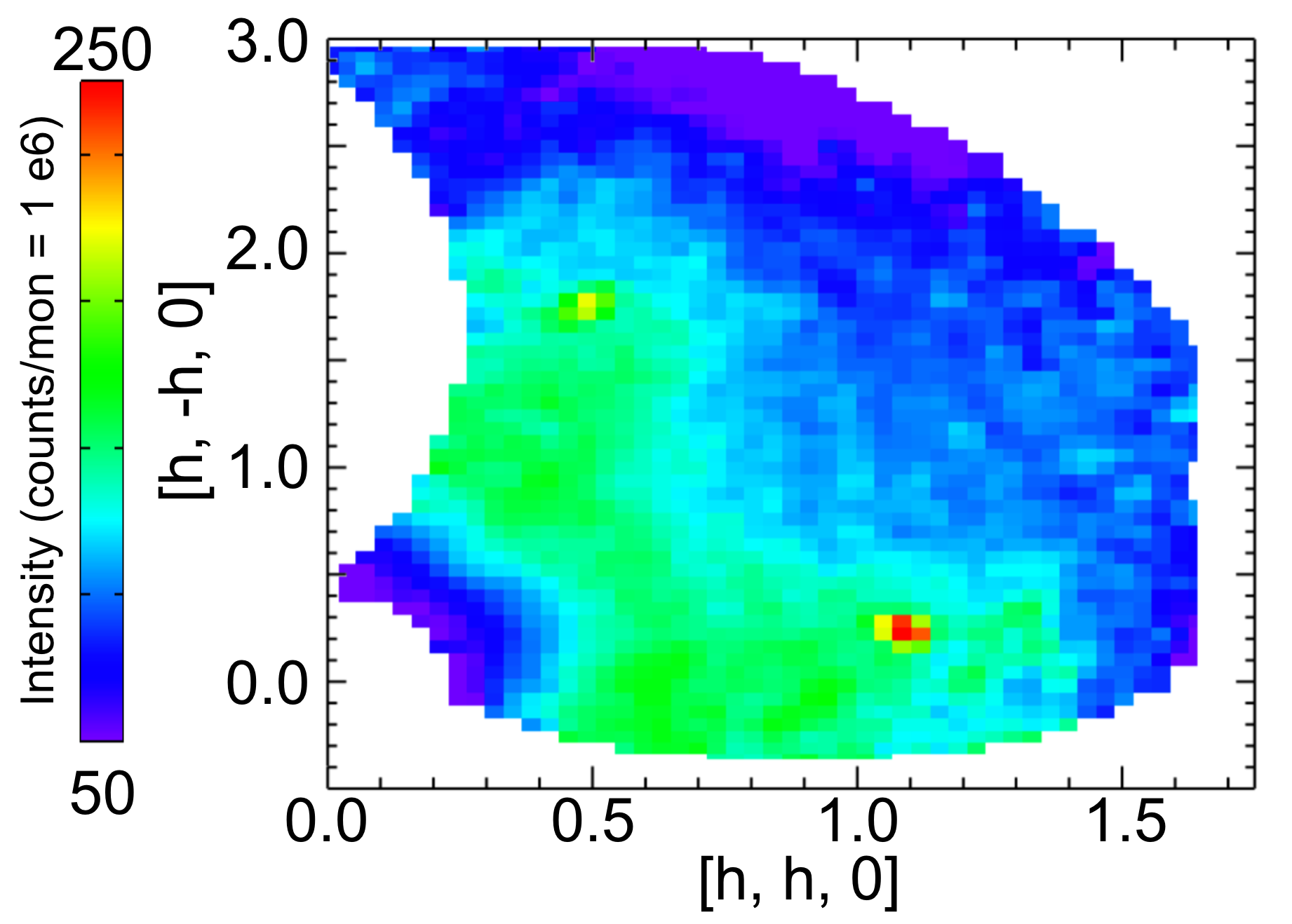}
		}
	\subfloat[ (f) $E =$ 1.15 meV \label{fig:MD_110meV}]{
  		\includegraphics[width=0.46\columnwidth]{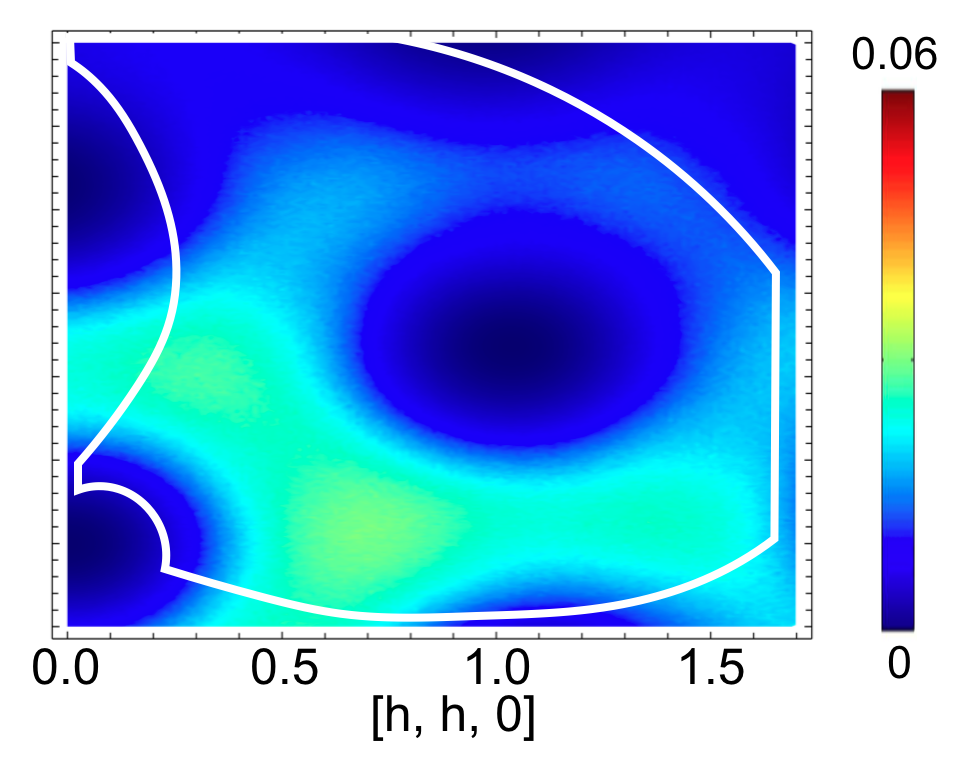}
	}
	\caption{
	Fluctuations in the spin-liquid phase of \CCO, showing 
	${\bf q}$--dependent structure on different energy scales.
	(a) INS data for \CCO\ at low energy, suggestive of a ``ring''
	of scattering, centered on $(2, 0, 0)$.      
 	(b) Equivalent results from molecular dynamics (MD) simulations.	
	(c) INS data for \CCO\ at intermediate energy, 
	showing ``bow-tie'' structures centered on $(1, 0, 0)$.    
        (d) Equivalent results from MD simulation.   
	(e) Cut through inelastic neutron 
	scattering (INS) data for \CCO\ at high energy, 
	also  showing ``bow-tie'' structure.
        (f) Equivalent results from MD simulation.   
        Experimental data are reproduced from 
        %Balz.~{\it et al.}~\cite{Balz2016}.   
        \cite{Balz2016}, with measurements 
        carried out at $T = 90\ \text{mK}$.
        MD simulations were carried out for a bilayer 
        breathing Kagome (BBK) model, at $T = 220\ \text{mK}$,  
        as described in the text.   
	Energies were chosen so as to compare corresponding 
	features in simulation and experiment.
 	}
	\label{fig:comparison.with.experiment}
\end{figure}

The first surprise in \CCO\ is a chemical one.
Instead of the usual $3+$ valance, Cr ions exhibit a highly 
unusual, $5+$ valence \cite{arcon98-JAmCeramSoc81}.
These Cr$^{5^+}$ ions are magnetic, with spin S=1/2, and occupy 
sites of the breathing bilayer--kagome (BBK) lattice [Fig.~\ref{fig:BBK.model}] 
\cite{Balz2017-JPCM29,balodhi17-PRM1}.
Curie--law fits to the magnetic susceptibility of \CCO\ reveal 
predominantly ferromagnetic (FM) interactions,  
with $\theta_{\sf CW} = 2.35\ \text{K}$  \cite{Balz2016,Balz2017-PRB95}.
This is accompanied by a broad peak in heat capacity 
at about $T \approx 3.1\ \text{K}$ \cite{Balz2016,Balz2017-PRB95,Sonnenschein2019}. 
However measurements of heat capacity, a.c. susceptibility 
and $\mu$SR asymmetry fail to find evidence of either,
magnetic order, or spin--glass freezing, down to $19\ \text{mK}$, two 
orders of magnitude lower than the scale of interactions \cite{Balz2016}.
Consistent with this, neutron scattering experiments find no magnetic 
Bragg peaks down to $90\ \text{mK}$ \cite{Balz2016,Balz2017-PRB95,Sonnenschein2019}.
Instead, scattering is predominantly inelastic and highly--structured, 
with results at $0.25\ \text{meV}$ showing hints of a ``ring'' centered 
on (2,0,0), while scattering at intermediate and high energies suggests  
``bow--tie'' like structures centered on \mbox{(1,0,0)}
\cite{Balz2016,Balz2017-PRB95} [Fig.~\ref{fig:comparison.with.experiment}].
With the application of magnetic field, these structures evolve into 
relatively--sharp, dispersing excitations, with measurements of heat capacity 
suggesting a qualitative change in behaviour for fields $B \gtrsim 1T$ 
\cite{Balz2016,Balz2017-PRB95}.
This particular combination of dynamic and thermodynamic 
properties sets \CCO\ apart from any spin liquid material yet 
studied, and presents an interesting challenge to theory.

%%%%%%%%%%%%%%%%%%%%%%%%%%%%%%%%%%%%%
%  Fig. 1 - BBK model
%%%%%%%%%%%%%%%%%%%%%%%%%%%%%%%%%%%%%

%%%%%%%%%%%%%%%%%%%%%%%%%%%%%%%%%%%%%

A range of different theoretical techniques have been applied to study \CCO.
Pseudo--fermion functional renormalisation group (PFFRG) calculations, for a spin--1/2 model 
parameterised from experiment, reproduce a ``ring'' in the static structure factor 
$S({\bf q}, \omega=0)$, and suggest  that the ground state of \CCO\ should be a 
quantum spin liquid \cite{Balz2016}.
This conclusion was supported by subsequent tensor--network calculations 
\cite{Kshetrimayum2020}.
%and exact--diagonalisation studies \cite{shimokawa-in-prep}.
%
Meanwhile the finite--temperature properties of \CCO\ have been explored 
through Monte Carlo simulations of a simplified, spin--3/2 honeycomb--lattice model \cite{biswas18}.
At high temperatures, these also reveal ``rings'' in the equal--time 
structure factor $S({\bf q})$, while at low temperatures a 3--state Potts transition is found 
into a nematic state which breaks lattice--rotation symmetries, but lacks long--range 
magnetic order \cite{Mulder2010, biswas18}.
And, intriguingly, the thermodynamic properties of \CCO\ have recently 
been argued to fit phenomenology based on spinons \cite{Sonnenschein2019}.

%%%%%%%%%%%%%%%%%%%%%%%%%%%%%%%%%%%%%

None of these approaches, however, shed light 
%on the qualitative difference between the dynamics observed 
%at low and at intermediate energies; 
on the nature of ``bow--tie'' structures observed at finite energies;
the evolution of the spin liquid in magnetic field; 
or the finite--temperature properties of the microscopically relevant, \mbox{spin--1/2}
BBK model.
And, most importantly, while there is agreement about the absence of 
conventional magnetic order, very little is known about the origins of the 
spin liquid which succeeds it.

%%%%%%%%%%%%%%%%%%%%%%%%%%%%%%%%%%%%%

This Article will be the first of two papers exploring the thermodynamics 
and dynamics of \CCO, starting from the spin--1/2 BBK model proposed 
by Balz {\it et al.} \cite{Balz2016,Balz2017-PRB95}
\begin{equation}
	{\cal H}_{\sf BBK} 
	= \sum_{\langle i j \rangle} J_{ij} {\bf S}_i \cdot {\bf S}_j - {\bf B} \cdot \sum_i {\bf S}_i \; ,
    \label{eq:H.BBK}
\end{equation}
where first--neighbour bonds $\langle i j \rangle$ are illustrated in Fig.~\ref{fig:BBK.model}, 
and parameters $J_{ij}$ can be extracted from fits to inelastic neutron scattering in 
high magnetic field $B \gg J$ [Table~\ref{tab:experimental.parameters}].  
In this paper, we extend the results of an earlier preprint \cite{pohle-arXiv}, 
making the approximation of treating spins as classical $O(3)$ vectors, 
and using a combination of classical Monte Carlo (MC) simulations 
and numerical integration of equations of motion (here referred to as 
``molecular dynamics'' or ``MD'' simulation \cite{moessner98-PRB58}), 
to evaluate their dynamics.
From this, we first establish a finite--temperature phase diagram 
for the BBK model of \CCO, and then track the evolution of its properties 
as a function of energy and magnetic field.
In the second Article, we will compare these findings with the results of exact
diagonalization and finite--temperature quantum--typicality calculations 
for a spin--1/2 BBK model~\cite{shimokawa-in-prep}.

%%%%%%%%%%%%%%%%%%%%%%%%%%%%%%%%%%%%%

In the classical limit, considered in this Article, we find that the BBK model 
supports a spin liquid state for a wide range of temperatures and parameter values.
At low energies, this is characterised by the slow, collective fluctuations 
of ferromagnetically aligned spins on triangular plaquettes.
These give rise to a ring--like structure in the dynamical structure factor 
$S({\bf q},\omega)$, at low energies, while fluctuations at higher energies have 
a qualitatively different character, reflecting the Kagome--like physics of 
individual spin--1/2 moments.
An added bonus of the (semi--)classical molecular dynamics simulations 
used is that both of these features can be visualised directly, through 
animations provided in the Supplemental Materials \cite{first.animation,second.animation}.  
At low temperatures, we find that this classical spin liquid undergoes 
a 3--state Potts transition into a phase breaking lattice--rotation 
symmetry (lattice nematic), consistent with results for an effective 
spin--3/2 model \cite{biswas18}.
For parameters taken from experiment, this transition occurs at 
$T \approx 66\ \text{mK}$.

%%%%%%%%%%%%%%%%%%%%%%%%%%%%%%%%%%
% Fig 2 - structured scattering in CCO; comparison with experiment
%%%%%%%%%%%%%%%%%%%%%%%%%%%%%%%%%%

%%%%%%%%%%%%%%%%%%%%%%%%%%%%%%%%%%%%%

\begin{figure}[t]
	\centering
	\captionsetup[subfigure]{labelformat=empty}
	\begin{minipage}[t]{0.99\columnwidth}
		\subfloat[\label{fig:PD.B}]{
  		\includegraphics[width=0.98\columnwidth]{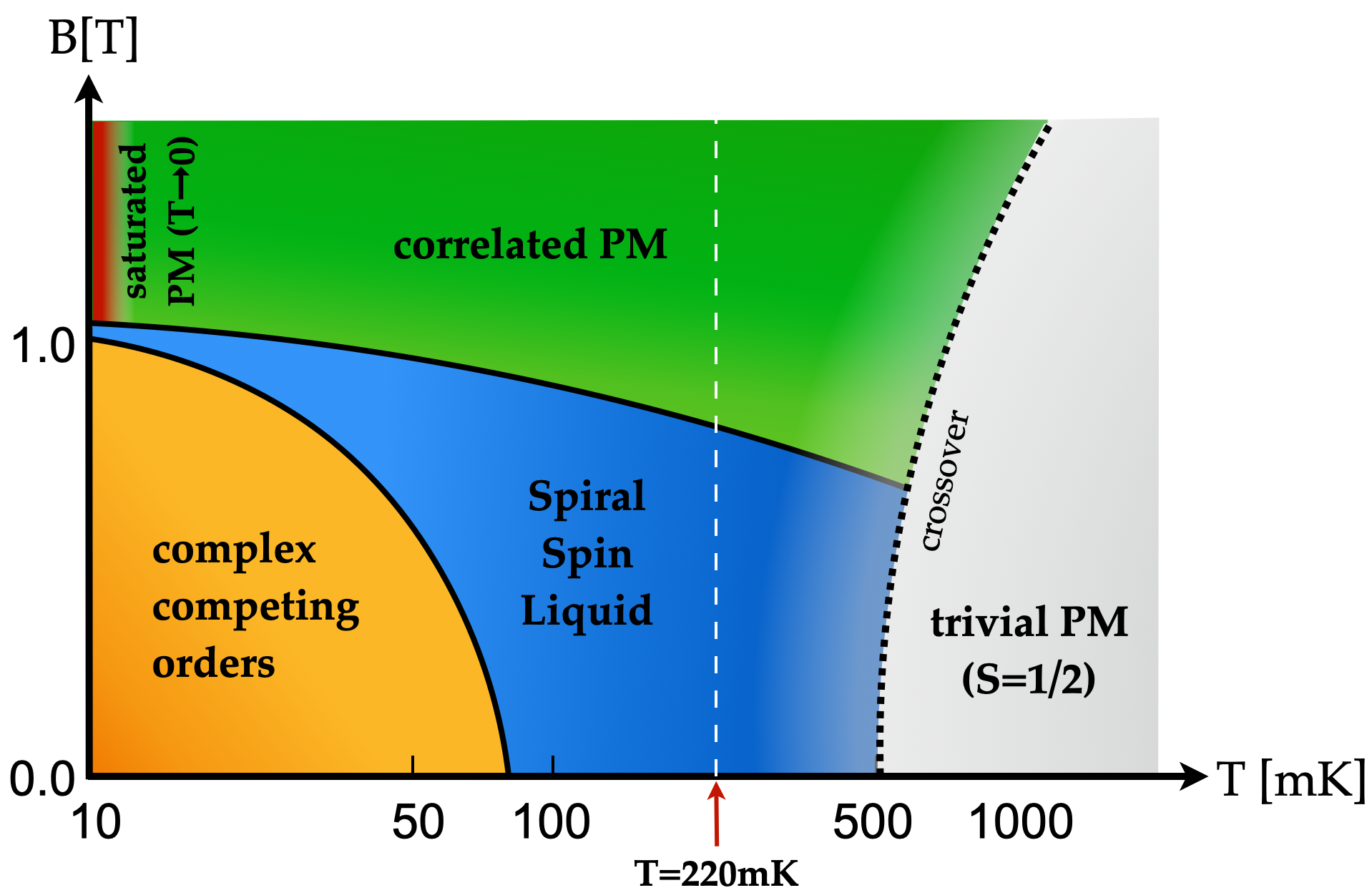}
		}
	\end{minipage}
	\caption{
	Schematic phase diagram of the classical bilayer breathing 
	Kagome (BBK) model of \CCO,  as function of temperature and magnetic field. 
	At high temperatures individual spin--1/2 moments
	fluctuate independently.   
	For $T \approx 500\ \text{mK}$ there is a crossover into a regime 
	in which spins on triangular plaquettes form effective spin--3/2 
	moments.
	For fields $B \lesssim 1\ \text{T}$ these form a spiral spin liquid, 
	characterised by a ring--like structure in the dynamical structure 
	factor $S({\bf q})$.
	At temperatures $T \lesssim 60\ \text{mK}$, this gives way 
	to states which break discrete symmetries of the lattice, 
	including a lattice nematic, and a multiple--q state.
	This hierarchy of temperature and field scales is derived from  
	%of ${\mathcal H}_{\sf BBK}$ [Eq.~\ref{eq:H.BBK}], 
	%for parameters taken from experiment [Table~\ref{tab:experimental.parameters}], 
	classical Monte Carlo (MC) simulations described 
	in Section~\ref{sec:thermodynamics} and 
	in Section~\ref{sec:thermodynamics.in.field}.
	The temperature for which dynamics have been characterised 
	in detail, $T = 220\ \text{mK}$, is shown with a white 
	dashed line [cf. Fig.~\ref{fig:comparison.with.experiment}].
	}
	\label{fig:schematic.phase.diagram.B}
\end{figure}

We also study the evolution of the dynamical and thermodynamical 
properties of the BBK model in applied magnetic field, concentrating
on parameters relevant to  \CCO\ [Fig.~\ref{fig:schematic.phase.diagram.B}].
%
%At finite temperature, the correlation of spins on FM plaquettes 
%converts the field--saturated state from a trivial paramagnet 
%into a phase with (algebraic) correlations of octupolar order.   
%

We find that the onset of the spin--liquid observed in experiment is 
associated with the closing of a gap to transverse spin excitations, 
at a field $B^*(T)$, with \mbox{$B^*(220\ \text{mK}) \approx 0.7\ \text{T}$}, 
and \mbox{$B^*(T\to 0) \to 1.1\ \text{T}$}.
From the nature of the spin excitations when this gap closes, 
we identify the low--field state as a gapless, ``spiral spin--liquid'', 
known for ring--like correlations in $S({\bf q})$
\cite{bergman07,okumura10-JPSJ79,benton15-JPSJ84,seabra16-PRB93,buessen18,Yao2021}. 
Simulations also reveal low--energy, longitudinal excitations, 
%initially associated with octupolar correlations at high field, 
which may explain the anomalously high specific heat measured at 
low temperatures.
At the lowest temperatures, we find a complex set of competing orders
including the finite--field extension of the lattice nematic and a multiple--q state.
Finally, we show how the field--saturated state provides an opportunity 
to study the ``half--moons'' recently discussed in the context of Kagome 
antiferromagnets \cite{Yan2018,Mizoguchi2018}.
Taken together, these results provide a broad characterisation of the 
BBK model of \CCO, within a \mbox{(semi--)classical} approximation which explains 
many of the features seen in experiment.

%%%%%%%%%%%%%%%%%%%%%%%%%%%%%%%%%%%%%
%   "Remainder of this article ..."
%%%%%%%%%%%%%%%%%%%%%%%%%%%%%%%%%%%%%

The remainder of this Article is structured as follows: 

%%%%%%%%%%%%%%%%%%%%%%%%%%%%%%%%%%%%%

In Sec.~\ref{section:CCO.v.BBK.model} we review the existing experimental 
and theoretical literature on \CCO, and introduce the spin--1/2 
bilayer breathing--Kagome (BBK) model used to interpret these results.

%%%%%%%%%%%%%%%%%%%%%%%%%%%%%%%%%%%%%

In Sec.~\ref{sec:thermodynamics} we present simulation results for the 
thermodynamics of the BBK model of \CCO\ in zero magnetic field.
Classical Monte Carlo (MC) results are used to construct a finite--temperature phase 
diagram which connects the spin liquid phase of \CCO\ with a domain 
of high classical ground--state degeneracy of the BBK model.

%%%%%%%%%%%%%%%%%%%%%%%%%%%%%%%%%%%%%

In Sec.~\ref{sec:dynamics} we show corresponding results for dynamics, 
taken by numerically integrating the equations of motion for states 
drawn from MC simulation (MD simulation).
Results are visualised through both animations and plots 
of the dynamical structure factor $S({\bf q}, \omega)$, 
which are used to connect with experiment.

%%%%%%%%%%%%%%%%%%%%%%%%%%%%%%%%%%%%%

In Sec.~\ref{sec:thermodynamics.in.field} we turn to the thermodynamic 
properties of the BBK model in applied magnetic field.
Classical MC simulation is used to establish a phase diagram 
as a function of field and temperature, for parameters appropriate to \CCO.

%%%%%%%%%%%%%%%%%%%%%%%%%%%%%%%%%%%%%

In Sec.~\ref{sec:dynamics.in.field} we explore the corresponding changes 
in spin dynamics as function of magnetic field, again through numerical integration 
of equations of motion (MD simulation).
Particular attention is paid to the way in which the spin liquid emerges
from the paramagnet found at high values of magnetic field.

%%%%%%%%%%%%%%%%%%%%%%%%%%%%%%%%%%%%%

In Sec.~\ref{sec:cco.v.bbk} we discuss the implication of these results 
for the understanding of \CCO.

%%%%%%%%%%%%%%%%%%%%%%%%%%%%%%%%%%%%%

Finally, in Sec.~\ref{sec:conclusions} we conclude with a brief summary 
of results and open questions.

%%%%%%%%%%%%%%%%%%%%%%%%%%%%%%%%%%%%%

Further technical information is provided in a short series of Appendices: 

%%%%%%%%%%%%%%%%%%%%%%%%%%%%%%%%%%%%%

Appendix~\ref{appendix:Numerics} contains technical details of the classical 
Monte Carlo (MC) and Molecular Dynamics (MD) techniques used in 
this study, as well as of the methods used to animate spin configurations.

%%%%%%%%%%%%%%%%%%%%%%%%%%%%%%%%%%%%%

Appendix~\ref{appendix:animation} contains technical 
details of the Animation of spin configurations.

%%%%%%%%%%%%%%%%%%%%%%%%%%%%%%%%%%%%%

Appendix~\ref{sec:CompExp} contains details of the structure factors 
and form factors used when comparing with experiment.

%%%%%%%%%%%%%%%%%%%%%%%%%%%%%%%%%%%%%

Appendix~\ref{sec:wilson.ratio} provides details of an estimate of the 
Wilson ratio for \CCO.

%%%%%%%%%%%%%%%%%%%%%%%%%%%%%%%%%%%%%
%  Fig. X - phase diagram in field
%%%%%%%%%%%%%%%%%%%%%%%%%%%%%%%%%%%%%

%%%%%%%%%%%%%%%%%%%%%%%%%%%%%%%%%%%%%
\section{$\text{Ca}_{10}\text{Cr}_7\text{O}_{28}$ 
and the BBK model}
%%%%%%%%%%%%%%%%%%%%%%%%%%%%%%%%%%%%%
\label{section:CCO.v.BBK.model}

While the study of \CCO\ has a short history, a wide range 
of different experimental and theoretical techniques have already been brought 
to bear on it.
In what follows we review attempts to unravel the properties of \CCO, 
starting from its chemistry and structure, and covering different aspects of 
its experimental characterisation, before surveying attempts to model it 
in terms of a bilayer breathing--Kagome (BBK) model.
A brief account is also given of the closely--related physics of the 
$J_1$--$J_2$ Heisenberg model on a honeycomb lattice.

%%%%%%%%%%%%%%%%%%%%%%%%%%%%%%%%%%%%%
\subsection{Chemistry and crystal structure}
%%%%%%%%%%%%%%%%%%%%%%%%%%%%%%%%%%%%%
\label{section:chemistry.cco}

The earliest motivation for studying \CCO\ came from chemistry.
In oxides, Cr~is typically found with a $3+$ valence, 
giving rise to magnets with spin--3/2 moments.
A typical example of such is the spinel CdCr$_2$O$_4$, 
a relatively classical magnet, interesting for the interplay between 
magnetic frustration and spin--lattice coupling --- see e.g. \cite{rossi19}.
\CCO, on the other hand, is one of a family of materials exhibiting the unusual, spin--1/2,  
Cr$^{5+}$ valence state \cite{gyepesova13-ActaCrysC69,arcon98-JAmCeramSoc81}.
And the spin--1/2 nature of the magnetic ions in \CCO\ brings with it physics of 
an altogether more quantum nature than is found in CdCr$_2$O$_4$.

%%%%%%%%%%%%%%%%%%%%%%%%%%%%%%%%%%%%%

Structurally, \CCO\ has much in common with SrCr$_2$O$_8$, 
a Mott insulator based on Cr$^{5+}$ ions, which 
has been studied for its quantum dimer ground state
%, and phase transitions in applied magnetic field 
\cite{wang16-PRL116}.
SrCr$_2$O$_8$ is composed of stacked, triangular--lattice 
bilayers, and has the high--temperature space group 
R3$\overline{m}$, \cite{cuno89}.
At \mbox{$T = 275\ \text{K}$}, it undergoes a structural phase transition
into a phase with space group C2/c, lifting the orbital degeneracy 
of the Cr$^{5+}$ ions \cite{islam10}.
This promotes a low--temperature phase in which Cr$^{5+}$
form a triangular lattice of singlet dimers, each arranged along 
the c-axis connecting the two planes of each bilayer \cite{islam10, quintero-castro10,wang16-PRL116}.

%%%%%%%%%%%%%%%%%%%%%%%%%%%%%%%%%%%%%
% Paragraph X  - chemical structure
%%%%%%%%%%%%%%%%%%%%%%%%%%%%%%%%%%%%%

\CCO\ differs from SrCr$_2$O$_8$ through the inclusion of non--magnetic 
Cr$^{6+}$ ions, at a ratio of 6 Cr$^{5+}$ ions to 1 Cr$^{6+}$ \cite{arcon98-JAmCeramSoc81}.
These convert the stacked, triangular bilayers of SrCr$_2$O$_8$ into 
weakly--coupled bilayers of a ``breathing'' kagome lattice, 
in which triangular plaquettes have alternating size 
\cite{Balz2017-JPCM29,balodhi17-PRM1} --- cf.~Fig.~\ref{fig:BBK.model}.  
This bilayer breathing--Kagome (henceforth, BBK) lattice has a very low 
symmetry, with the space group identified as R3c \cite{gyepesova13-ActaCrysC69,Balz2017-JPCM29}.  
Within this space group, the magnetic Cr$^{5+}$ ions have a 6--site 
unit cell, and the Cr$^{5+}$ site is located within a (distorted) %%
CrO$_4$ tetrahedron.
The crystal field at this site is sufficiently low that the  
degeneracy of the $e_g$ orbitals is quenched, leaving 
a single 3d electron in a single orbital, i.e. a spin--1/2 moment.

%%%%%%%%%%%%%%%%%%%%%%%%%%%%%%%%%%%%%
%  Table 1-  comparison with Heavy Fermions
%%%%%%%%%%%%%%%%%%%%%%%%%%%%%%%%%%%%%

\begin{table}
	\centering
	\begin{tabular}{|P{1.7cm}|P{2.2cm}|P{2.0cm}|P{1.0cm}|P{0.8cm}|}
		\hline	
		& $\gamma $  &   $\chi_0$ & $\mu_{\sf eff}$ & $R_{\sf W}$ \\
		& [mJ\ mol$^{-1}$\ K$^{-2}$] &    [emu mol$^{-1}$ Oe$^{-1}$] & [$\mu_B$] &\\
		\hline
%		Cu 	& $0.695$  			& $-5.46\ \times\ 10^{-6}$	& & n/a	\\ 
		Cu 	& $0.695$  			& weakly diagmagnetic	& 1.7 & n/a	\\ 
		\hline
		CeCu$_6$ \cite{Ott1987,Amato1987}	&  $1550$ 	& $0.3$	& 2.5 & $4$\\ 
		\hline
		\CCO\ 	\cite{Sonnenschein2019,Kshetrimayum2020}	& $13500$  	&  $3.0$	& 1.7  &16.2\\ 
		\hline
	\end{tabular} 
	\caption{ 
	Comparison of the low temperature thermodynamic properties of \CCO\ 
	with the conventional metal Cu, and the heavy fermion material CeCu$_6$.
	The parameters shown are the linear coefficient of specific heat $C(T\to 0) = \gamma T$; 
	the paramagnetic susceptibility  $\chi(T\to 0) = \chi_0$, effective moment 
	$\mu_{\sf eff}$, %[Eq.~(\ref{eq:effective.moment})], 
	and the Wilson ratio $R_W$ [Eq.~(\ref{eq:wilson.ratio})].
	%
	% $W = \frac{ \pi^2 k_B^2 }{ J(J+1) (g\mu_B)^2 } \frac{\chi_0}{\gamma}$.
	}
	\label{tab:heavy.fermions}
\end{table}

%%%%%%%%%%%%%%%%%%%%%%%%%%%%%%%%%%%%%
\subsection{Thermodynamic properties}
%%%%%%%%%%%%%%%%%%%%%%%%%%%%%%%%%%%%%
\label{section:thermodynamics.cco}

The thermodynamic properties of \CCO\ distinguish it as a frustrated
magnet in which spins interact, but continue to fluctuate down to 
very low temperatures.
The magnetic susceptibility of \CCO\ displays a Curie--law 
behaviour 
\begin{eqnarray}
\chi^{-1}(T)  \approx \frac{T - \theta_{\sf CW}}{C}
\end{eqnarray}
down to temperatures of a few Kelvin.
A positive Curie--Weiss temperature of  
\begin{eqnarray}
\theta_{\sf CW} = 2.35\ \text{K} \; , 
\end{eqnarray}
consistent with dominant FM interactions, was reported by 
Balz {\it et al.}~\cite{Balz2016,Balz2017-PRB95}, 
with the slightly higher value of $\theta_{\sf CW} = 4.1(6)\ \text{K}$ 
being reported by Balodhi and Singh~\cite{balodhi17-PRM1}.
Both groups find a value of $C$ consistent with an effective moment 
\begin{eqnarray}
\mu_{\sf eff} \approx 1.7\ \mu_B
\end{eqnarray}
at each Cr$^{5+}$ site, as would be expected for a spin--1/2 moment, 
assuming a Land\'e factor $g = 2$.

%%%%%%%%%%%%%%%%%%%%%%%%%%%%%%%%%%%%%

At low temperatures, the magnetisation of \CCO\ rises rapidly 
in applied magnetic field, consistent with a gapless 
ground state \cite{Balz2017-PRB95,Kshetrimayum2020}.  
At low fields, the magnetisation is found to be nearly linear in $B$, 
with an associated susceptibility
$$
\chi_0  = 3.0\ \text{emu/mol\ Oe} \; ,
$$
at $T= 1.8\ \text{K}$; a large value even by comparison with 
heavy Fermion materials [Table~\ref{tab:heavy.fermions}].
% \cite{hewson-book,fulde-book}.
%
This behaviour stands in marked contrast with SrCr$_2$O$_8$, 
where the energy gap from the dimerized ground--state to the 
lowest--lying triplet excitation ensures that the magnetisation remains 
zero up to a field $H_{c1} = 30.4\ \text{T}$ \cite{quintero-castro10,wang16-PRL116}.
The magnetization of \CCO\ is also broadly independent of the direction
in which field is applied, and saturates at a relatively 
low field, with a sharp kink in $M(H)$ observed at a scale of $H \sim 1\ \text{T}$, 
and complete saturation is observed for 
fields no greater than $H \sim 10\ \text{T}$, at a temperature of 
$1.8\ \text{K}$ \cite{Balz2016,Balz2017-PRB95}.

%%%%%%%%%%%%%%%%%%%%%%%%%%%%%%%%%%%%%

Heat capacity measurements carried out in zero field by Balz {\it et al.} 
\cite{Balz2016,Balz2017-PRB95} offer a 
consistent picture, with a dramatic kink in $C(T)$ at $T \approx 500\ \text{mK}$, 
a broad maximum at $T \approx 3.1\ \text{K}$, and no evidence for either 
magnetic order, or the opening of a spin--gap, down to 
\mbox{$T = 300\ \text{mK}$}.
The measured values of $C(T)$ at this temperature are consistent with a 
high density of low--lying excitations, with $C/T$ achieving values 
\mbox{$\sim 14\ \text{J\ mol$^{-1}$\ K$^{-2}$}$}  \cite{Balz2017-PRB95} 
which, again, are large even by the standard of heavy--fermion materials 
[Table~\ref{tab:heavy.fermions}].  
Later experiments, reported by Sonnenschein {\it et al.}~\cite{Sonnenschein2019}, 
extended measurements down to $T = 37\ \text{mK}$, finding a nearly 
linear specific heat $C(T) \approx \gamma T$ over the temperature range 
$100\ \text{mK} \lesssim T \lesssim 500\ \text{mK}$, 
with 
$$
\gamma = 13.5\ \text{J\ mol$^{-1}$\ K$^{-2}$} \; .
$$
For $37\ \text{mK} < T \lesssim 100\ \text{mK}$, measurements find  
$C(T) < \gamma T$, showing a slight suppression relative to a purely 
linear behaviour, but no evidence for a gap to excitations.
Meanwhile, at higher $T$, in the absence of magnetic field, $C/T$ 
is a monotonically--decreasing function of temperature \cite{Balz2017-PRB95}.
Qualitatively similar results for $C(T)$ at higher temperatures 
were also reported by Balodhi and Singh \cite{balodhi17-PRM1}, with 
the caveat that the measured values differ by a numerical factor 
$\sim 2$ between the two groups.
%
% {\bf [This \ldots]}

%%%%%%%%%%%%%%%%%%%%%%%%%%%%%%%%%%%%%

In applied magnetic field, the values of $C/T$ found at low
temperatures steadily decrease, and plots of $C/T$ acquire a shoulder
at $T \lesssim 1\ \text{K}$ \cite{Balz2017-PRB95}. 
A qualitative change occurs for $B \approx 1\ \text{T}$, when a downturn 
become visible in $C/T$ at low temperatures, consistent with suppression
of low--lying excitations by the opening of a gap.
Attempts to model the field--temperature dependence of $C(T)$
as the sum of $T^3$ contribution from phonons, and Shottky anomaly 
(broad peak) coming from spin excitations, meet with some success.
However this approach fails to explain the relatively high density of 
low--lying excitations seen in experiment, especially if the Shottky peak
is associated with the gap measured in inelastic neutron scattering, 
as described below \cite{Balz2017-PRB95}.

%%%%%%%%%%%%%%%%%%%%%%%%%%%%%%%%%%%%%
\subsection{Macroscopic dynamics}
%%%%%%%%%%%%%%%%%%%%%%%%%%%%%%%%%%%%%
\label{section:macroscopic.dynamics.cco}

Meanwhile, measurements of AC susceptibility $\chi(\nu, T)$, for 
frequencies $\nu \lesssim 20\ \text{kHz}$, exhibit a broad peak 
at a temperature \mbox{$T^* \sim 330\ \text{mK}$} \cite{Balz2016}.
Under other circumstances, this might hint at spin--glass freezing.
However Cole--Cole plots of the real against imaginary parts of 
$\chi$ remain semi--circular for temperatures both above and below 
$T^*$, suggesting that a single timescale governs the macroscopic 
relaxational dynamics of \CCO, even at low temperatures \cite{Balz2016}.
Consistent with this, $\mu$SR measurements reveal persistent spin 
fluctuations down to $19\ \text{mK}$, with the measured 
relaxation rates increasing with decreasing temperature, and 
saturating for $T < T^*$ \cite{Balz2016}.

%%%%%%%%%%%%%%%%%%%%%%%%%%%%%%%%%%%%%

The picture painted by these experiments is one of a magnet 
whose moments continue to fluctuate down to temperatures two orders of 
magnitude smaller than the characteristic scale of exchange interactions.
A clear hierarchy of other temperature and field scales emerges, 
with both $\theta_{\sf CW}$, and the broad maximum in $C(T)$, picking
out a scale of $T \approx 3\ \text{K}$; AC susceptibility and $\mu$SR 
revealing changes in dynamics at $T \approx 330\ \text{mK}$;  
and magnetisation and heat capacity suggesting a change of phase
at $B \approx 1\ \text{T}$.
This behaviour would be hard to reconcile with any simple 
paramagnet, and is entirely consistent with a quantum spin liquid.
But from these measurements alone, it is difficult to confirm  
the collective nature of spin fluctuations, or to say what kind of spin liquid 
might be found in \CCO.

%%%%%%%%%%%%%%%%%%%%%%%%%%%%%%%%%%%%%
\subsection{Neutron scattering}
%%%%%%%%%%%%%%%%%%%%%%%%%%%%%%%%%%%%%
\label{section:neutron.scattering.cco}

More insight into the nature of magnetic correlations in \CCO\ can 
be gained through the structure factor measured in neutron scattering.
Neutron scattering has been carried out on both powder and 
single--crystal samples of \CCO, using a variety of different 
neutron instruments \cite{Balz2016,Balz2017-PRB95,Sonnenschein2019}.
Elastic scattering fails to reveal any magnetic Bragg peaks down to 
$90\ \text{mK}$, consistent with the absence of any other signals of 
long--range magnetic order.
At the lowest energies experiments reveal instead a 
quasi--elastic signal, extending up to $\sim 0.2\ \text{meV}$.
This quasi--elastic signal is essentially independent of ${\bf q}$,  
on the scale of the BZ, and has been attributed to incoherent scattering 
from randomly distributed nuclear isotopes \cite{Balz2016}.
Magnetic scattering is inelastic in character, and highly structured,
confirming the collective nature of spin fluctuations.
Strong scattering for $\omega \lesssim 0.4\ \text{meV}$ echoes 
the characteristic temperature scale seen in thermodynamic 
measurements.
However this is clearly not the only energy scale in the problem; 
a further strong signal is seen at $\omega \sim 1\ \text{meV}$, 
accompanied by a broad background of scattering extending up to 
$\omega \sim 1.5\ \text{meV}$.
Consistent with the lack of magnetic Bragg peaks, no hint 
is found of the spin waves which would be associated 
with the breaking of spin--rotation symmetry.

%%%%%%%%%%%%%%%%%%%%%%%%%%%%%%%%%%%%%

The strong ${\bf q}$-- and $\omega$--dependence of scattering is 
evident in energy cuts through the measured dynamical structure factor 
$S({\bf q}, \omega)$, reproduced in Fig.~\ref{fig:comparison.with.experiment}.
Scattering at low energies provides the most information about correlations
in the ground state, but unfortunately is obscured by the incoherent signal 
for $\omega \lesssim 0.2\ \text{meV}$.
None the less, measurements at $0.25\ \text{meV}$ [Fig.~\ref{fig:Exp_025meV}] 
reveal that low--energy fluctuations are strongly ${\bf q}$--dependent, with 
hints of a ring--like structure centered on $(2, 0, 0)$.
(Much stronger scattering seen at other zone centers reflects 
phonons \cite{Balz2016}).
The scattering at an intermediate energy of $0.65\ \text{meV}$ [Fig.~\ref{fig:Exp_065meV}] is also strongly ${\bf q}$--dependent, 
but reveals a completely different kind of correlation.
In this case, instead of a ``ring'' at $(2, 0, 0)$, experiments
suggest a ``bow-tie'' centered on $(1, 0, 0)$.    
And the same bow--tie pattern is more clearly visible
at the relatively high energy of $0.9\ \text{meV}$ 
[Fig.~\ref{fig:Exp_090meV}].
%

%%%%%%%%%%%%%%%%%%%%%%%%%%%%%%%%%%%%%

It is 	worth noting, that energy values in MD simulations have been slightly shifted, 
to allow for a qualitative comparison of the scattering pattern to INS experiments. 
While the main features, namely ``rings'' at low energy and ``bow-ties'' at higher energy,
could be reproduced well within our \mbox{(semi-)classical} method,
they occur at slightly different energies.
This renormalisation is presumably due in the approximations inherent 
in both the BBK model of \CCO\ (which neglects anisotropic exchange 
interactions), and our \mbox{(semi--)classical} treatment of its dynamics.
We will revisit this last point in a coming work \cite{shimokawa-in-prep}.

%%%%%%%%%%%%%%%%%%%%%%%%%%%%%%%%%%%%%
\subsection{The BBK model}%: a minimal microscopic model of magnetism in \CCO}
%%%%%%%%%%%%%%%%%%%%%%%%%%%%%%%%%%%%%
\label{section:bbk.model.cco}

Taken together, both thermodynamic and dynamical measurements
of \CCO\ are consistent with the existence of a gapless (or nearly gapless) 
quantum spin liquid at low temperatures.   
This phenomenology contains elements which are familiar from the 
behaviour of other magnets, such as the bow--tie patterns observed 
in scattering at higher energies, reminiscent of the pinch--points 
observed in Coulombic phases \cite{Henley2010}.
However, while there are plenty of examples of studies of two--dimensional
quantum spin liquids \cite{balents10-Nature464,Savary_2016,anderson73-MatResBull8,lee08-Science321,shimizu03-PRL91,knolle19,zhou17,masutomi04-PRL92,han12-Nature492,kitaev06-AnnPhys321,jackeli09-PRL102,banerjee16-NatMater15,baek17-PRL119,winter17,hermanns17}, 
no model or material provides a complete analogue 
to the behaviour of \CCO, even at a qualitative level.
And the fact that \CCO\ displays such different behaviour 
on different energy scales means that a low--energy effective 
theory alone cannot unlock all of its secrets.
To make further progress in understanding \CCO, 
a microscopic model is therefore needed.

%%%%%%%%%%%%%%%%%%%%%%%%%%%%%%%%%%%%%

The simplest model one can consider for \CCO\ is one 
in which both orbital effects, and spin--orbit coupling, are neglected, 
so that each Cr$^{5+}$ ion is treated as a spin--1/2 moment, 
interacting through Heisenberg interactions.
The neglection of other terms allowed by lattice symmetry, such as 
Dzyaloshinskii--Moriya (DM) interactions, finds some justification 
in the 3d nature of the magnetic electrons, and the lack of 
magnetic anisotropy observed in experiment.
However even a minimal, $SU(2)$--symmetric model, will 
have many different parameters, since the BBK lattice 
supports seven inequivalent first--neighbour bonds 
\cite{Balz2016,Balz2017-PRB95}.
And ultimately, the extent to which one can parameterise such 
a complex model from experiment, and use it to 
understand the novel physics of \CCO, becomes an empirical question.

%%%%%%%%%%%%%%%%%%%%%%%%%%%%%%%%%%%%%

Fortunately, the low saturation field of \CCO\ means that it is 
possible to parameterise a minimal, microscopic model for its 
magnetism from inelastic neutron scattering (INS) experiments 
on its field--polarised state.
These reveal gapped, two--dimensional spin--wave excitations 
%[Fig.~\ref{fig:DynamicsHighField}] 
(discussed in Section~\ref{sec:dynamics.in.high.field}, below) which, within 
experimental resolution, are adequately described by a Heisenberg 
model for a single bilayer \cite{Balz2016,Balz2017-PRB95}, 
\begin{equation}
	{\cal H}_{\sf BBK} 
	= \sum_{\langle i j \rangle} J_{ij} {\bf S}_i \cdot {\bf S}_j -  B \sum_i S^z_i 
	\; , 
\end{equation}
introduced as Eq.~(\ref{eq:H.BBK}).
%with parameters shown in Table~\ref{tab:experimental.parameters}.
%
This spin--1/2 BBK model %of \CCO\ introduced by Balz {\it et al.} 
has 5 first--neighbour couplings within a single bilayer [Fig.~\ref{fig:BBK.model}], 
of which 3 are ferromagnetic (FM) and 2 antiferromagnetic (AF) 
\mbox{[Table~\ref{tab:experimental.parameters}]}.
The strongest interactions, $J_{21} \approx -0.8\ \text{meV}$
and $J_{22} \approx -0.3\ \text{meV}$, are FM and occur within  
the triangular plaquettes of the BBK lattice.
Meanwhile, the couplings between these plaquettes are 
FM in the interplane direction, $J_0 \approx -0.1\ \text{meV}$,
and AF within the breathing--Kagome planes; $J_{31},  
J_{32} \approx 0.1\ \text{meV}$.

%%%%%%%%%%%%%%%%%%%%%%%%%%%%%%%%%%%%%

The picture of \CCO\ which emerges is therefore one of strongly--coupled FM plaquettes, 
which are in turn coupled AF %antiferromagnetically 
within a single breathing--Kagome plane, and FM %ferromagnetically 
between the two layers. %planes of the bilayer. %BBK lattice.

%%%%%%%%%%%%%%%%%%%%%%%%%%%%%%%%%%%%%
%   Table 1 - exchange paramters
%%%%%%%%%%%%%%%%%%%%%%%%%%%%%%%%%%%%%

\begin{table}
	\centering
	\begin{tabular}{|P{1.5cm}|P{2.7cm}|P{1.5cm}|}
		\hline
		    ${\cal H}_\text{\sf BBK}$   [Eq.~(\ref{eq:H.BBK})] 	
		&  \CCO\ \cite{Balz2016,Balz2017-PRB95} 
		&  ${\cal H}_\text{\sf HC}$  [Eq.~(\ref{eq:H.HC})] \\ 
		\hline
		$ J_\text{0} $ 	& $-0.08(4)\ \text{meV}$ 			& $J_\text{1}$ 		\\ 
		\hline
		$ J_\text{21} $ 	& $-0.76(5)\ \text{meV}$ 			& 	-- 		 	\\ 
		\hline 
		$ J_\text{22} $ 	& $-0.27(3)\ \text{meV}$ 			&	--  			\\ 
		\hline 
		$ J_\text{31} $ 	& $\phantom{-}0.09(2)\ \text{meV}$ 	&	$J_\text{2}$ 	\\ 
		\hline 
		$ J_\text{32} $ 	& $\phantom{-}0.11(3)\ \text{meV}$ 	&	$J_\text{2}$ 	\\ 
		\hline 
	\end{tabular} 
	\caption{ 
	Exchange interactions within the bilayer breathing--Kagome (BBK) 
	model of \CCO, [Eq.~(\ref{eq:H.BBK})], as estimated from inelastic neutron 
	scattering in high magnetic field \cite{Balz2016,Balz2017-PRB95}.
	Bond indices $J_{ij}$, following the conventions of Balz {\it et al.} 
	\cite{Balz2016,Balz2017-PRB95}, are defined in Fig.~\ref{fig:BBK.model}.
	Also listed are the corresponding  interactions within an effective 
	spin--3/2 honeycomb--lattice model [Eq.~(\ref{eq:H.HC})], with 
	bond indices defined in Fig.~\ref{fig:HC.model}.
	}
	\label{tab:experimental.parameters}
\end{table}

%%%%%%%%%%%%%%%%%%%%%%%%%%%%%%%%%%%%%
\subsection{Theoretical work based on the BBK model}
%%%%%%%%%%%%%%%%%%%%%%%%%%%%%%%%%%%%%
\label{section:theory.bbk.cco}

While even this, minimal model of \CCO\ may seem alarmingly complicated, 
it does provide a concrete microscopic starting point for understanding 
experiment, and here some good progress has already been made.

%%%%%%%%%%%%%%%%%%%%%%%%%%%%%%%%%%%%%

A straightforward, but informative exercise, is to use linear spin wave (LSW)
theory to track the evolution of $S({\bf q},\omega)$ with magnetic field 
\cite{Balz2017-PRB95}.
Mean--field theory for Eq.~(\ref{eq:H.BBK}), 
using experimental parameters \mbox{[cf. Table~\ref{tab:experimental.parameters}]}, 
predicts that the saturated state is stable for $B > 1.1\ \text{T}$.
Meanwhile LSW for Eq.~(\ref{eq:H.BBK}) gives a qualitatively reasonable 
description of the scattering observed in experiment, and its field evolution 
for $B \gtrsim 1\ \text{T}$  \cite{Balz2017-PRB95}.

%%%%%%%%%%%%%%%%%%%%%%%%%%%%%%%%%%%%%

More sophisticated methods have been applied to the spin--liquid 
ground state found for $T=0$.
Pseudofermion functional renormalisation group (PFFRG) calculations 
for the BBK model Eq.~(\ref{eq:H.BBK}) \cite{Balz2016}, 
find a disordered ground state for a range of parameters centered 
on those found in experiment. %\mbox{[cf. Table~\ref{tab:experimental.parameters}]}. 
And, encouragingly, PFFRG calculations of the static structure factor, 
$S({\bf q},\omega =0)$, exhibit a ring--like structure, similar to that observed
in experiment.
The stability of this spin liquid state is, interestingly, found to depend  
on the differences in parameters between the different planes of the 
bilayer, with symmetric choices leading to ordered ground states.

%%%%%%%%%%%%%%%%%%%%%%%%%%%%%%%%%%%%%

Subsequent tensor--network calculations, based on ``projected entangled 
simplex states'' (PESS)  \cite{Kshetrimayum2020}, also predict a spin--liquid
ground state.
This approach has been used to estimate the ground--state  
magnetisation %the BBK model as a function of magnetic field, 
for parameters taken from \CCO\ \mbox{[cf. Table~\ref{tab:experimental.parameters}]}. 
At low fields, the magnetisation is found to be nearly linear in $B$, 
with associated susceptibility
$$
\chi_{\sf  PESS} = 1.1731\ \text{$\mu_B$/Cr$^{5+}$ T} = 3.93\ \text{emu/mol\ Oe} \; ,
$$
about 30 \% larger than the value observed in experiments
carried out at $T = 1.8\ \text{K}$ \cite{Balz2017-PRB95}.
The same calculations find a saturation 
field of $B \approx 1\ \text{T}$ \cite{Kshetrimayum2020}.

%%%%%%%%%%%%%%%%%%%%%%%%%%%%%%%%%%%%%
%  Fig. 3 - honeycomb lattice model
%%%%%%%%%%%%%%%%%%%%%%%%%%%%%%%%%%%%%

\begin{figure}[t]
	\centering
  		\includegraphics[width=0.26\textwidth]{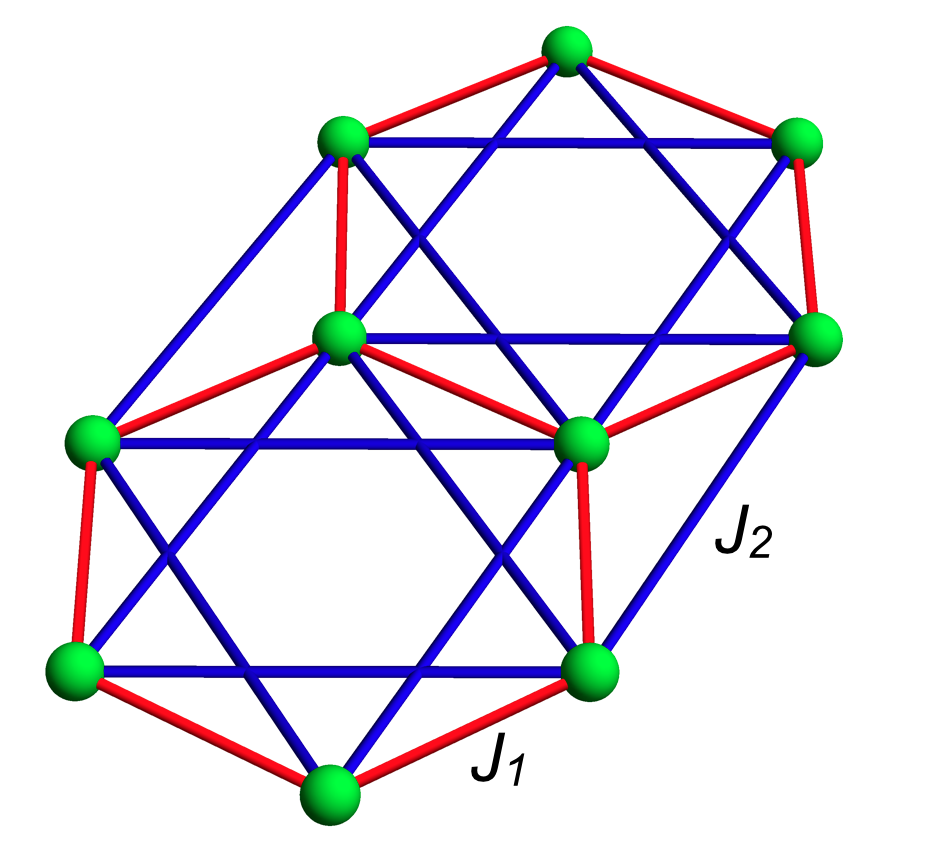}
	\caption{ 
	Honeycomb lattice realised by triangular plaquettes of 
	spins in \CCO, cf. Fig.~\ref{fig:BBK.model}.
	Labelling of bonds identifies parameters of the effective spin--3/2 
	model, Eq.~(\ref{eq:H.HC}).
	Experimental estimates of these parameters can be found in 
	Table~\ref{tab:experimental.parameters}.
	}
	\label{fig:HC.model}
\end{figure}

%%%%%%%%%%%%%%%%%%%%%%%%%%%%%%%%%%%%%

A phenomenological approach to the low--temperature properties 
of \CCO\ has also been developed by 
Sonnenschein {\it et al.} \cite{Sonnenschein2019}.
Taking inspiration from the broad continuum found in inelastic neutron 
scattering \cite{Balz2016}, and the (nearly) linear specific 
heat at low temperatures, these authors introduce a model of 
non--interacting Fermionic spinons hopping on a (decorated) 
honeycomb lattice.  
This model is not derived directly from Eq.~(\ref{eq:H.BBK}), but 
respects the symmetries of the BBK lattice, and is parameterised 
so as to reproduce the energy scales and some of the key  qualitative 
features of the scattering 
%and energy scales of $S({\bf q}, \omega)$, as 
seen in experiment.
It has three  two--fold degenerate bands; the lower  occupied 
band has a nearly circular  hole--like Fermi surface, while 
the unfilled high--energy bands mirror the dispersion of graphene.
This approach which corresponds to a $U(1)$ QSL, reproduces the 
rings of scattering in $S({\bf q}, \omega)$ at low energy 
$\hbar \omega \sim 0.15\ \text{meV}$, and suggests features 
at intermediate energy $\hbar \omega \sim 0.85\ \text{meV}$ 
which contain at least relics of pinch--point structure.

%%%%%%%%%%%%%%%%%%%%%%%%%%%%%%%%%%%%%

By introducing further phenomenological parameters for pairing of  
spinons, Sonnenschein {\it et al.} are also able 
to model the deviation from $T$--linear specific heat found for 
$35\ \text{mK} < T < 100\ \text{mK}$  \cite{Sonnenschein2019}.
The best fits are found for an f--wave gap, 
leading to a Fermi surface with Dirac points, and implying 
a $\mathbb{Z}_2$ spin liquid ground state \cite{Senthil2000}.  
This spinon pairing also ``cures'' a seeming 
contradiction with experiment, since a $U(1)$ QSL would show 
additional, divergent, contributions to $C(T)$ coming from gapless gauge 
fluctuations \cite{Senthil2004,Montrunich2005}.

%%%%%%%%%%%%%%%%%%%%%%%%%%%%%%%%%%%%%
\subsection{Effective %$J_1$--$J_2$ 
honeycomb lattice model}
%%%%%%%%%%%%%%%%%%%%%%%%%%%%%%%%%%%%%
\label{section:J1J2.model.cco}

Phenomenology aside, published theory for quantum effects 
in \CCO\ are limited to its ground state. Also, very little is known 
about the properties of the spin--1/2 BBK model at finite temperature.
Some progress has however been made in the classical limit, by 
considering a simplified, spin--3/2 model.

%%%%%%%%%%%%%%%%%%%%%%%%%%%%%%%%%%%%%

The strongest couplings in the BBK model of \CCO\ are the FM interactions 
within the triangular plaquettes of the BBK lattice, 
\mbox{$J_\text{21}  \approx -0.76\ \text{meV}$} 
and 
\mbox{$J_\text{22}  \approx -0.27\ \text{meV}$} 
(cf. shaded green triangles in Fig.~\ref{fig:BBK.model}).
This immediately suggests a simplification, namely treating each plaquette 
as a spin--3/2 moment on the medial, honeycomb lattice
\begin{equation}
{\cal H}_{\hexagon}	=  J_1 \sum_{\langle ij \rangle_1} \ {\bf S}_i \cdot {\bf S}_j 
				 + J_2 \sum_{\langle ij \rangle_2}  {\bf S}_i \cdot {\bf S}_j 
				  - B^z  \sum_i S^z_i  
				   \; , 
\label{eq:H.HC}
\end{equation}
where the first--neighbour coupling $J_1 \approx -0.08\ \text{meV}$ corresponds
to the FM inter--layer coupling $J_\text{0}$ in the original BBK model, 
while the second--neighbour coupling $J_2  \approx 0.10\ \text{meV}$ 
can be taken to be the mean of the AF intra--layer interactions $J_{\sf eff}$
as
\begin{equation}
	J_{\sf eff} \equiv (J_{31} + J_{32})/2 \; .
\label{eq:Jeff}
\end{equation}

Classically at least, this simplified model can be expected to give 
a reasonable account for the properties of \CCO\ for 
$T, \omega < J_\text{21}, J_\text{22}$. 

%%%%%%%%%%%%%%%%%%%%%%%%%%%%%%%%%%%%%

\begin{figure}[t]
	\centering
  		\includegraphics[width=0.8\columnwidth]{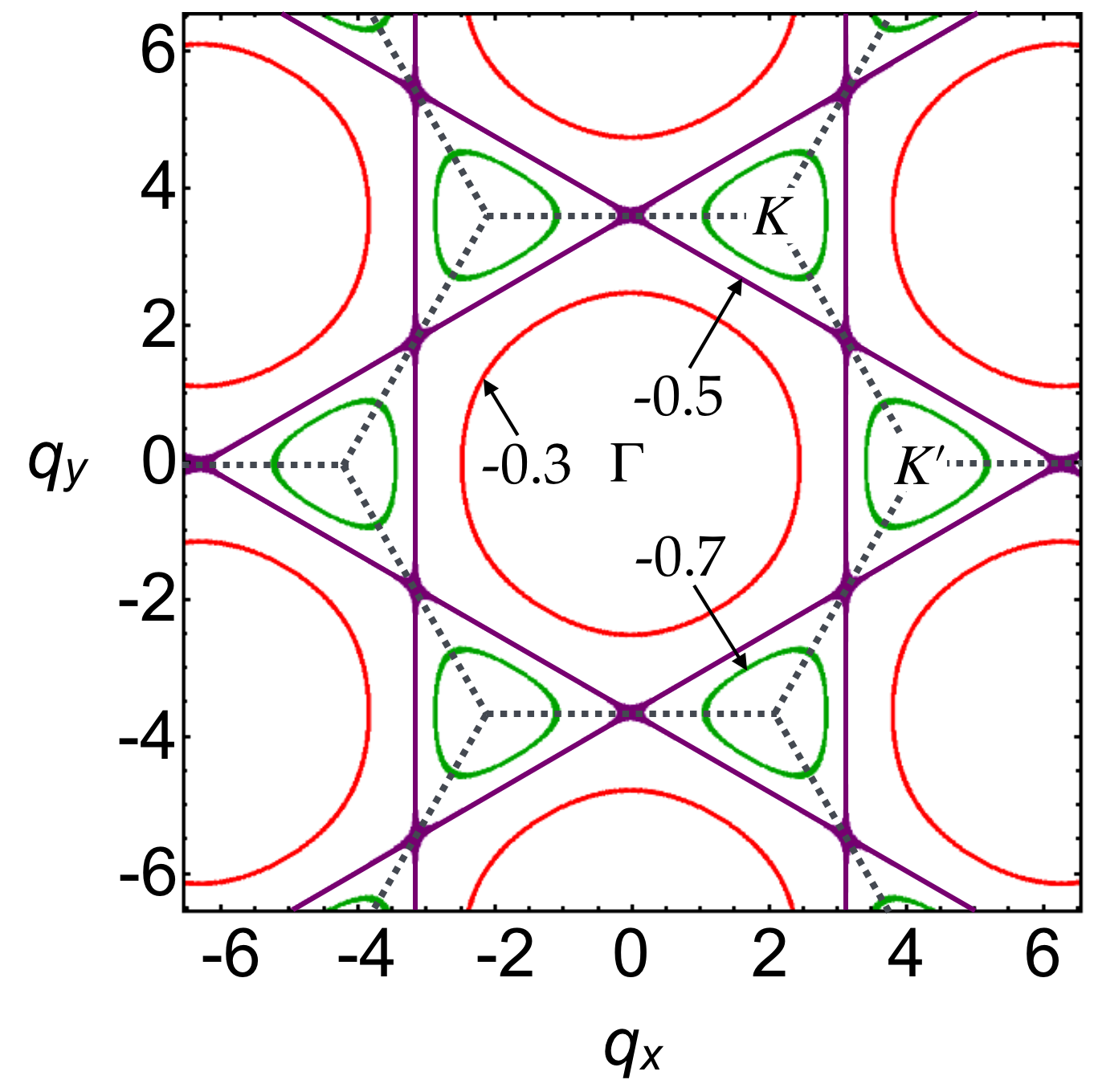}
	\caption{ 
	Loci of degenerate classical ground states in the 
	$J_1$--$J_2$ Heisenberg model on a honeycomb lattice, Eq.~(\ref{eq:H.HC}).
	Depending on the ratio of parameters, these loci may 
	take the form of ``large'' rings centered on $\Gamma$ 
	\cite{rastelli79-PhysicaB+C97,fouet01-EPJB20,okumura10-JPSJ79}, 
	or ``small'' rings centered on $K$ \cite{Mulder2010,biswas18}.
	Here, results are shown for $J_2/J_1 = -0.3$ (red line), $-0.5$ (purple) 
	and $-0.7$ (green).
}
	\label{fig:ring.degeneracy}
\end{figure}

Working with a spin--3/2 model on a honeycomb lattice has the 
added advantage that it connects \CCO\ with an established literature on 
unconventional magnetic phases in honeycomb--lattice models 
with competing interactions \cite{rastelli79-PhysicaB+C97,fouet01-EPJB20,Mulder2010,okumura10-JPSJ79}.
The $J_1$--$J_2$ Heisenberg model Eq.~(\ref{eq:H.HC}), is a special
case of the $J_1$--$J_2$--$J_3$ Heisenberg model on honeycomb lattice, 
where third--neighbour interactions are also taken into account.  
The classical ground states of this parent model are distinguished 
by the competition between the different forms of coplanar spiral order.
And, for $J_3 = 0$, where spirals with different ordering wave vectors meet, 
a continuous manifold of ground states is formed, with wave vectors belonging 
to a ring--like locus of points in $\mathbf{q}$--space [Fig.~\ref{fig:ring.degeneracy}].

%%%%%%%%%%%%%%%%%%%%%%%%%%%%%%%%%%%%%

Depending on parameters, this ring can be centered 
on the $\Gamma$ point at the center of the Brilluoin zone (BZ) 
\cite{rastelli79-PhysicaB+C97,fouet01-EPJB20,okumura10-JPSJ79}, 
or on the K--point at the corner of the BZ \cite{Mulder2010}.
Both of these cases have already been studied for AF $J_1$.
In the case of the ``large'' ring centered on $\Gamma$, spin liquids
with ``ring'' or ``pancake'' motifs are found at high temperatures, while 
at very low temperatures, a phase transition is identified into a state 
breaking the 3--fold rotation symmetry of the lattice \cite{okumura10-JPSJ79}.
Very similar results are found for the ``small'' ring centered on $K$,
where the transition into the ordered phase was identified as 
a 3--state Potts transition \cite{Mulder2010}.

%%%%%%%%%%%%%%%%%%%%%%%%%%%%%%%%%%%%%
%  Fig. 3 - honeycomb lattice model
%%%%%%%%%%%%%%%%%%%%%%%%%%%%%%%%%%%%%

%%%%%%%%%%%%%%%%%%%%%%%%%%%%%%%%%%%%%

Biswas and Damle have extended the analysis of Eq.~(\ref{eq:H.HC})
to the case of FM interactions $J_1 < 0$, through a combination 
of classical MC simulation, MD simulation, spin--wave theory, 
and large--N calculations 
\cite{biswas18}. 
Considering the limit $B=0$, a ratio of parameters $J_1/J_2$
motivated by \CCO\ \mbox{[cf. Table~\ref{tab:experimental.parameters}]},
they find that fluctuations select a discrete set of spiral states from the 
``small'' ring of ground states centered on ${\bf q} = K$.
Corresponding MC simulations of Eq.~(\ref{eq:H.HC}) reveal a 3--state
Potts transition into a state breaking lattice rotation symmetry at 
$T^* \approx 66\ \text{mK}$, consistent with \cite{Mulder2010}.
This is accompanied by critical slowing down in the dynamics found in MD simulation,
and a large but finite spiral-correlation length. 
Close to $T^*$, the equal--time structure factor $S({\bf q})$ shows 
a locus of highly--degenerate states near the K-points of the Brillouin zone, 
consistent with the ``small'' ring.
At higher temperatures, $S({\bf q})$ instead shows a larger ring
of scattering, near the zone boundary, similar to that observed 
in \CCO\ \cite{Balz2016}.

%%%%%%%%%%%%%%%%%%%%%%%%%%%%%%%%%%%%%

Quantum effects within the $J_1$--$J_2$ honeycomb lattice model 
have only been studied in detail for $S=1/2$.
In this case the phase breaking lattice rotation symmetry 
is proposed to be a valence bond solid (VBS) \cite{Mulder2010}, 
and QSL states have been proposed elsewhere 
for both FM \cite{fouet01-EPJB20} and AF $J_1$ \cite{fouet01-EPJB20}.
In the case of $S=3/2$, spin--wave estimates suggest that spiral 
order is unstable for parameters relevant to \CCO\ \cite{biswas18}, 
but do not reveal the nature of any competing QSL.

%%%%%%%%%%%%%%%%%%%%%%%%%%%%%%%%%%%%%%
%  Fig X - finite-temperature phase diagram
%%%%%%%%%%%%%%%%%%%%%%%%%%%%%%%%%%%%%%

\begin{figure*}[t]
	\centering
  		\includegraphics[width=0.99\textwidth]{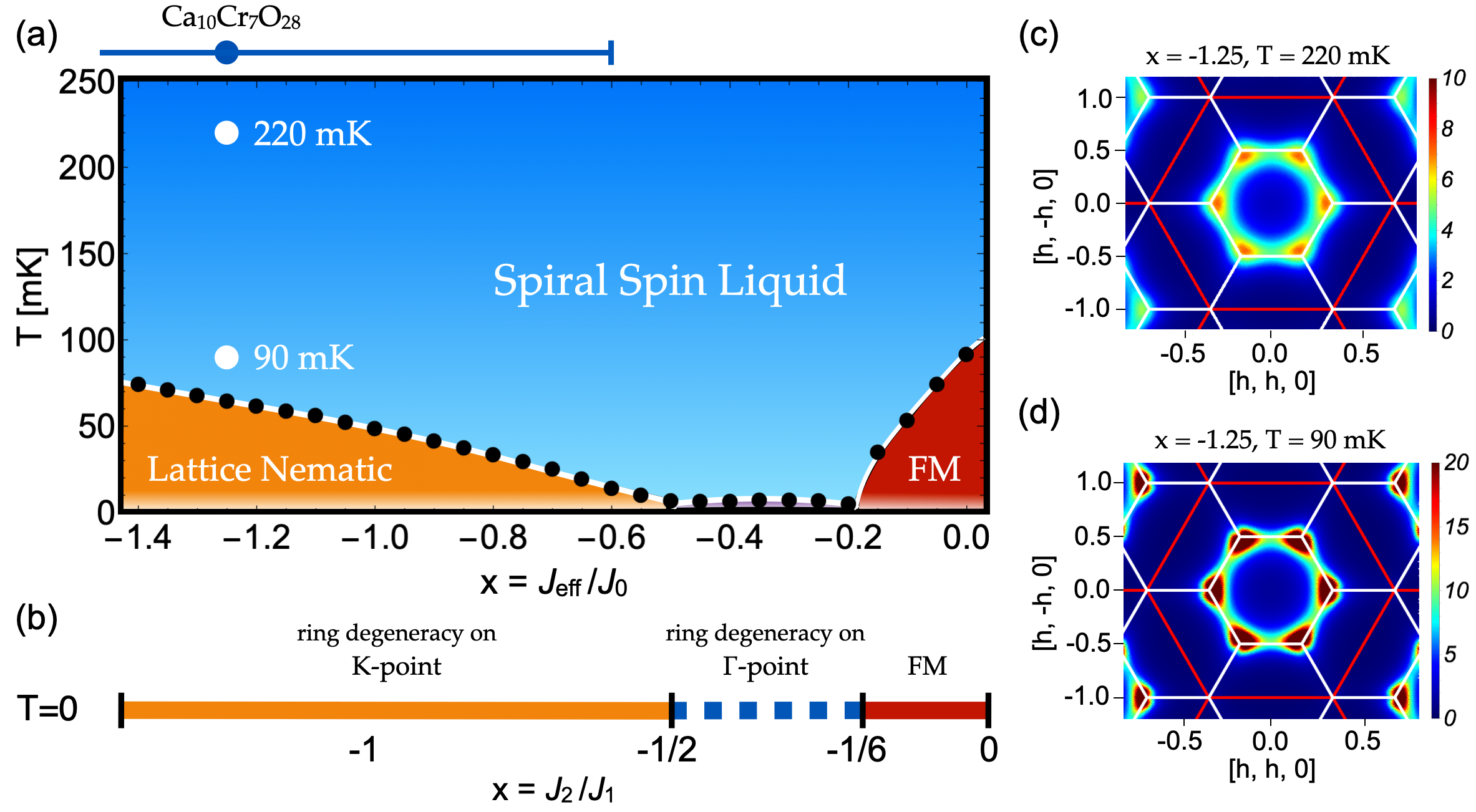}
	\caption{
		(a) Finite--temperature phase diagram of the BBK model, 
		Eq.~(\ref{eq:H.BBK}), showing wide extent of spin liquid phase.  
		Result are shown as a function of $x = J_{\sf eff}/J_0$ [Eq.~(\ref{eq:Jeff})], 
		and all other 
		parameters taken from experiment [Table~\ref{tab:experimental.parameters}].	
		The temperatures at which inelastic neutron scattering (INS) 
		experiments were carried out on \CCO\ are shown with white dots, 
		at $x = -1.25$ \cite{Balz2016, Balz2017-PRB95}. 
		(b) Classical ground--state phase diagram of the corresponding  
		honeycomb-lattice model, Eq.~(\ref{eq:H.HC}), as a function of 
		$x = J_2/J_1$, following 
		[\onlinecite{rastelli79-PhysicaB+C97,fouet01-EPJB20, Mulder2010}]. 
		(c) Equal--time structure factor $S({\bf q})$ at for $x=-1.25$,  
		$T = 220\ \text{mK}$; the boundary of the Brillouin Zone (BZ) 
		is shown with   white lines.
		Strong correlations are observed near the boundary of the $1^{st}$ BZ, 
		with the highest intensity centered on the K--points at zone corners.
		Form factors associated with magnetic correlations have the 
		periodicity of the $4^{th}$ BZ, shown here with red lines.
		(d) Equivalent results at $x = -1.25$, $T = 90\ \text{mK}$, showing 
		how scattering from the ``small'' ring centered on K-points becomes 
		sharper at low temperatures.
		All results were obtained from classical Monte Carlo (MC) simulation of 
		Eq.~(\ref{eq:H.BBK}), as described in Sec.~\ref{sec:thermodynamics} and 
		Appendix~\ref{appendix:Numerics}, for a cluster of linear dimension 
		L = 48 (N = 13,824).
		Phase boundaries in (a) were determined from the  
                 sharp peak in specific heat $C(T)$ associated with either the breaking of 
                 lattice--rotation symmetry, or the onset of ferromagnetic (FM) fluctuations. 
		}
	\label{fig:thermodyn}
\end{figure*}

%%%%%%%%%%%%%%%%%%%%%%%%%%%%%%%%%%%%%
\subsection{Open questions}
%%%%%%%%%%%%%%%%%%%%%%%%%%%%%%%%%%%%%
\label{section:open.questions.cco}

While considerable progress has been made, the complex phenomenology 
of \CCO\ has yet to find any complete, or microscopically--grounded explanation.
On the experimental side, further tests of spin liquid properties  through, e.g.
thermal transport, could bring new insights.
And improvements in the resolution of inelastic scattering would also 
be very valuable, making it possible to better--constrain microscopic or 
phenomenological models, and to more accurately test proposals about the 
spin--liquid state.

%%%%%%%%%%%%%%%%%%%%%%%%%%%%%%%%%%%%%
% open questions
%%%%%%%%%%%%%%%%%%%%%%%%%%%%%%%%%%%%%

Meanwhile, obvious challenges for theory include: 

\begin{enumerate}[label=(\roman*)]

\item connecting the 
finite--energy and finite--temperature properties of \CCO\ with the spin--1/2 
BBK model, Eq.~(\ref{eq:H.BBK});

\item extending the analysis of this model 
to finite magnetic field ;

\item identifying the mechanism driving its 
low--temperature spin liquid state;

\item identifying interesting 
properties of the BBK model which may, as yet, be obscure in 
experimental data for \CCO.

\end{enumerate}

%%%%%%%%%%%%%%%%%%%%%%%%%%%%%%%%%%%%%
% comment on methods
%%%%%%%%%%%%%%%%%%%%%%%%%%%%%%%%%%%%%

In this Article, and the one which follows \cite{shimokawa-in-prep}, we
continue the project, begun in \cite{pohle-arXiv}, of addressing the points 
\mbox{(i)--(iv)} above.
The results in this first Article are drawn exclusively from (semi--)classical methods:
Monte Carlo simulation; linear spin-wave theory; and numerical intergration of 
equations of motion for spins, which we refer to as molecular dynamics (MD) simulation.
And before embarking on this journey, some comment is due on the validity
of using classical methods 
%to simulate a quantum magnet. 
to address questions such as these, in a quantum magnet.

%%%%%%%%%%%%%%%%%%%%%%%%%%%%%%%%%%%%%

\CCO\ is a highly--frustrated, quasi--two dimensional system, with spin--1/2 
moments, and a strong candidate for a quantum spin liquid (QSL) \cite{Balz2016}.   
So it might at first seem that there was little to be learnt from 
classical techniques.
None the less, experience with other models that support QSL, 
where exact (or numerically exact) quantum results are available for comparison ---
notably the Kitaev model \cite{Samarakoon2017}, and quantum 
spin ice \cite{Taillefumier2017,Benton2018-PRL121})  
--- teaches that, suitably interpreted, classical approaches yield a surprising 
amount of insight into both the correlations and dynamics of quantum 
spin liquids at finite temperature.
Moreover, classical simulations always bring meaningful advantage in terms 
of the size of the system that can be simulated, and the ease with which results 
can be interpreted.

%%%%%%%%%%%%%%%%%%%%%%%%%%%%%%%%%%%%%

Our approach in this Article will therefore be to pursue classical simulations 
of the BBK model, cautiously, correcting for bias where we can, and noting 
it where we can't.
To this end, we benchmark simulation results against both experiment, 
and known soluble limits of the model. 
The strength of this approach, as well as its ultimate limitations, will become 
apparent in the second Article, when we compare explicitly with the results 
of quantum simulations of the BBK model \cite{shimokawa-in-prep}.

%%%%%%%%%%%%%%%%%%%%%%%%%%%%%%%%%%%%%
\section{Thermodynamic properties of ${\cal H}_{\sf BBK}$}		
%%%%%%%%%%%%%%%%%%%%%%%%%%%%%%%%%%%%%
\label{sec:thermodynamics}

We begin our anlaysis of the spin--1/2 BBK model 
[Eq.~(\ref{eq:H.BBK})], by exploring its thermodynamic 
properties in the absence of magnetic field, using 
classical Monte Carlo (MC) simulation.
Here the goal is to better understand experiments carried out 
at finite temperature on \CCO, as described in 
Section~\ref{section:CCO.v.BBK.model}, and to link them 
with known theoretical results for the honeycomb lattice model 
[Eq.~(\ref{eq:H.HC})].

%%%%%%%%%%%%%%%%%%%%%%%%%%%%%%%%%%%%%%

Except where otherwise stated, simulations were carried out for parameters 
taken from \CCO\ [Table~\ref{tab:experimental.parameters}], for rhombohedral 
clusters of 
\begin{eqnarray}
N = 6 \times L^2 
\label{eq:N}
\end{eqnarray}
spins, subject to periodic boundary conditions.   
Simulations employed a local Metropolis update, within the heat--bath 
method, augmented by both over relaxation and parallel tempering steps.
Further details of the numerical techniques used can be found in 
Appendix~\ref{appendix:Numerics}.

%%%%%%%%%%%%%%%%%%%%%%%%%%%%%%%%%%%%%%

Key results are summarised in the finite--temperature phase diagram, 
Fig.~\ref{fig:thermodyn}.

%%%%%%%%%%%%%%%%%%%%%%%%%%%%%%%%%%%%%
\subsection{Symmetry--breaking at low temperatures} 		
%%%%%%%%%%%%%%%%%%%%%%%%%%%%%%%%%%%%%
\label{sec:3.state.Potts}

The first obvious questions to address are (i) how the different energy 
scales found in the BBK model manifest themselves in thermodynamic 
properties, such as heat capacity, and (ii) whether the model exhibits any 
kind of long--range order at low temperature.
This second question is very clearly motivated by the work of 
Biswas and Damle on an effective honeycomb--lattice model for \CCO, 
where a 3--state Potts transition into a state with broken lattice 
rotation symmetry is found for \mbox{$T^* \sim 64\ \text{mK}$} \cite{biswas18}.   

%%%%%%%%%%%%%%%%%%%%%%%%%%%%%%%%%%%%%%
%  Fig X - heat capacity
%%%%%%%%%%%%%%%%%%%%%%%%%%%%%%%%%%%%%%

\begin{figure}[t]
	\centering
	\captionsetup[subfigure]{labelformat=empty}
		\centering
		\subfloat[]{
  		\includegraphics[width=0.90\columnwidth]{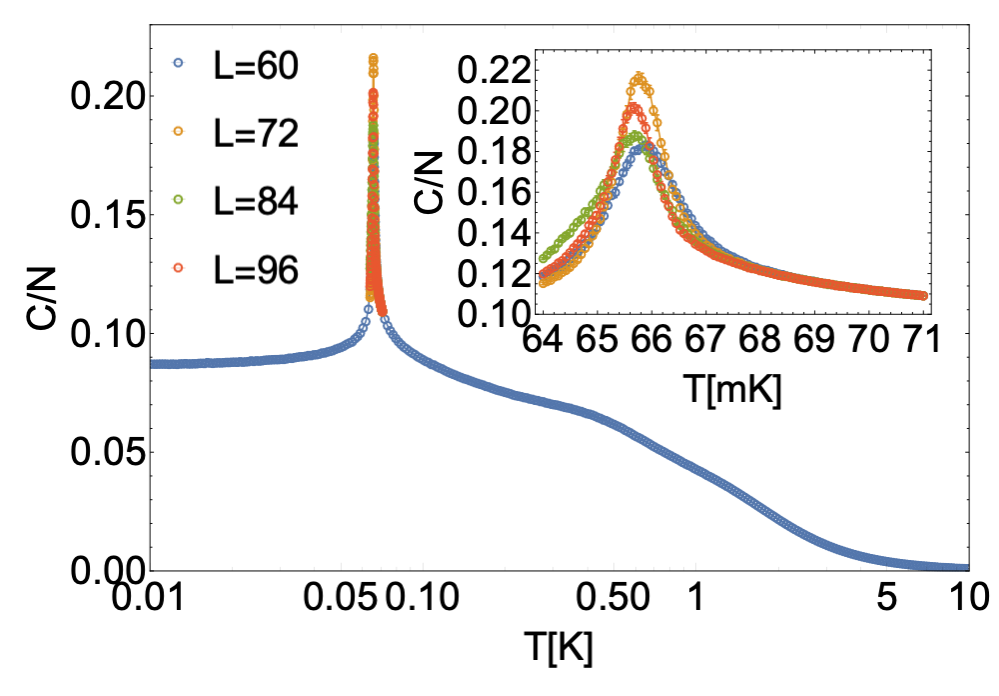}
		}
	\caption{
		Normalized specific heat $c(T) = C(T)/N$ of spin--1/2 BBK model, for parameters 
		taken from \CCO\ [cf.~Table~\ref{tab:experimental.parameters}], 
		with sharp peak at \mbox{$T^*\ \approx 66\ \text{mK}$} signaling 
		a phase transition into a low--temperature ordered state. 
		Here $c(T)$ has been plotted on a log--linear scale, for direct 
		comparison with \cite{biswas18}.
		Inset: detail of $c(T)$ near transition on a linear scale.
		Results are taken from classical Monte Carlo (MC) simulation
		of Eq.~(\ref{eq:H.BBK}), for clusters with linear dimension 
		$L = 60, 72, 84, 96$, as defined in the text.
		%
		%Non--monotonic scaling of $c(T)$ with $L$, for \mbox{$T \approx T^*$}, 
		%reflects the difficulty of thermalising simulations using a MC update 
		%based on individual spins.
		}
	\label{fig:heat.capacity}
\end{figure}

%%%%%%%%%%%%%%%%%%%%%%%%%%%%%%%%%%%%%%

In Fig.~\ref{fig:heat.capacity}, we present MC results for normalized specific heat $c(T) = C(T)/N$ 
evaluated for parameters taken from \CCO\ [Table~\ref{tab:experimental.parameters}].
The dominant features of these results are a shoulder 
at $T \approx 500\ \text{mK}$, consistent with a crossover into a phase 
with collective spin fluctuations, and a sharp peak at 
\mbox{$T^* \approx 66\ \text{mK}$}, suggestive of a finite--temperature 
phase transition.   

%%%%%%%%%%%%%%%%%%%%%%%%%%%%%%%%%%%%%%

These temperature scales should be compared with the 
parameters of the BBK model [Table~\ref{tab:experimental.parameters}].
Both are less than the scale of the FM coupling within triangular plaquettes 
($J_{22} \sim -2.3\  \text{K}$, 
$J_{21} \sim -6.5\ \text{K}$), 
but comparable with the interactions between spins in different plaquettes
($J_0 \sim -69\ \text{mK}$, 
$J_{31} \sim 770\ \text{mK}$, 
$J_{32} \sim 950\ \text{mK}$).
Thus we expect the thermodynamics of the BBK model to be determined
by the collective excitations of groups of three spins on FM--coupled plaquettes 
--- the regime described the effective honeycomb--lattice model, Eq.~(\ref{eq:H.HC}).

%This peak has a width $\sim 1\ \text{mK}$, and therefore
%appears very sharp when viewed 
%%on a scale of $10\text{--}100\ \text{mK}$.}
%on the logrithmic salce of Fig.~\ref{fig:heat.capacity}.

%%%%%%%%%%%%%%%%%%%%%%%%%%%%%%%%%%%%%%

By analogy with earlier work on the honeycomb--lattice model, reviewed
in Section~\ref{section:J1J2.model.cco}, we expect the peak at 
\mbox{$T^* \approx 66\ \text{mK}$} to originate in a continuous transition 
into a state with broken lattice--rotation symmetry.
Closer examination, %of this peak, on the scale $64\text{--}71\ \text{mK}$, 
shown in the inset of Fig.~\ref{fig:heat.capacity}, reveals a 
nearly--symmetric peak, whose height has a weakly 
non--monotonic dependence of system size.
The weakly non--monotonic behaviour seen in $c(T)$ presumably 
reflects the difficulty of simulating the critical behaviour of a system 
with dynamics based on triads of spins, using an update based on a single spin.
Away from the critical region, i.e. for $T \neq T^*$, this does not present 
a problem.  
However it does prevent us from analysing the nature of any 
phase transition on the basis of $c(T)$ alone.

%%%%%%%%%%%%%%%%%%%%%%%%%%%%%%%%%%%%%%

\begin{figure}[t]
	\centering
	\captionsetup[subfigure]{labelformat=empty}
	\subfloat[\label{fig:NemOrder.lattice1}]{
  		\includegraphics[width=0.8\columnwidth]{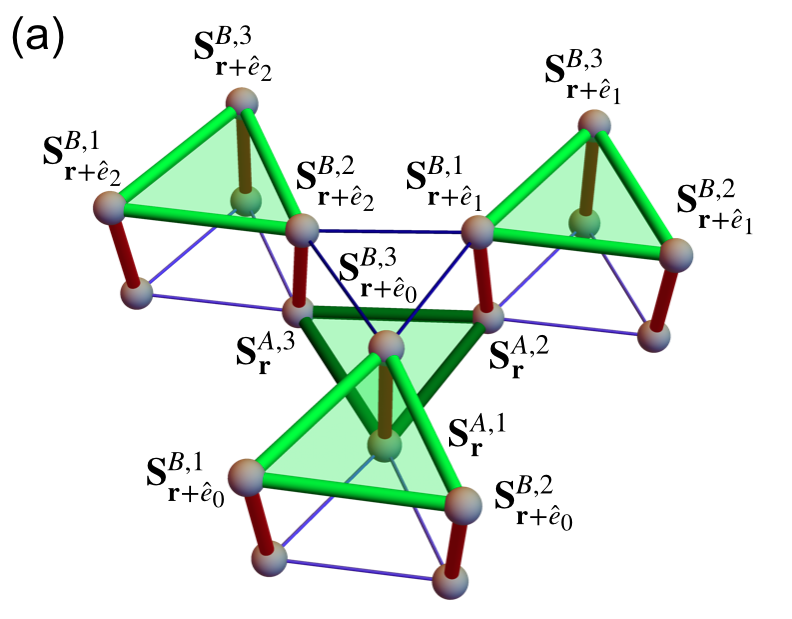}
	} \\
	\subfloat[\label{fig:NemOrder.lattice2}]{
  		\includegraphics[width=0.8\columnwidth]{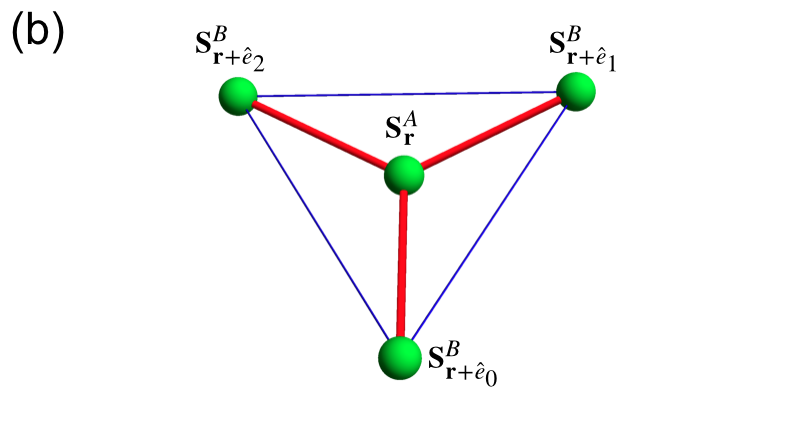}
	} 
	\caption{
		Illustration of labels used in definition of lattice--nematic order 
		parameter, Eq.~(\ref{eq:chi.phi}), associated with low--temperature 
		ordered phase.
		%[cf. Fig.~\ref{fig:order.parameter.scusceptibility}].
		%
		(a) Four neighbouring triangular plaquettes of the BBK lattice, 
		here shaded green. 
		(b) Equivalent set of neighbouring sites within the honeycomb lattice 
		considered in [\onlinecite{Mulder2010, biswas18}].
		}
	\label{fig:order.parameter}
\end{figure}

Instead, we now turn to an order parameter sensitive to the breaking 
of lattice rotation symmetry, of the type considered in \cite{Mulder2010,okumura10-JPSJ79,biswas18}.
We write this as
\begin{eqnarray}
\phi &=& \sum_{{\bf r} \in \hexagon} \phi ({\bf r}) \; ,
\label{eq:phi} 
\end{eqnarray}
where the sum upon ${\bf r}$ runs over all sites 
of a honeycomb lattice (equivalently, all triangular plaquettes of the BBK lattice),
\begin{eqnarray}
\phi ({\bf r}) &=& \frac{1}{S^2} \left[ 
		{\bf S}_A({\bf r}) {\bf S}_B({\bf r} + \hat{e}_0)  
		 + \omega {\bf S}_A ({\bf r}) {\bf S}_B ({\bf r} + \hat{e}_1)  \right. \nonumber\\
		 && \qquad \left .+ \omega^2 {\bf S}_A ({\bf r}) {\bf S}_B ({\bf r} + \hat{e}_2)
		 \right] \; ,
\label{eq:phi.r} 		 
\end{eqnarray}
with	
\begin{eqnarray}
     \omega &=& e^{i 2\pi / 3} \; ,
\end{eqnarray}
and
\begin{equation}
	{\bf S}_{(A,B)} ({\bf r}) = \sum_{\delta = 1}^{3} {\bf S}_{(A,B), \delta} ({\bf r}) \; ,
\end{equation}
is the total spin of an individual triangular plaquette, where the label A (B) 
corresponds to the lower (upper) plane of the BBK lattice.
The vectors $\hat{e}_i$ defining the bonds between plaquettes in 
Eq.~(\ref{eq:phi.r}) are illustrated in Fig.~\ref{fig:order.parameter}.

%%%%%%%%%%%%%%%%%%%%%%%%%%%%%%%%%%%%%%

\begin{figure}[t]
	\centering
	\captionsetup[subfigure]{labelformat=empty}
		\subfloat[\label{fig:NemOrder.specH}]{
  		\includegraphics[width=0.9\columnwidth]{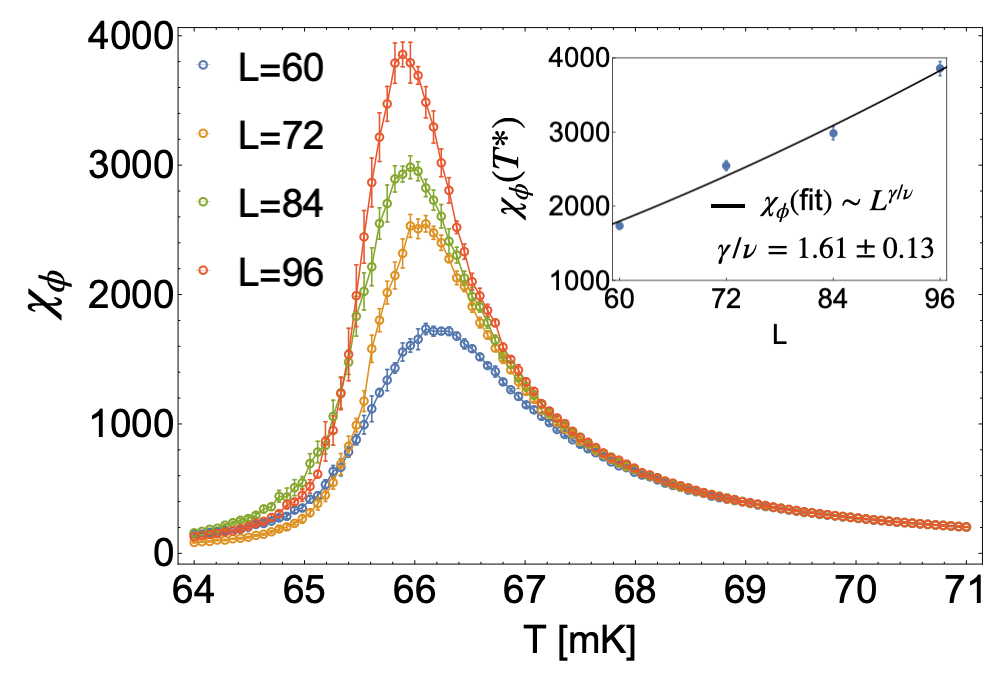}
		}
	\caption{
		Order--parameter susceptibility %$\chi_\phi$ 
		associated with breaking of lattice--rotation 
		symmetry, Eq.~(\ref{eq:chi.phi}), showing evidence for a 
		finite--temperature phase transition 
		at \mbox{$T^*\ \approx 66\ \text{mK}$}.	
		Inset: maximum value of susceptibility scales as 
		$\chi_\phi \sim L^{\gamma/\nu}$, with critical exponent 
		$\gamma/\nu = 1.61 \pm 0.13$. 
		Within error bars, this is consistent with a three--state Potts 
		transition ($\gamma/\nu = 26/15 \approx 1.733$), 
		as discussed in [\onlinecite{Mulder2010, biswas18}].
		Results are taken from classical Monte Carlo simulation
		of Eq.~(\ref{eq:H.BBK}), for the same clusters used to determine 
		heat capacity [Fig.~\ref{fig:heat.capacity}].
		}
	\label{fig:order.parameter.scusceptibility}
\end{figure}

In Fig.~\ref{fig:order.parameter.scusceptibility}, we present results 
for the order--parameter susceptibility corresponding to Eq.~(\ref{eq:phi}),
\begin{equation}
	\chi_\phi = \frac{1}{NT} \big( \langle | \phi |^2 \rangle - \langle | \phi | \rangle^2 \big)    \ .
	\label{eq:chi.phi}
\end{equation}
A sharp peak is found at $T^*\ \approx 66\ \text{mK}$, confirming 
that the anomaly found in specific heat [Fig.~\ref{fig:heat.capacity}], 
is associated with a phase transition into a state with broken 
lattice--rotation symmetry.
For system sizes $L \geq 60$, $\chi_\phi$ this peak shows a regular scaling 
with system size,  
\begin{eqnarray}
\chi_\phi \sim L^{\gamma/\nu}  \qquad \gamma/\nu = 1.61 \pm 0.13		\ .
\end{eqnarray}
Within error bars, this is consistent with the critical exponent of a 
three-state Potts model transition in 2D ($\gamma/\nu = 26/15 \approx 1.733$), 
as already discussed for the effective spin--3/2 honeycomb--lattice model \cite{Mulder2010, biswas18}.
The critical temperature $T_{\text{c}} \approx 65~mK$ is also consistent with the 
value, found for the effective model \cite{biswas18}.
Thus, while the BBK model of \CCO, Eq.~(\ref{eq:H.BBK}), is 
considerably more complicated than the effective honeycomb 
lattice model, Eq.~(\ref{eq:H.HC}), for parameters taken from 
experiment, Table~\ref{tab:experimental.parameters}, it exhibits 
exactly the same 3--state Potts transition, at a very similar 
temperature \cite{biswas18}.

%%%%%%%%%%%%%%%%%%%%%%%%%%%%%%%%%%%%%%
%  Fig 7 - definition of Z3 order parameter
%%%%%%%%%%%%%%%%%%%%%%%%%%%%%%%%%%%%%%
%

%%%%%%%%%%%%%%%%%%%%%%%%%%%%%%%%%%%%%%

Building on the analogy with the honeycomb lattice, we can 
extend this analysis, from the parameters currently associated 
with \CCO, to a parameter set equivalent to varying the ratio $J_2/J_1$ 
in Eq.~(\ref{eq:H.HC}).
We do this by varying the ratio $x = J_{\sf eff}/J_0$, where $J_{\sf eff}$ [Eq.(\ref{eq:Jeff})]
plays the role of $J_2$, and $J_0$ plays the role of $J_1$. 
Doing so, we arrive at the finite--temperature phase diagram 
shown in Fig.~\ref{fig:thermodyn}(a), where estimates 
of $T^*$ have been taken from the peak in heat capacity.

%%%%%%%%%%%%%%%%%%%%%%%%%%%%%%%%%%%%%%
%  Fig X - order parameter susceptibility
%%%%%%%%%%%%%%%%%%%%%%%%%%%%%%%%%%%%%%

%%%%%%%%%%%%%%%%%%%%%%%%%%%%%%%%%%%%%%
%  Fig X - Sq over a wide range of parameters
%%%%%%%%%%%%%%%%%%%%%%%%%%%%%%%%%%%%%%

\begin{figure*}[t]
	\centering
	\includegraphics[width=0.98\textwidth]{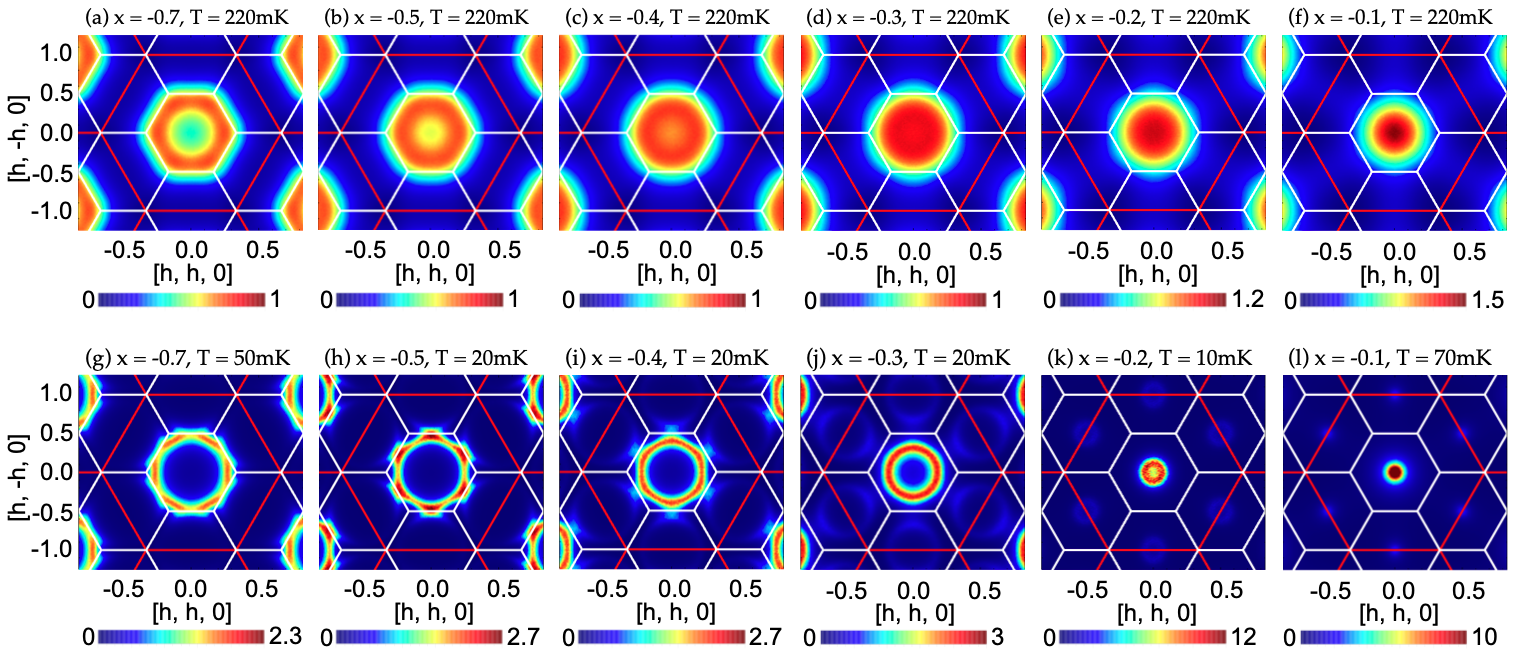}
%	\subfloat[$T=200\ \text{mK}$]{\includegraphics[width=0.999\textwidth]{Thermodyn/Sq-high-T.png}}\\
%	\subfloat[$T=T^*(x) + \epsilon$]{\includegraphics[width=0.999\textwidth]{Thermodyn/Sq-low-T.png}}
	\caption{
		Evolution of the equal time structure factor $S({\bf q})$ 
		as a function of $x = J_{\sf eff}  / J_0$ [Eq.~(\ref{eq:Jeff})],
		showing the impact 
		of classical ground state degeneracy on correlations.
		(a)--(f) Results at $T=220\ \text{mK}$, within the spin--liquid phase, 
		showing structured diffuse scattering, with correlations strongest on 
		the zone boundary for $x < -0.5$.
%		%
%		For $x = -0.2$, $x = -0.1$, these resemble the correlations of the 
%		``pancake--liquid'' identified in \cite{okumura10-JPSJ79}.
		%
		(g)--(l) Results at $T = T^*(x) + \epsilon$, just above the phase transition
		into an ordered state, showing the imprint of the different ``ring'' degeneracies. 
		The transition from a ``large'' ring of degenerate spirals centered 
		on $\Gamma$, to a ``small'' ring of degenerate spirals centered 
		on $K$, occurs for $x = -0.5$.
		The asymmetry of scattering within the ``small'' ring, with 
		suppression of weight at large $q$, reflects the form--factor 
		of the BBK lattice.
		All results were obtained from classical Monte Carlo simulation of 
		Eq.~(\ref{eq:H.BBK}), for a cluster of linear dimension 
		L = 60 \mbox{(N = 21,600)}, with $J_{\sf eff} \equiv (J_{31} + J_{32})/2 $, 
		and other parameters taken from experiment 
		[Table~\ref{tab:experimental.parameters}].	
		Magnetic scattering has the periodicity of the $4^{th}$ 
		Brillouin zone, shown here with red lines.
		}
	\label{fig:Sq}
\end{figure*}

%%%%%%%%%%%%%%%%%%%%%%%%%%%%%%%%%%%%%

We find that the ordering temperature, $T^*$, takes on a significantly 
higher value for parameters associated with a ``small'' ring of degenerate 
spiral states, for $x < -1/2$, than for parameters associated with 
a ``large'' ring of spiral ground states $-1/2 < x < -1/6$ [cf. Fig.~\ref{fig:ring.degeneracy}].
This appears to be consistent with transition temperatures found in 
earlier studies of the honeycomb lattice \cite{Mulder2010,okumura10-JPSJ79,biswas18}.
And, naively, it suggests that the entropy associated with the ``large''
ring of spirals, centered on $\Gamma$, is greater than the entropy 
associated with the small ring of spirals, centered on $K$.
We return to this point in the context of the discussion of spin--liquid 
properties at finite temperature, below.

%%%%%%%%%%%%%%%%%%%%%%%%%%%%%%%%%%%%%%

We note that, for $x> -1/6$, the anomaly seen in $c(T)$, and 
the corresponding estimate of a critical temperature
in Fig.~\ref{fig:thermodyn}, should be associated with the 
onset of strong FM fluctuations, rather than lattice--symmetry breaking.
However, as this case does not appear to have any bearing 
on the physics of \CCO, we shall not consider it further here.

%%%%%%%%%%%%%%%%%%%%%%%%%%%%%%%%%%%%%%

Published estimates of the exchange parameters of \CCO\ 
[Table~\ref{tab:experimental.parameters}] suggest a ratio of effective 
honeycomb--model interactions $x \approx 1.25$.
This places \CCO\ within the region where the honeycomb model, 
Eq.~(\ref{eq:H.HC}), has a small ring of ground states 
centered on ${\bf q} = K$.
However the relatively large uncertainty in the estimated values of 
$J_1$ and $J_2$, leads to considerable uncertainty in ratio,  
reflected in the error bar in the placement of \CCO\ in Fig.~\ref{fig:thermodyn}.

%%%%%%%%%%%%%%%%%%%%%%%%%%%%%%%%%%%%%
\subsection{Correlations at finite temperatures} 		
%%%%%%%%%%%%%%%%%%%%%%%%%%%%%%%%%%%%%
\label{sec:spiral.spin.liquid}

We next turn  to the nature of spin correlations in the regime 
$T^* = 66\ \text{mK} < T \lesssim 500\ \text{mK}$, where results for heat capacity 
are suggestive of collective behaviour.
This is the range of temperatures where spin--liquid 
behaviour is most likely to be found in the classical limit of the 
BBK model.
It is also the temperature regime relevant to published neutron 
scattering data for the quantum spin liquid in \CCO\ \cite{Balz2016,Balz2017-PRB95,Sonnenschein2019}.

%%%%%%%%%%%%%%%%%%%%%%%%%%%%%%%%%%%%%

In Fig.~\ref{fig:thermodyn}, we show MC results for the equal--time 
structure factor $S({\bf q})$, calculated for parameters relevant to \CCO.
At a temperature of \mbox{$T = 220\ \text{mK}$}, 
a broad ``ring'' of strong fluctuations is observed for ${\bf q}$ near to the 
Brillouin zone (BZ) boundary, with the strongest signal occurring 
near the $K$ points at BZ corners  [Fig.~\ref{fig:thermodyn}(c)].
On lowering temperature to \mbox{$T = 90\ \text{mK}$},  spectral weight is 
transferred from the zone boundary to diffuse, U--shaped structures at the 
zone corners [Fig.~\ref{fig:thermodyn}(d)].

%%%%%%%%%%%%%%%%%%%%%%%%%%%%%%%%%%%%%

Very similar results for $S({\bf q})$ have been found in large--N 
and classical MC calculations for the honeycomb--lattice model 
at \mbox{$T = 100\ \text{mK}$} \cite{biswas18}.
The preponderance of scattering near the zone corners can be 
understood in terms of the ``small'' ring of classical ground states
in the honeycomb--lattice model for $|J_1| \sim J_2$ 
[cf. Fig.~\ref{fig:ring.degeneracy}].
Meanwhile the asymmetry visible in $S({\bf q})$ 
near the BZ corners reflects the form factor 
of the BBK lattice, which has the periodicity of 
the $4^{th}$ BZ, shown here with red lines.

%%%%%%%%%%%%%%%%%%%%%%%%%%%%%%%%%%%%%

When combined with the heat capacity [Fig.~\ref{fig:heat.capacity}], 
these results suggest that fluctuations of spins 
in \CCO\ at temperatures \mbox{$T \lesssim 500\ \text{mK}$} 
are both collective and highly structured, involving a set of {\bf q} 
vectors which bears the imprint of nearby (classical) ground state degeneracies.
This is consistent with a spin--liquid state at finite 
temperatures, where entropy predominates.
And it is broadly similar to what has been observed in a number of models 
supporting ``spiral spin liquids'' 
%including the honeycomb--lattice model discussed 
%in Section~\ref{section:CCO.v.BBK.model} 
\cite{bergman07,okumura10-JPSJ79,benton15-JPSJ84,seabra16-PRB93,buessen18,Yao2021}. 

%%%%%%%%%%%%%%%%%%%%%%%%%%%%%%%%%%%%%

With this in mind, it is interesting to further explore the analogy with the  
honeycomb lattice model, by varying the values of the effective 
parameters $J_1$ and $J_2$, in such a way as to tune between different 
kinds of ground state degeneracies.
In Fig.~\ref{fig:Sq} we present results for the evolution of  $S({\bf q})$
as a function of $x = J_{\sf eff} /J_0$, [Eq.~(\ref{eq:Jeff})], 
at temperatures well within the 
spin--liquid phase, and just above the transition 
into the ordered ground state [cf. Fig.~\ref{fig:thermodyn}].
We consider a range of parameters $-0.7 \le x \le -0.1$ which span both 
the ``large'' and ``small' ring degeneracies of the honeycomb lattice model
[cf. Fig.~\ref{fig:ring.degeneracy}].

%%%%%%%%%%%%%%%%%%%%%%%%%%%%%%%%%%%%%

At $T = 220\ \text{mK}$ (upper panels), for $x < - 0.3$, correlations are strongest 
near to the BZ boundary.
Meanwhile, for $x \ge -0.3$, they start to resemble the 
``pancake--liquid'' identified in \cite{okumura10-JPSJ79}, with stronger scattering 
in the zone center.
At temperatures just above the peak observed in specific heat, 
$T = T^* + \epsilon$ (lower panels), the imprint of the ground--state 
degeneracy of the equivalent honeycomb--lattice model, Eq.~(\ref{eq:H.HC}), 
is evident in rings of high intensity.
These are centered on $K$ for  $x < - 0.5$ and on $\Gamma$ for $x > - 0.5$
\mbox{[cf. Fig.~\ref{fig:ring.degeneracy}]}.

%%%%%%%%%%%%%%%%%%%%%%%%%%%%%%%%%%%%%

Returning to experimentally--motivated parameters, \mbox{$x = 1.25$}, it is 
interesting to compare these results with published calculations of the 
structure factor of the BBK model coming from PFFRG \cite{Balz2016}
and phenomenological parton approaches \cite{Sonnenschein2019}.
PFFRG results for the static structure factor $S({\bf q}, \omega = 0)$ at $T=0$ 
suggest strong correlations near to the BZ boundary, with additional weight near 
to zone corners $K$  \cite{Balz2016}.
This is qualitatively very similar to MC results at $T = 220\ \text{mK}$, within 
the spiral spin liquid phase [Fig.~\ref{fig:thermodyn}(c)].
The $T=0$ parton phenomenology, meanwhile, shows energy--dependent 
ring---like structure in $S({\bf q}, \omega)$, centered on $\Gamma$  \cite{Sonnenschein2019}.
And to meaningfully compare with this, it is necessary to analyse dynamics.

%%%%%%%%%%%%%%%%%%%%%%%%%%%%%%%%%%%%%
%% Fig. 10 - S(q, \omega) at B = 0T -- rings at low, diffuse scattering 
%%                at high energy
%%%%%%%%%%%%%%%%%%%%%%%%%%%%%%%%%%%%%

\begin{figure}
	\centering	
	\subfloat[$ \tilde{S}({\bf q}, \omega),  B=0$
	\label{fig:S.zero.field}]{
  		\includegraphics[width=0.9\columnwidth]{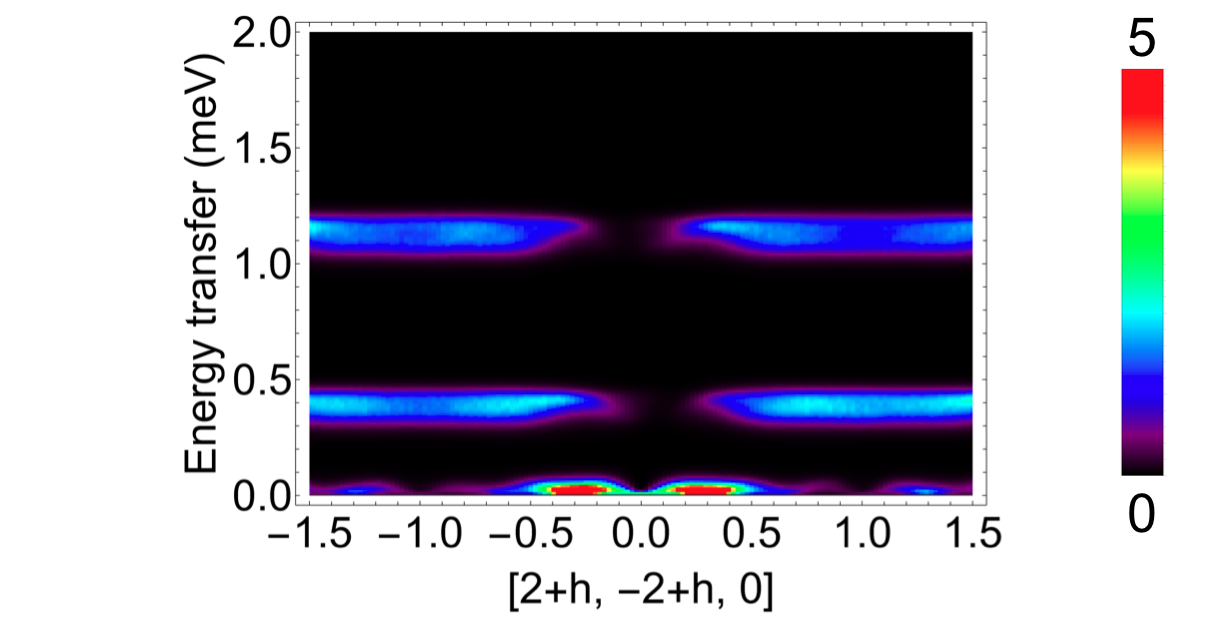}
	}\\
	\subfloat[$\tilde{S}({\bf q}, \omega = 0.01\ \text{meV})$ 
	\label{fig:Swq.1.meV}]{
  		\includegraphics[width=0.9\columnwidth]{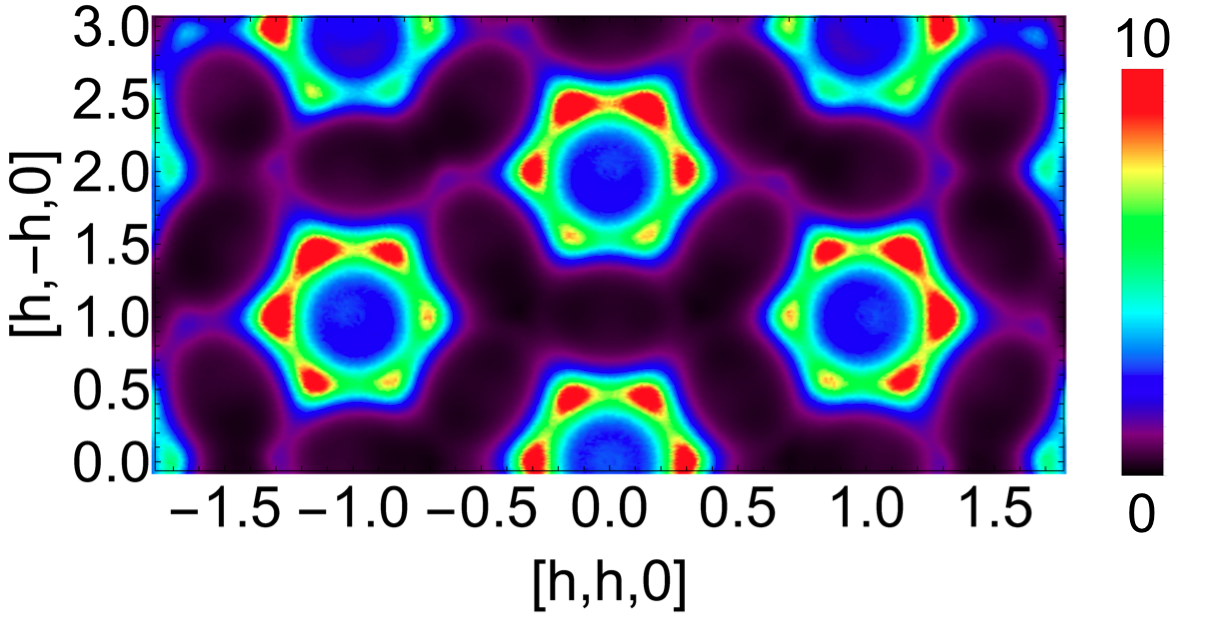}
	}\\
	\subfloat[$\tilde{S}({\bf q}, \omega = 0.35\ \text{meV}) $ \label{fig:Swq.35.meV}]{
  		\includegraphics[width=0.9\columnwidth]{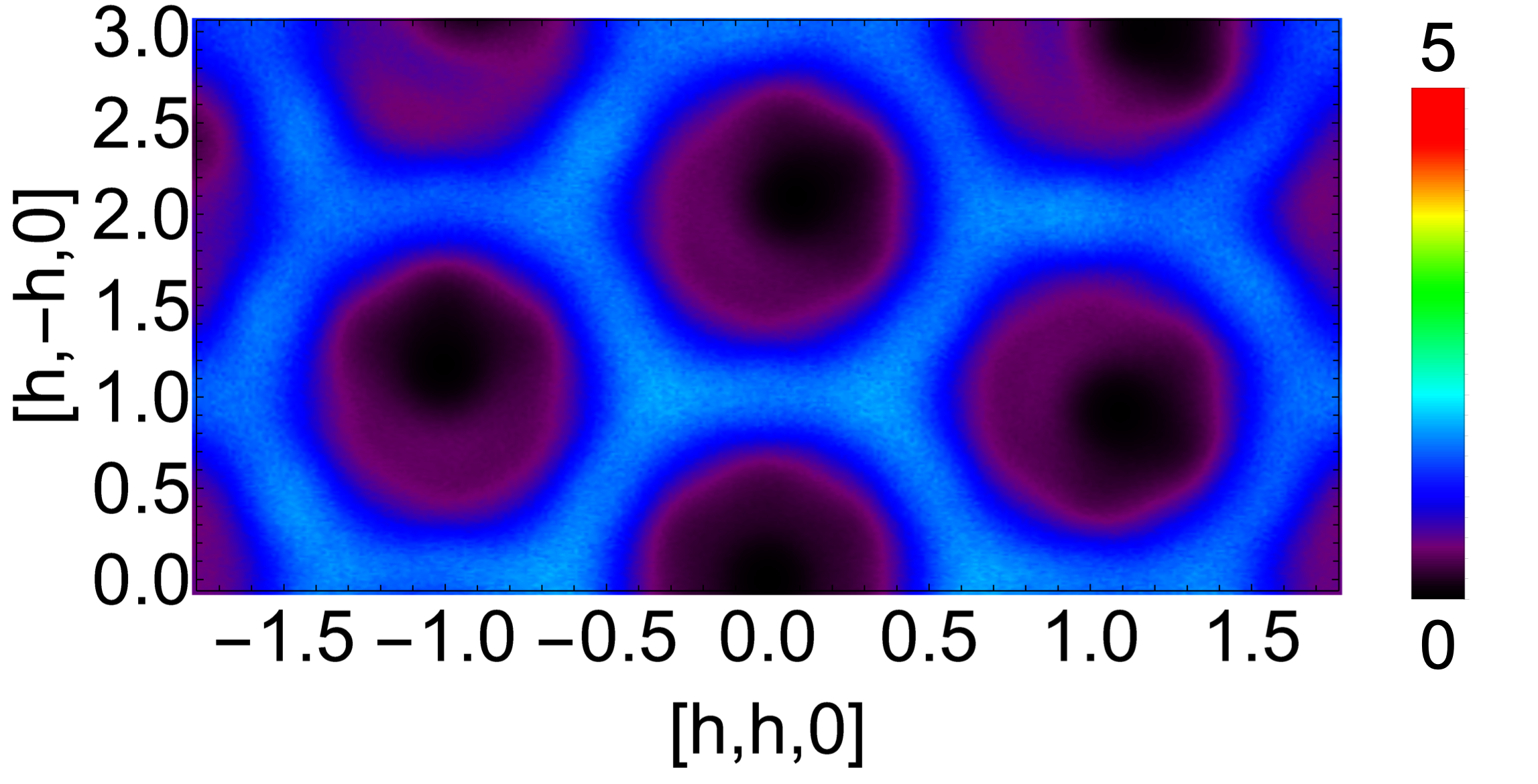}
	}\\
	\subfloat[$\tilde{S}({\bf q}, \omega = 1.13\ \text{meV}) $ \label{fig:Swq.113.meV}]{
  		\includegraphics[width=0.9\columnwidth]{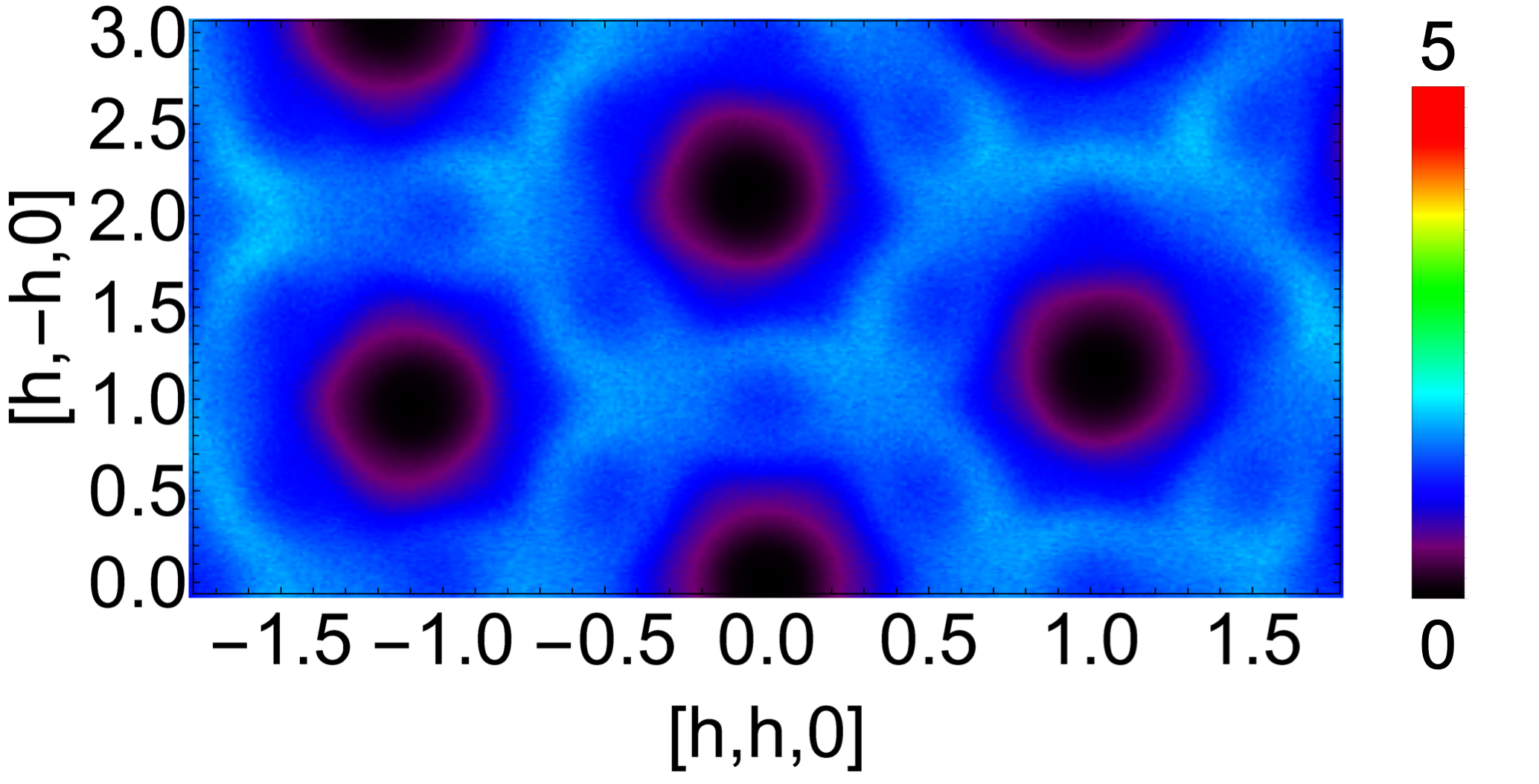}
	}
	\caption{
	Spin dynamics in zero field, showing the different correlations found on 
	different energy scales.
	The first moment of the dynamical structure factor 
	$\tilde{S}({\bf q}, \omega)$ [Eq.~(\ref{eq:S.tilda})] was obtained from molecular 
	dynamics (MD) simulation of the BBK model, Eq.~(\ref{eq:H.BBK}), at \mbox{$T = 220\ \text{mK}$}, 
	for a cluster of linear dimension $L = 48$ \mbox{($N = 13,824$)}, 
	with parameters taken from Experiment [Table.~\ref{tab:experimental.parameters}].
	Results are presented with energy resolution of FWHM $= 0.02\ \text{meV}$, and no
	Cr$^{5+}$ form factor. 
	}
	\label{fig:dynamics.in.zero.field}
\end{figure}

%%%%%%%%%%%%%%%%%%%%%%%%%%%%%%%%%%%%%
\section{Dynamical properties of ${\cal H}_{\sf BBK}$}		
%%%%%%%%%%%%%%%%%%%%%%%%%%%%%%%%%%%%%
\label{sec:dynamics}

The thermodynamic properties of the BBK model, reported in 
Section~\ref{sec:thermodynamics}, are consistent with the experimental 
observation that \CCO\ enters a spin--liquid state at temperatures 
\mbox{$T \lesssim 500\ \text{mK}$}.
To learn more about this spin liquid, we now turn to dynamical simulations.
These will provide us with a basis for comparison with inelastic 
neutron scattering (INS) experiments, which reveal dynamics 
at a range of different energy scales 
\cite{Balz2016,Balz2017-PRB95,Sonnenschein2019}.

We adopt a (semi--)classical approach, in which 
spin configurations are drawn from a Boltzmann distribution,   
generated by MC simulation, and then evolved according to the 
Heisenberg equation of motion 
\cite{moessner98-PRL80,conlon10-PRB81,Taillefumier2014}   
%
%%%%%%%%%%%%%%%%%%%%%%%%%%%%%%%%%%%%%
%
\begin{eqnarray}
	\frac{d {\bf S}_i}{d t} 	= \frac{i}{\hbar} [{\bf S}_i, \mathcal{H}_{\sf BBK}]
	= \bigg( \sum_j J_{ij} {\bf S}_j -  B^z\hat{\mathbf{z}} \bigg) \times {\bf S}_i		\; .
\label{eq:E.of.M}
\end{eqnarray}
%
%%%%%%%%%%%%%%%%%%%%%%%%%%%%%%%%%%%%%
%
%where, initially, we shall consider the case of zero magnetic field, $B^z = 0$.
%%
%We consider a single bilayer, with parameters taken from experiment 
%[cf. Table~\ref{tab:experimental.parameters}].
%
This approach to the dynamics of spin liquids was popularised by 
Moessner and Chalker~\cite{moessner98-PRB58,moessner98-PRL80}, 
who dubbed it {\it ``Molecular Dynamics''} (MD) simulation, by analogy with 
Monte Carlo techniques used in the simulation of fluids.

We note that this ``MD'' approach has much in common 
with %``micormagnetic'' simulation obtained by 
the numerical solution of  the  Landau--Lifschitz--Gilbert equation
\cite{Gilbert2004}
\begin{eqnarray}
\frac{\partial {\bf M}}{\partial t} 
	= - \gamma^*{\bf M} \times {\bf H}  
	- \frac{\alpha}{M} {\bf M} \times \frac{\partial {\bf M}}{\partial t}
  \; .
\end{eqnarray}
However there are also some important differences,  
% in the way in which the method is implemented, 
notably the use of microscopic spin variables ${\bf S}_i$, 
rather than a course--grained magnetisation ${\bf M}$; the absence of a
phenomenological damping term $\alpha$; and the use of an ensemble 
of initial states drawn from classical MC simulation.
Further details of our numerical methods can be found in 
Appendix~\ref{appendix:Numerics}.

%%%%%%%%%%%%%%%%%%%%%%%%%%%%%%%%%%%%%%

In the context of \CCO, the great advantage of the MD approach is that it is possible to simulate 
the finite--temperature spin dynamics of clusters of more than N=$10,000$ spins, 
facilitating a quantitative comparison with experiment, even in the absence of magnetic 
order, and in models which would be subject to a sign problem in quantum Monte Carlo simulation.
And even though the MD method fails to take into account either entanglement, \nic{or} quantum 
statistics of excitations, it has  previously been found to give a good  account of the qualitative features 
of quantum models for problems as diverse as the Kitaev model \cite{Samarakoon2017}; quantum spin ice 
\cite{Taillefumier2017}; the spin liquid phase of NaCaNi$_2$F$_7$ \cite{Zhang2019},
and the semi--classical dynamics of spin density waves \cite{chern18}.

%%%%%%%%%%%%%%%%%%%%%%%%%%%%%%%%%%%%%
\subsection{Comparison of simulation with experiment}
%%%%%%%%%%%%%%%%%%%%%%%%%%%%%%%%%%%%%
\label{sec:dynamics.v.experiment}

We start by comparing simulations of dynamics carried out in the spin--liquid 
phase of the BBK model with results found in experiment. 
Key results have already been summarised in Fig.~\ref{fig:comparison.with.experiment}, 
where we show MD results for the dynamical structure factor $S({\bf q}, \omega)$,
side-by-side with results from INS \cite{Balz2016}.
To facilitate comparison, simulation results have been convoluted with 
a Gaussian mimicking experimental resolution, and a magnetic 
form factor appropriate for a Cr$^{5+}$ ion [cf.  Appendix~\ref{sec:CompExp}].
Both experiment and theory show  rings of scattering at low energy,
consistent with the discussion in Section~\ref{sec:spiral.spin.liquid}.
They also agree on a broad network of scattering at intermediate to high energy, 
which is concentrated near to the zone boundary, with hints of ``bow--tie''
like structures near to $(1, 0, 0)$.
(Bright ``spots'' seen in INS near to zone centers 
represent scattering from phonons, and do not form part 
of the magnetic signal \cite{Balz2016}).  

%%%%%%%%%%%%%%%%%%%%%%%%%%%%%%%%%%%%%

Clearly, the MD simulations capture important elements of 
the physics of \CCO.   
Moreover, the ``ring'' found at low energies in MD simulation corresponds 
very closely to the ring found in the static structure factor $S({\bf q}, \omega=0)$ 
in PFFRG calculations \cite{Balz2016}, and a related parton 
phenomenology \cite{Sonnenschein2019}.
This suggests that important aspects of quantum treatments 
of the BBK model are reproduced.
The question which remains, is what this tells us about the nature, 
and origin, of the spin-liquid in \CCO\ ?

%%%%%%%%%%%%%%%%%%%%%%%%%%%%%%%%%%%%%
\subsection{Evolution of spin configurations in spin liquid as a 
function of time [First Animation]}
%%%%%%%%%%%%%%%%%%%%%%%%%%%%%%%%%%%%%
\label{sec:first.animation}
 
One very direct route into this question is to look at the evolution of spin 
configurations in real--space, as a function of time.
This is accomplished in the First Animation provided in the 
Supplementary Material \cite{first.animation}.
The simulations shown in this animation were carried out for $N=5400$ 
spins at a temperature of $T = 220\ \text{mK}$, deep inside the classical 
spin--liquid regime [Fig.~\ref{fig:thermodyn}(a)], where the equal--time 
structure factor $S({\bf q})$ reveals a diffuse ``ring'' structure 
[Fig.~\ref{fig:thermodyn}(c)].
%, in the absence of magnetic field ($B = 0\ \text{T}$).
%
As the Animation shows, the spins continue to fluctuate, even at this low 
temperature.
And on closer inspection, the simulation reveals that the spins exhibit both 
slow and fast dynamics, and they have very different characters. 
The slow precession of locally collinear spins is mixed with fast 
fluctuations of seemingly-uncorrelated spins.
For clarity, spins which rotate quickly have been coloured red, while 
spins which rotate slowly  have been coloured green; further details of
the animation can be found in Appendix~\ref{appendix:animation}.

%%%%%%%%%%%%%%%%%%%%%%%%%%%%%%%%%%%%%
\subsection{Correlations as a function of energy and momentum}
%%%%%%%%%%%%%%%%%%%%%%%%%%%%%%%%%%%%%
\label{sec:animations}

To better understand the dynamics on different time scales observed in the  
spin liquid, we now return to the dynamical structure factor.
This time however, in order to compensate for the classical statistics 
of the MC simulations, we do not plot the dynamical structure factor 
$S({\bf q}, \omega)$ directly, but rather its first moment, divided by the 
temperature at which the simulations are carried out  
\cite{pohle-arXiv, Zhang2019, remund-in-preparation}:
\begin{eqnarray}
   \tilde{S}({\bf q}, \omega)  
  %=  S^{\sf quantum}_{\sf T = 0}({\bf q}, \omega)  
   					 = \frac{1}{2} \frac{\omega}{k_B T} S_{\sf MD} 
   ({\bf q}, \omega) \; .
   \label{eq:S.tilda}
\end{eqnarray}
Further details of this approach can be found in Appendix~\ref{sec:MD.corrected}.

%%%%%%%%%%%%%%%%%%%%%%%%%%%%%%%%%%%%%

In Fig.~\ref{fig:S.zero.field}, we present results for parameters
equivalent to the First Animation, for a cluster of $N=13, 824$ spins.
Results are shown for the $[2+h,-2+h,0]$ 
plane in reciprocal space, considered in \cite{Balz2017-PRB95}; 
the atomic form factor has been set equal to unity, 
in order to make it easier to distinguish correlations over a broad area 
of reciprocal space.
Bands of excitations are observed with three different energy scales, 
\mbox{ $\omega \sim 0.0$, $0.4$ and $1.1\ \text{meV}$}.
The upper bands are diffuse, with no sharp features, and only very weakly 
dispersing, suggesting that excitations are nearly localised.
The characteristic energy scale of the upper (lower) ``flat'' mode is set by the 
ferromagnetic coupling strengths $J_{22}$ ($J_{21}$), reflecting 
the transition from a high--spin to a low--spin state on an individual triangular 
plaquette.
Each ``mode'' comprises two distinct bands, with band--width determined 
by the antiferromagnetic coupling $J_{31}$ ($J_{32}$).

These quasi--localised excitations are a robust feature of the bilayer 
breathing kagome (BBK) model of \CCO, and are also seen in quantum 
simulations \cite{shimokawa-in-prep}.
They are an echo, at finite energy, of the ``Coulombic'' spin--liquid found in the  
Heisenberg antiferrmognet on a Kagome lattice, and support both 
``pinch--point'' and ``half--moon'' features in neutron scattering \cite{Yan2018}.
%
%In this respect, they are analagous to the ``dynamical spin liquid'' found in 
%Nd$_2$Zr$_2$O$_7$ \cite{Benton2016}.
%
As required for a rotationally symmetric Hamiltonian like Eq.~(\ref{eq:H.BBK}), 
spectral weight vanishes for ${\bf q} = 0$ at all finite $E$.
The lower band, meanwhile, is much more structured, with spectral weight 
predominantly 
found in high-intensity patches centered around ${\bf q} = 0$.

%%%%%%%%%%%%%%%%%%%%%%%%%%%%%%%%%%%%%

In Fig.~\ref{fig:Swq.1.meV}, we show a cross section through 
$S({\bf q}, \omega)$ for $\omega = 0.01\ \text{meV}$, once again 
choosing our plane in reciprocal space to match equivalent results in 
\cite{Balz2016,Balz2017-PRB95}.
The high-intensity patches centered around ${\bf q} = 0$ are immediately 
recognisable as the ``ring'' observed in $S({\bf q})$ 
[Fig.~\ref{fig:thermodyn}(c)] with strong intensity near the K-points at 
the zone-corners.
And further ``rings'' are observed at $(2, 0, 0)$, etc., reflecting 
the periodicity of the 4$^{th}$ BZ.
These are connected by a diffuse web of scattering which preserves 
the overall  6-fold rotation symmetry of the lattice.

%%%%%%%%%%%%%%%%%%%%%%%%%%%%%%%%%%%%%

In Fig.~\ref{fig:Swq.35.meV}, we present equivalent results for 
\mbox{$\omega = 0.35\ \text{meV}$}, the characteristic energy scale of 
the intermediate band of excitations in Fig.~\ref{fig:S.zero.field}.
The structure observed is utterly different.
The ``Ring'' like features found in the lower band 
are conspicuously absent, being replaced by a broad web of 
correlations tracking the boundary of the extended (4$^{th}$) BZ.
Superimposed on this web, we find a blurred but regular array 
of triangular features, which meet to form ``bow--ties'' centered 
on reciprocal lattice points $(1, 0, 0)$, etc.

%%%%%%%%%%%%%%%%%%%%%%%%%%%%%%%%%%%%%

Finally, in Fig.~\ref{fig:Swq.113.meV}, we present results for 
\mbox{$\omega = 1.13\ \text{meV}$}, the characteristic energy scale of 
the highest band of excitations in Fig.~\ref{fig:S.zero.field}.
Here once again we find a broad web of correlations tracking the 
boundary of the extended BZ.
And once again this has structure superimposed on it.
But at this particular energy, that structure takes the form of 
crescent features centered on the same reciprocal lattice 
vectors as the bow--ties described above, i.e.  $(1, 0, 0)$, etc.

%%%%%%%%%%%%%%%%%%%%%%%%%%%%%%%%%%%%%
\subsection{Evolution of spin configurations in spin liquid as 
a function of time, revisited [Second Animation]}
%%%%%%%%%%%%%%%%%%%%%%%%%%%%%%%%%%%%%
\label{sec:second.animation}

From these results it is clear that i) dynamics occur on three different 
times scales, and ii) that the dynamics on long time scales (low--energy 
band) is qualitatively different from that on short time scales 
(intermediate-- and high--energy bands).
With these lessons in mind, we can revisit the time--evolution of spin 
configurations in real space, and apply a filter to separate dynamics 
into slow, intermediate and fast bands of excitations.

%%%%%%%%%%%%%%%%%%%%%%%%%%%%%%%%%%%%%

In the Second Animation provided in the Supplementary 
Material \cite{second.animation}, we show the separate time 
evolution of slow, intermediate and fast spin fluctuations, 
in three different panels.
Viewed at ``normal'' speed, only the slow fluctuations are 
clearly intelligible, as collective rotation of spins which are 
locally collinear on each of the ferromagnetic plaquettes 
of the BBK lattice.
To make comparison easier, in the second part of the Animation, 
we adjust the ``clock'' for each panel, speeding up the slow fluctuations,
and slowing down the fast ones, such that all processes occur at (roughly)
the same subjective speed.
At the same time, we reintroduce the color cues for speed, with 
rapdily--rotating spins appearing in red. 
Further details of the entire procedure are given 
in Appendix~\ref{appendix:animation}.

%%%%%%%%%%%%%%%%%%%%%%%%%%%%%%%%%%%%%

Once the time series coming from simulations has been processed in 
this way, the contrasting character of excitations at different timescales 
is obvious.
Slow fluctuations, once speeded up, are more obviously collective, with 
triads of spins on neighbouring FM plaquettes moving in unison.
Intermediate and fast fluctuations, meanwhile, are seen to have the same 
character, and to comprise two, seemingly uncorrelated processes.
The first of these is the collective rotation of spins on the AF plaquettes of the 
lattice, within each of which they (approximately) maintain a condition of net 
zero spin, familiar from the Kagome--lattice AF \cite{Chalker1992, zhitomirsky08-PRB78, 
Taillefumier2014}.
Superimposed on this are extremely fast spin--flips of individual spins, 
which appear to propagate in pairs around the lattice.

%%%%%%%%%%%%%%%%%%%%%%%%%%%%%%%%%%%%%

In summary, the dynamics of the classical spin liquid found in 
the BBK model of \CCO\ are complicated, unusual, and interesting.
The degree of complexity is perhaps not a surprise, given the large 
unit cell and low--symmetry of the BBK lattice.
None the less, it is possible to interpret simulation results for dynamics, 
through a combination of spectral functions, and animations of real--space 
dynamics.
What makes the dynamics unusual, is that they appear to be qualitatively 
different on different timescales, and to combine aspects of two very 
different spin liquids, the ``spiral spin liquids'' studied in models with complex 
competing interactions \cite{bergman07,okumura10-JPSJ79,benton15-JPSJ84,seabra16-PRB93,buessen18,Yao2021}, 
and the celebrated spin liquid found in Kagome lattice AF \cite{Chalker1992,zhitomirsky08-PRB78,Robert2008,Taillefumier2014}.  
This multiple--scale dynamics would be interesting by itself, but what makes it 
compelling is that many aspects of these dynamics have already been 
seen in INS measurements on \CCO, a point which we return to 
in Section~\ref{sec:cco.v.bbk}, below.

%%%%%%%%%%%%%%%%%%%%%%%%%%%%%%%%%%%%%
\section{Thermodynamic properties of BBK model in applied magnetic field} 		
%%%%%%%%%%%%%%%%%%%%%%%%%%%%%%%%%%%%%
\label{sec:thermodynamics.in.field}

%%%%%%%%%%%%%%%%%%%%%%%%%%%%%%%%%%%%%
%  Fig. X - phase diagram in field
%%%%%%%%%%%%%%%%%%%%%%%%%%%%%%%%%%%%%

\begin{figure}[t]
	\centering
	\captionsetup[subfigure]{labelformat=empty}
	\begin{minipage}[t]{0.99\columnwidth}
		\subfloat[\label{fig:PD.B2}]{
  		\includegraphics[width=0.98\columnwidth]{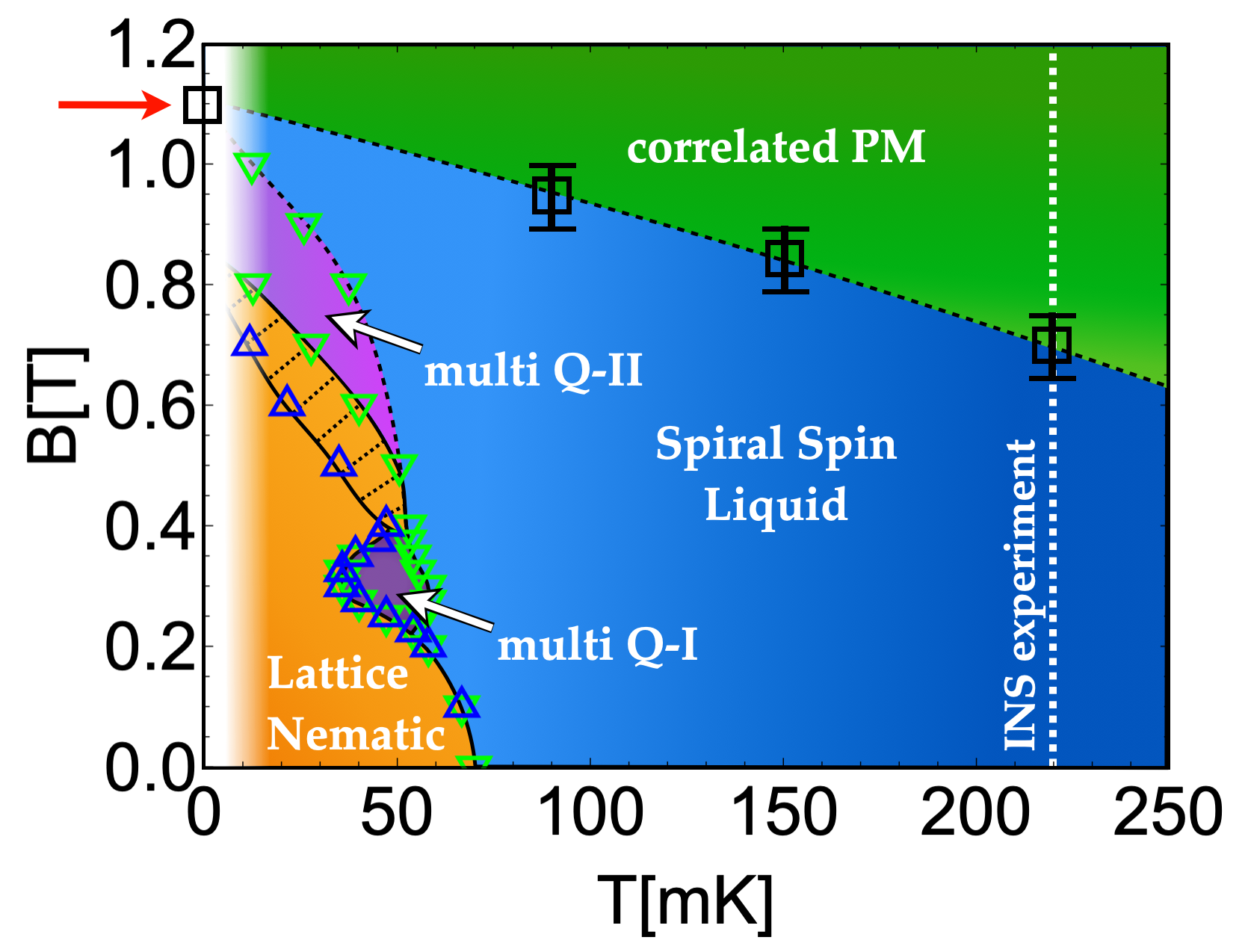}
		}
	\end{minipage}
	\caption{
	Low--temperature phase diagram found in classical Monte Carlo simulations of the 
	bilayer breathing Kagome (BBK) model of \CCO, for a cluster of $N = 1944$ spins.
	The closing of a gap to transverse spin excitations at a critical field 
	$B \lesssim 1\ \text{T}$ converts the correlated paramagnet into 
	a spiral spin liquid, with characteristic ``ring'' in $S({\bf q})$ 
	[Section~\ref{sec:dynamics.in.field}].
	A small domain of lattice--nematic and multiple--q orders is found at 
	temperatures below $70\ \text{mK}$ [Section~\ref{sec:thermodynamics.in.field}].
	Results are taken from classical Monte Carlo simulation of ${\mathcal H}_{\sf BBK}$ 
	[Eq.~(\ref{eq:H.BBK})], as described in Section~\ref{sec:thermodynamics.in.field}.
%	for cluster of $N= xxx$ spins, with parameters from experiment 
%	[Table~\ref{tab:experimental.parameters}], as described in Section~\ref{sec:thermodynamics.in.field}.
%	%
	%
	The temperature associated with simulation results for dynamics, 
	 $T = 220\ \text{mK}$, [Section~\ref{sec:dynamics}, 
	 Section~\ref{sec:dynamics.in.field}] is shown with a white dashed 
	 line [cf. Fig.~\ref{fig:comparison.with.experiment}].
	The critical fields indicating the closing of the energy gap to spin 
	excitations, as obtained from MD simulations  at $T=90,150,220\ \text{mK}$
	[cf. Fig~\ref{fig:BandGapEvolutionField}], and linear spin wave theory at $T=0$, 
	are shown with black squares ($\square$).
	Low--temperature anomalies in specific heat $c(T)$ are denoted with 
	green triangles (\textcolor{green}{$\triangledown$}); peaks in the nematic 
	order parameter susceptibility [cf. Eq.~(\ref{eq:phi.perp})], 
	with blue triangles (\textcolor{blue}{$\triangle$}).
	}
	\label{fig:phase.diagram.B}
\end{figure}

Studies of \CCO\ in applied magnetic field have already proved 
very useful in, e.g. providing estimates of the microscopic exchange 
parameters of the BBK model [cf. Section~\ref{section:bbk.model.cco}].
In what follows we use magnetic field as a tool to learn more about the 
nature and origin of its spin--liquid phase, starting with the thermodynamic 
properties found in MC simulations, for parameters 
corresponding to \CCO\ [Table~\ref{tab:experimental.parameters}].
Key results are summarised in the phase diagram, Fig.~\ref{fig:phase.diagram.B}.

%%%%%%%%%%%%%%%%%%%%%%%%%%%%%%%%%%%%%
%  Fig. X - magnetisation as function of field
%%%%%%%%%%%%%%%%%%%%%%%%%%%%%%%%%%%%%

\begin{figure}[t]
	\centering
	\captionsetup[subfigure]{labelformat=empty}
	\begin{minipage}[t]{0.99\columnwidth}
		\subfloat[\label{fig:specH_B}]{
  		\includegraphics[width=0.95\columnwidth]{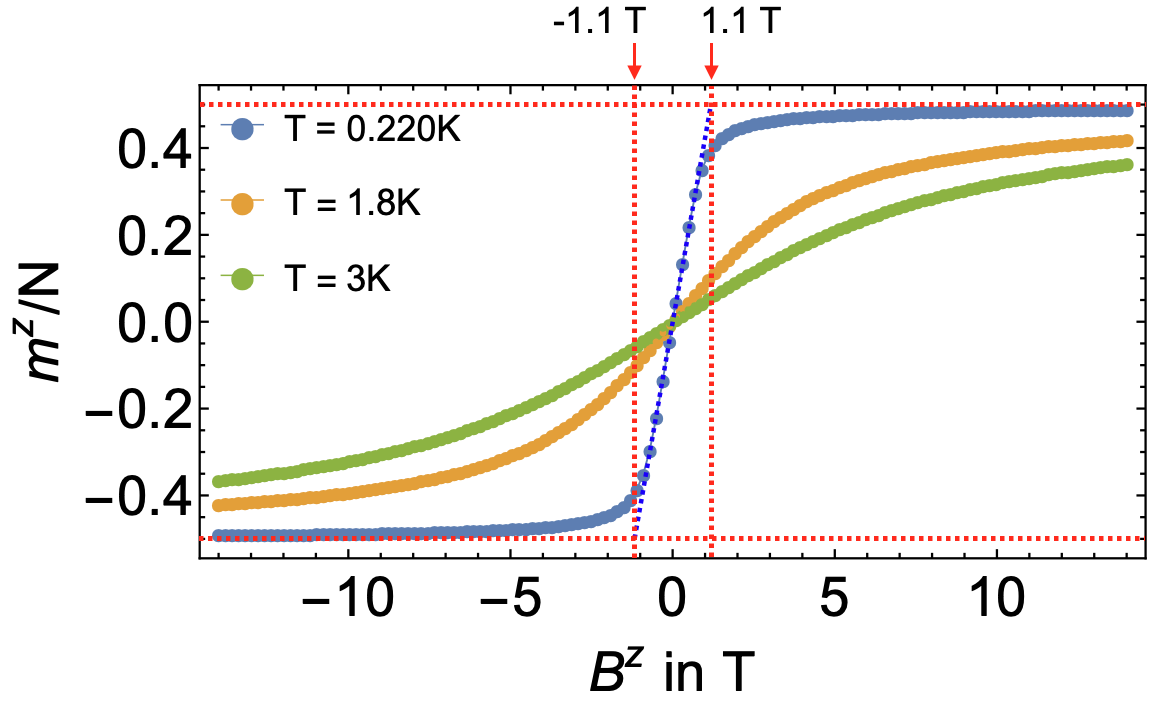}
		}
	\end{minipage}
	\caption{
	Magnetisation of the classical BBK model of \CCO\ as a function of 
	magnetic field, showing strong dependence on temperature.
	At the lowest temperatures, magnetisation approaches saturation, $|m^z| \to 0.5$
	(horizontal dashed line) for $|B| > 1.1\ \text{T}$, the saturation field predicted 
	by linear spin-wave theory at $T=0$ (vertical dashed line).
	At higher temperatures very large fields are needed to 
	saturate the magnetization, consistent with a high density 
	of thermally--excited spin excitations.
	Results are taken from classical Monte Carlo simulation of 
	${\mathcal H}_{\sf BBK}$ [Eq.~(\ref{eq:H.BBK})], for a cluster of linear dimension 
	L = 48 \mbox{(N = 13,824)}, with parameters taken 
	from experiment [Table~\ref{tab:experimental.parameters}].	
	}
	\label{fig:M.as.function.B}
\end{figure}

%%%%%%%%%%%%%%%%%%%%%%%%%%%%%%%%%%%%%
%  Fig. X - magnetisation as function of field
%%%%%%%%%%%%%%%%%%%%%%%%%%%%%%%%%%%%%

\begin{figure}[t]
	\centering
	\captionsetup[subfigure]{labelformat=empty}
	\begin{minipage}[t]{0.99\columnwidth}
		\subfloat[\label{fig:specH_B2}]{
  		\includegraphics[width=0.95\columnwidth]{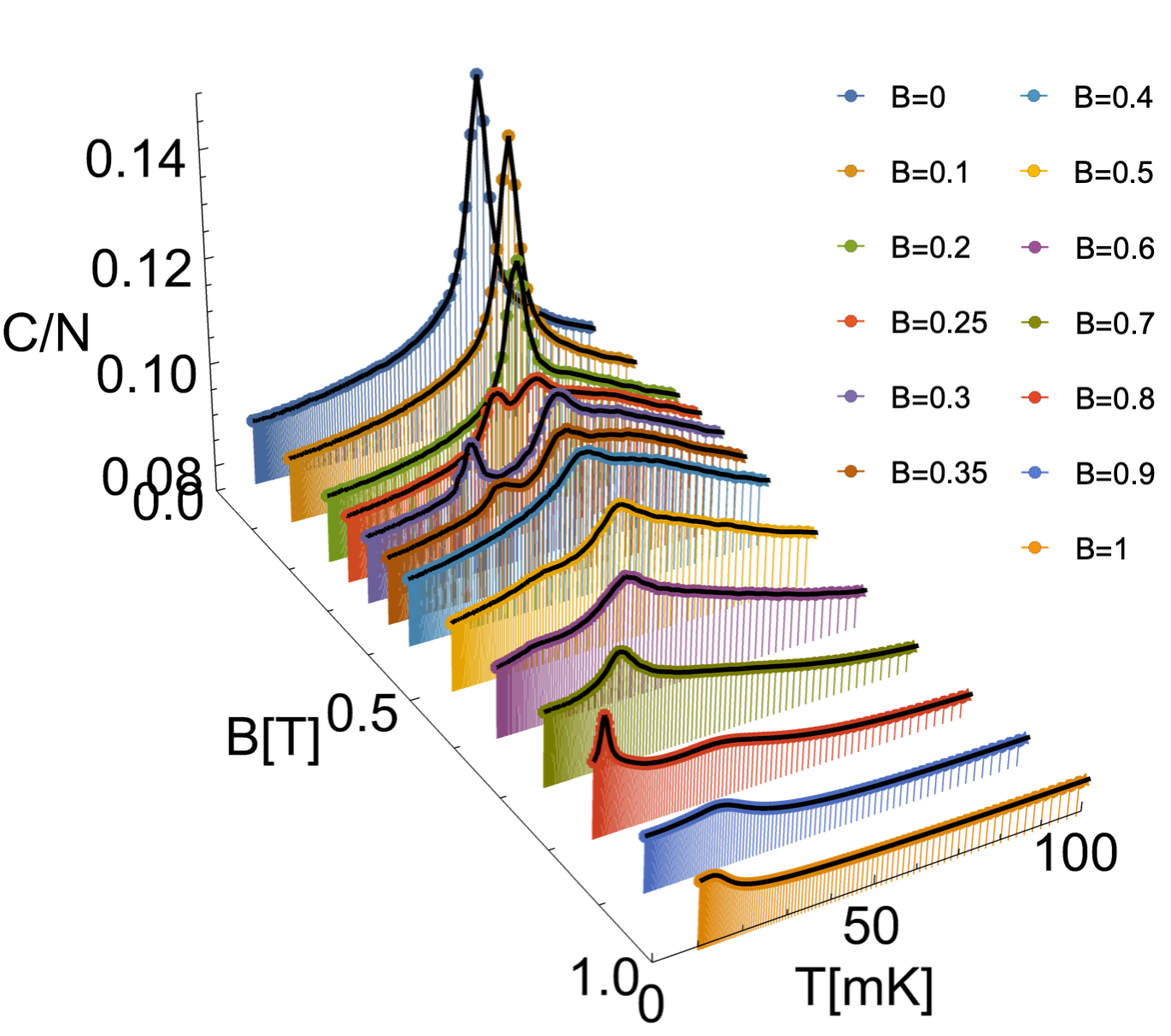}
		}
	\end{minipage}
	\caption{
	Heat capacity of BBK model of \CCO\ as a function of 
	magnetic field.
	Results are shown for $T < 100\ \text{mK}$, with magnetic field ranging 
	from $B=0\ \text{T}$ [blue curve, top of panel] to $B=1.0\ \text{T}$ 
	[orange curve, bottom of panel].
	A number of different anomalies are observed, 
	distinguishing different (quasi--)ordered phases, as described in the 
	main text.
	Results are taken from classical Monte Carlo simulation of 
	${\mathcal H}_{\sf BBK}$ [Eq.~(\ref{eq:H.BBK})], for a cluster of linear dimension 
	L = 18 \mbox{(N = 1944)}, with parameters taken 
	from experiment [Table~\ref{tab:experimental.parameters}].	
	}
	\label{fig:c.as.function.B}
\end{figure}

%%%%%%%%%%%%%%%%%%%%%%%%%%%%%%%%%%%%%
\subsection{Magnetisation in field}
%%%%%%%%%%%%%%%%%%%%%%%%%%%%%%%%%%%%%
\label{sec:magnetisation.in.field}

We consider first the magnetisation, $m(B)$.
In Fig.~\ref{fig:M.as.function.B} we show MC simulation results for $m(B)$, 
at temperatures of $T = 220\ \text{mK}$ (within the spin liquid) and 
$T = 1.8\ \text{K}$ and $3\ \text{K}$, (within a high-temperature 
paramagnetic phase) [cf. Fig.~\ref{fig:phase.diagram.B}].
MD simulations find a spin liquid which is gapless in zero field [Fig.~\ref{fig:S.zero.field}], 
consistent with both expectations for an O(3)--invariant classical 
model, and experiment on \CCO\ \cite{Balz2016}.
In keeping with this, simulations at $T = 220\ \text{mK}$, reveal 
that the spin liquid has a finite magnetic susceptibility, with a nearly linear 
behavior of $m(B)$ up to a field $B \sim 0.7\ \text{T}$.
Interestingly, the susceptibility within the spin liquid at this temperature 
is very close to the value found by dividing the saturated moment $m=0.5$ 
by the zero--temperature saturation field found in spin wave theory, 
$B = 1.1\ \text{T}$ (vertical dashed line).
For $B > 0.7\ \text{T}$, $m(B)$ is more ``rounded'', and tends towards the 
full saturated moment (horizontal dashed line) for $B \approx 10\ \text{T}$.

%%%%%%%%%%%%%%%%%%%%%%%%%%%%%%%%%%%%%

Results for $m(B)$ in the paramagnetic phase, for $T = 1.8\ \text{K}$ and $3\ \text{K}$, 
also show a finite magnetic susceptibility, but no hint of saturation up to the highest fields
simulated.
We note that the failure of the magnetisation to saturate at these temperatures is an 
artifact of classical statistics, and is not reproduced  by quantum 
simulations \cite{shimokawa-in-prep}.

%%%%%%%%%%%%%%%%%%%%%%%%%%%%%%%%%%%%%
\subsection{Heat capacity in field}
%%%%%%%%%%%%%%%%%%%%%%%%%%%%%%%%%%%%%
\label{sec:heat.capacity.in.field}

More can be learned by looking at the heat capacity, $c(T)$ under applied 
magnetic field.
Results taken from MC simulations are shown in Fig.~\ref{fig:c.as.function.B}.
Many of the large--scale features observed in the absence of field persist; 
in particular, the broad shoulder at $T \sim 500\ \text{mK}$ (not shown) is little 
affected by fields $B \lesssim 1\ \text{T}$, suggesting that spin excitations 
below this temperature scale remain collective.
And, for magnetic fields $B \lesssim 0.7\ \text{T}$, we can identify a sharp peak
in specific heat in $c(T)$ which connects smoothly with the phase transition 
from spin--liquid to $Z_3$ lattice nematic observed in zero field [Fig.~\ref{fig:heat.capacity}].

%%%%%%%%%%%%%%%%%%%%%%%%%%%%%%%%%%%%%

However, for fields \mbox{$ B \gtrsim 0.2\ \text{T}$}, new features start to emerge.
In particular, for \mbox{$0.225 < B < 0.375\ \text{T}$}, the sharp anomaly 
in $c(T)$ associated with the onset of lattice--rotation symmetry splits into two peaks.
Moreover for \mbox{$B \gtrsim 0.5\ \text{T}$}, a further weak anomaly (shoulder) 
becomes visible in $c(T)$, at a temperature higher than any sharp peak.
This new feature moves steadily to lower temperature with increasing field, 
finally interpolating to $T \to 0$ at the critical field found in spin wave theory 
$B_c = 1.1\ \text{T}$.   

%%%%%%%%%%%%%%%%%%%%%%%%%%%%%%%%%%%%%

These features divide the low--temperature phase diagram into several  
distinct regions, which we characterise below.

%%%%%%%%%%%%%%%%%%%%%%%%%%%%%%%%%%%%%
\subsection{Evolution of $Z_3$ lattice--nematic order in field}
%%%%%%%%%%%%%%%%%%%%%%%%%%%%%%%%%%%%%
\label{sec:Z3.in.field}

\begin{figure}[t]
	\centering
	\captionsetup[subfigure]{labelformat=empty}
	\begin{minipage}[t]{0.99\columnwidth}
		\subfloat[\label{fig:specH_chi_B0.3}]{
  		\includegraphics[width=0.95\columnwidth]{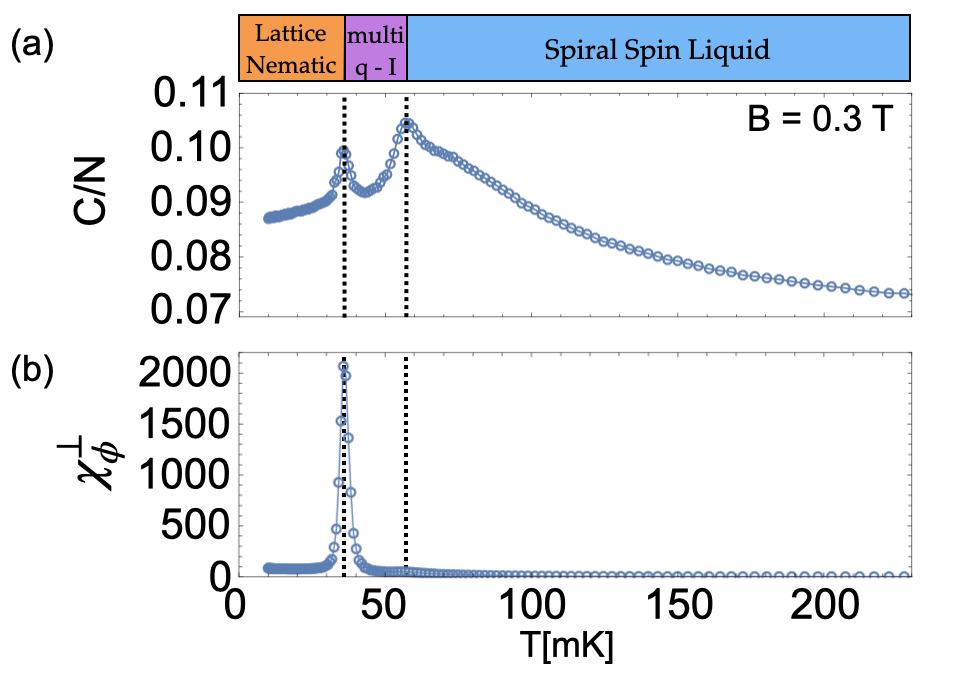}} \\
		\vspace{-1cm}
		\subfloat[\label{fig:Sq_B0.3a}]{
  		\includegraphics[width=0.95\columnwidth]{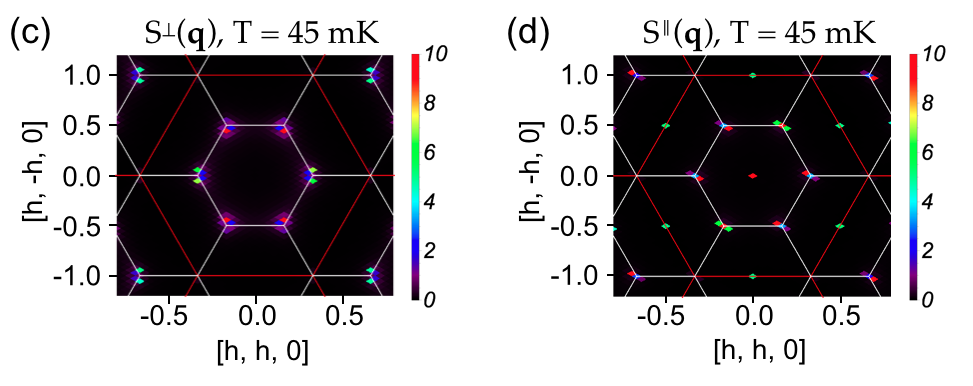}}\\
		\vspace{-1cm}
		\subfloat[\label{fig:Sq_B0.3b}]{
  		\includegraphics[width=0.95\columnwidth]{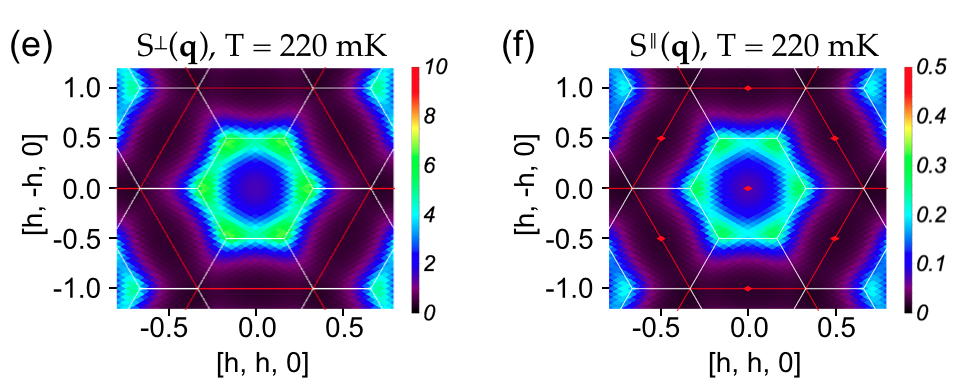}}
	\end{minipage}
	\caption{
	Thermodynamic properties of BBK model of  \CCO\ at \mbox{$B = 0.3$T}
	showing evidence for a multiple--q state separating the $Z_3$--ordered 
	lattice--nematic from the spiral spin liquid. 
	(a) Heat capacity $C(T)/N$, showing distinct peaks at 
	$35(2)\ \text{mK}$ and  
	$58(2)\ \text{mK}$.
	(b) Susceptibility associated $Z_3$ order parameter, $\chi^{\perp}_\phi(T)$ 
	[Eqs.~(\ref{eq:chi.phi}), (\ref{eq:phi.perp})], showing sharp peak at $35(2)\ \text{mK}$.
	(c) Transverse structure factor $S_\perp ({\bf q})$  at \mbox{$T = 45$mK}, within the 
	``multiple--q (I)'' phase, showing a matrix of points associated with a multiple--q state 
	near to  zone corners.
	(d) Longitudinal structure factor $S_\parallel ({\bf q})$ at the same temperature.
	Zone-center peaks reflect the finite magnetisation of the system.
	(e) Transverse structure factor $S_\perp ({\bf q})$  at \mbox{$T = 220$mK}, showing 
	diffuse scattering features, as observed in zero field [see Fig.\ref{fig:thermodyn} (c)]. 
	(f) Longitudinal structure factor $S_\parallel ({\bf q})$ at \mbox{$T = 220$mK}, showing the 
	same diffuse scattering features as in (e) with additional Bragg peaks at Brillouin zone 
	centers, accounting for spin polarization in field. 
		%	
%	The sharp anomaly in $C(T)/N$ 
%	splits into two peaks, with the peak at lower temperature tracking the maximum in 
%	$\chi^{\perp}_\phi(T)$, well separating the $Z_3$ lattice nematic from a novel multi-q ordered 
%	state. 
	%
	Results are taken from classical Monte Carlo simulation of 
	the BBK model Eq.~(\ref{eq:H.BBK}), for a cluster of linear dimension 
	\mbox{L = 18 \mbox{(N = 1944)}}, with parameters taken 
	from experiment [Table~\ref{tab:experimental.parameters}].	
	To distinguish multiple--q states, simulations have been performed in absence 
	of parallel tempering [Appendix ~\ref{appendix:Numerics}]. 
	}
	\label{fig:inField_0.3}
\end{figure}

We first consider the phase found at low temperature and low field,  
which connects with the $Z_3$ lattice--nematic order found for $B=0$. 
The order parameter for lattice--nematic order, Eq.~(\ref{eq:phi}), can 
be generalised for finite magnetic fields, as
\begin{eqnarray}
\phi_\nu &=& \sum_{{\bf r} \in \hexagon} \phi_\nu ({\bf r}) \; ,
\label{eq:phi.nu} 
\end{eqnarray}		 
where
\begin{eqnarray}
\phi_\parallel ({\bf r}) &=& \frac{1}{S^2} \left[ 
		S^z_A({\bf r}) S^z_B({\bf r} + \hat{e}_0)  
		 + \omega S^z_A ({\bf r}) S^z_B ({\bf r} + \hat{e}_1)  \right. \nonumber\\
		 && \qquad \left .+ \omega^2 S^z_A ({\bf r}) S^z_B ({\bf r} + \hat{e}_2)
		 \right] \; , \\
\phi_\perp ({\bf r}) &=& \frac{1}{S^2} \left[ 
		{\bf S}^\perp_A({\bf r})  {\bf S}^\perp_B({\bf r} + \hat{e}_0)  
		 + \omega {\bf S}^\perp_A ({\bf r})  {\bf S}^\perp_B ({\bf r} + \hat{e}_1)  \right. \nonumber\\
		 && \qquad \left .+ \omega^2 {\bf S}^\perp_A ({\bf r})  {\bf S}^\perp_B ({\bf r} + \hat{e}_2)
		 \right] \; , 
		 \label{eq:phi.perp}
\end{eqnarray}		 
where the conventions for labelling sites within the plaquettes A and B
are given in Fig.~\ref{fig:order.parameter}, and
\begin{eqnarray}		 
	{\bf S}^\perp ({\bf r}) &=& (S^x({\bf r}), S^y({\bf r}))	
	\label{eq:S.perp}	 
\end{eqnarray}
We find that the dominant contribution comes from $\phi_\perp$, 
and the associated order parameter susceptibility, $\chi^\perp_\phi(T)$, 
shows a sharp peak which tracks the anomaly in $c(T)$, 
confirming that this originates in the breaking of 
lattice--rotation symmetry [Fig.~\ref{fig:specH_chi_B0.3}].

%%%%%%%%%%%%%%%%%%%%%%%%%%%%%%%%%%%%%
%  Fig. X - multiple-q state
%%%%%%%%%%%%%%%%%%%%%%%%%%%%%%%%%%%%%

%%%%%%%%%%%%%%%%%%%%%%%%%%%%%%%%%%%%%
%  Fig. X - "pink" phase
%%%%%%%%%%%%%%%%%%%%%%%%%%%%%%%%%%%%%

\begin{figure}[t]
	\centering
	\captionsetup[subfigure]{labelformat=empty}
	\begin{minipage}[t]{0.99\columnwidth}
		\subfloat[\label{fig:specH_chi_B0.8}]{
  		\includegraphics[width=0.95\columnwidth]{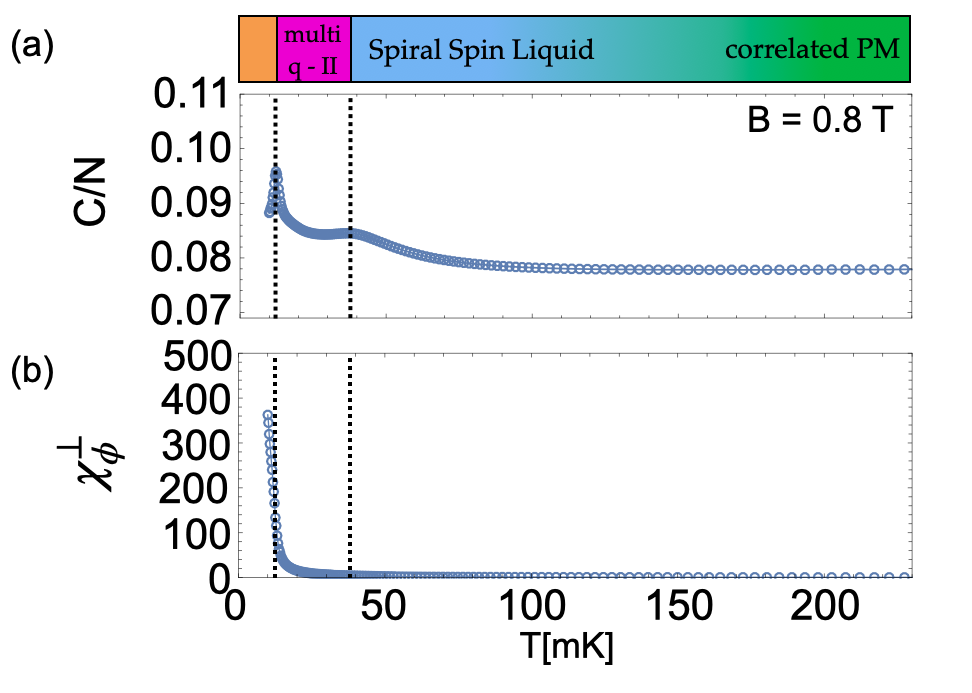}} \\
		\vspace{-1cm}
		\subfloat[\label{fig:Sq_B0.8a}]{
  		\includegraphics[width=0.95\columnwidth]{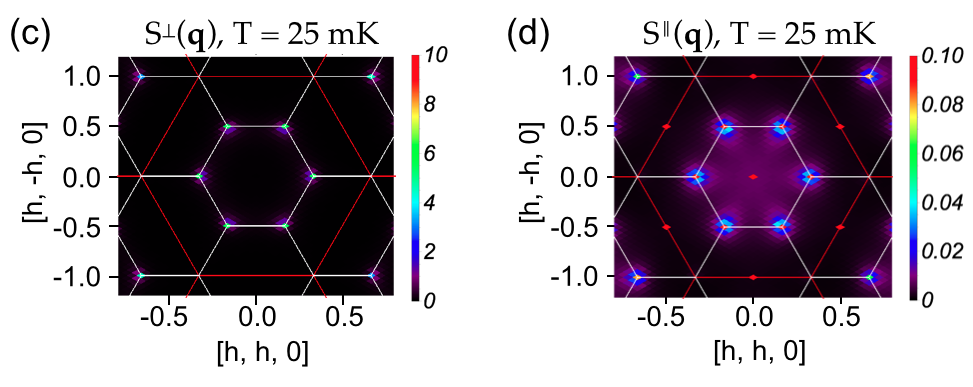}}\\
		\vspace{-1cm}
		\subfloat[\label{fig:Sq_B0.8b}]{
  		\includegraphics[width=0.95\columnwidth]{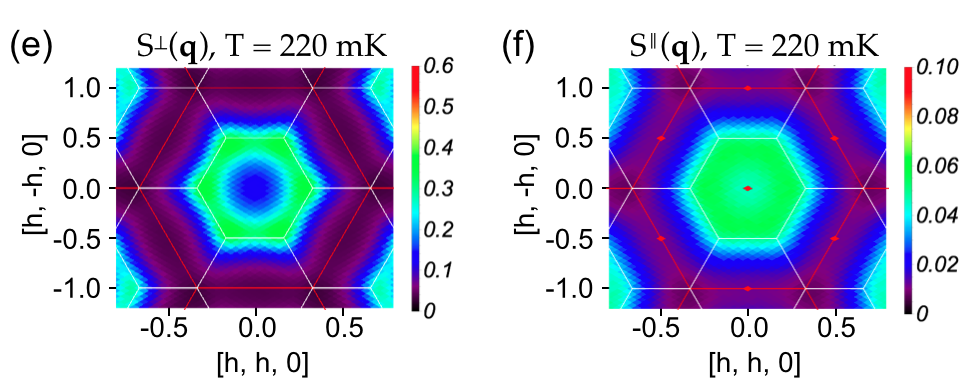}}
	\end{minipage}
	\caption{
	Thermodynamics of BBK model of \CCO\ at \mbox{$B = 0.8$T}.
	%showing features associated with spin-3/2 correlated paramagnet (PM); 
	%spiral-spin liquid;  '``pink phase''; and $Z_3$--ordered lattice--nematic.
	%
	(a) Heat capacity $C(T)/N$, showing shoulder around  
	$\sim 40\ \text{mK}$, associated with the onset of the ``multiple--q (II)'' phase, 
	and sharp peak at $11(2)\ \text{mK}$, marking phase transition into 
	$Z_3$ lattice nematic.
	The transition between the correlated paramagnet and spiral spin liquid 
	is not accompanied by any visible anomaly in $C(T)/N$.
	(b) Susceptibility associated with $Z_3$ order parameter, 
	$\chi^{\perp}_\phi(T)$ [Eqs.~(\ref{eq:chi.phi}), (\ref{eq:phi.perp})], 
	showing sharp peak at $11(2)\ \text{mK}$.
	(c) Transverse structure factor $S_\perp ({\bf q})$  at \mbox{$T = 25$mK} within the 
	``multiple--q (II)'' phase, showing points associated with a multiple--q state near to  zone corners.
	(d) Longitudinal structure factor $S_\parallel ({\bf q})$ at the same temperature.
	(e) Transverse structure factor $S_\perp ({\bf q})$  at \mbox{$T = 220$mK}, showing 
	diffuse scattering features, as observed in Fig.\ref{fig:inField_0.3}(e).
	(f) Longitudinal structure factor $S_\parallel ({\bf q})$ at \mbox{$T = 220$mK}, showing 
	qualitatively different features, compared to Fig.\ref{fig:inField_0.3}(f), emphasizing the 
	correlated nature of the paramagnetic phase. 
	Results are taken from classical Monte Carlo simulation of 
	the BBK model Eq.~(\ref{eq:H.BBK}), for a cluster of linear dimension 
	L = 18 \mbox{(N = 1944)}, with parameters taken 
	from experiment [Table~\ref{tab:experimental.parameters}].	
	To facilitate the understanding of order, measurements have been 
	performed in absence of parallel tempering.
	}
	\label{fig:inField_0.8}
\end{figure}

%%%%%%%%%%%%%%%%%%%%%%%%%%%%%%%%%%%%%%%%
% Fig. X - dynamics in high field
%%%%%%%%%%%%%%%%%%%%%%%%%%%%%%%%%%%%%%%%

\begin{figure*} 
	\centering
  	\subfloat[INS, T = 90mK \label{fig:HighField_a}]{
	\includegraphics[width=0.3\linewidth]{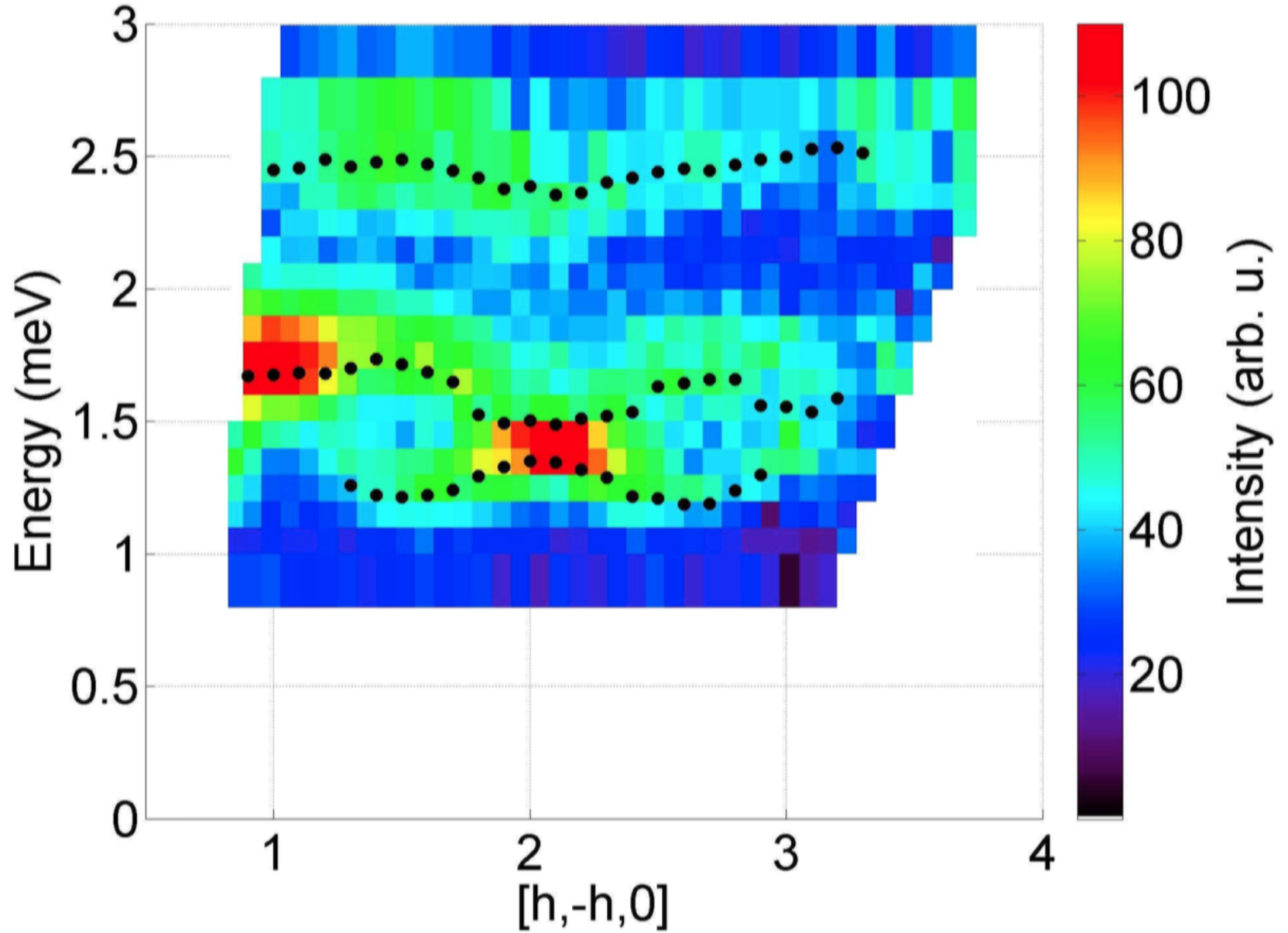}
	}
	\quad
	\subfloat[$S^{\sf LSW}_{\sf T = 0}({\bf q}, \omega)$   \label{fig:HighField_b}]{
	\includegraphics[width=0.275\linewidth]{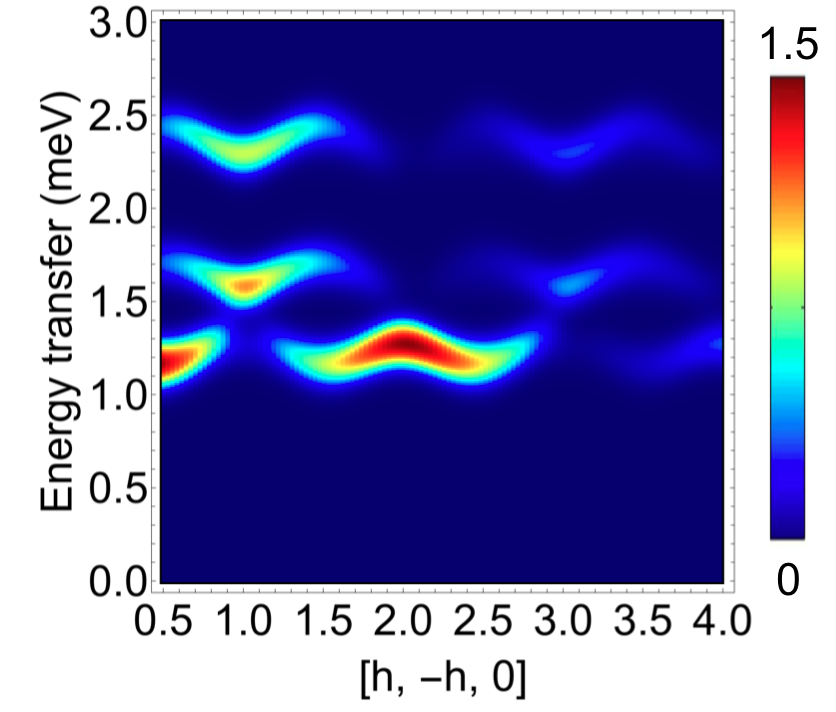}
	} 
	\quad
	\subfloat[$\tilde{S}^{\sf MD}_{\sf T = 10mK}({\bf q}, \omega)$  \label{fig:HighField_c}]{
	\includegraphics[width=0.275\linewidth]{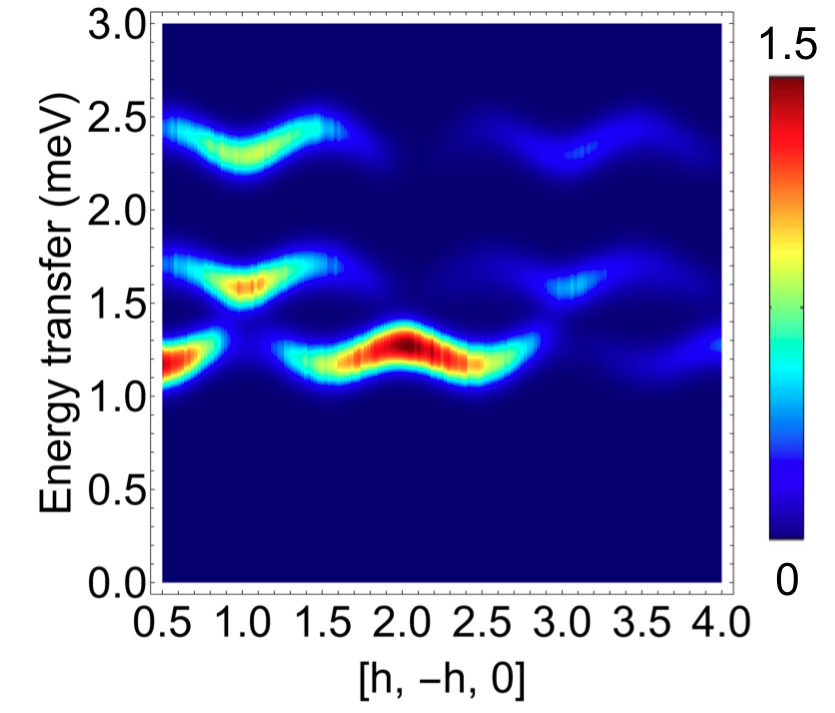}
	} 
	\\
	\subfloat[INS, T = 90mK\label{fig:HighField_d}]{
	\includegraphics[width=0.3\linewidth]{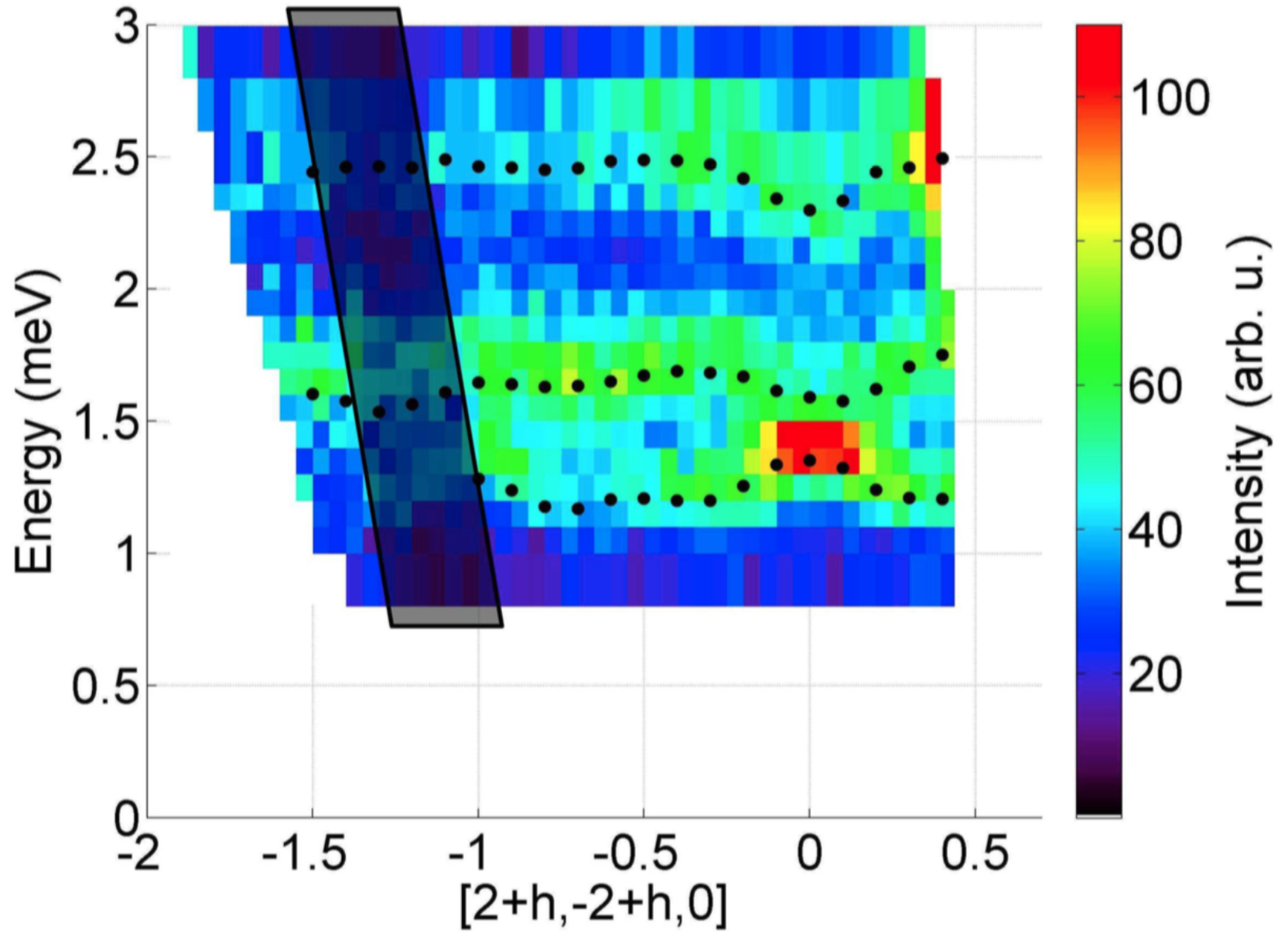}
	}
	\quad
	\subfloat[$S^{\sf LSW}_{\sf T = 0}({\bf q}, \omega)$  \label{fig:HighField_e}]{
	\includegraphics[width=0.275\linewidth]{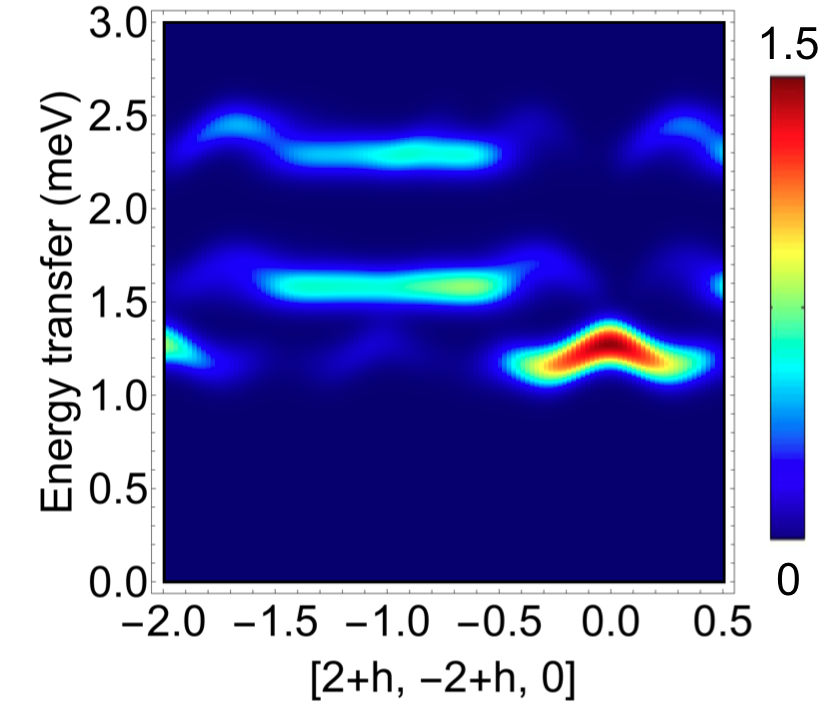}
	} 
	\quad
	\subfloat[$\tilde{S}^{\sf MD}_{\sf T = 10mK}({\bf q}, \omega)$  \label{fig:HighField_f}]{
	\includegraphics[width=0.275\linewidth]{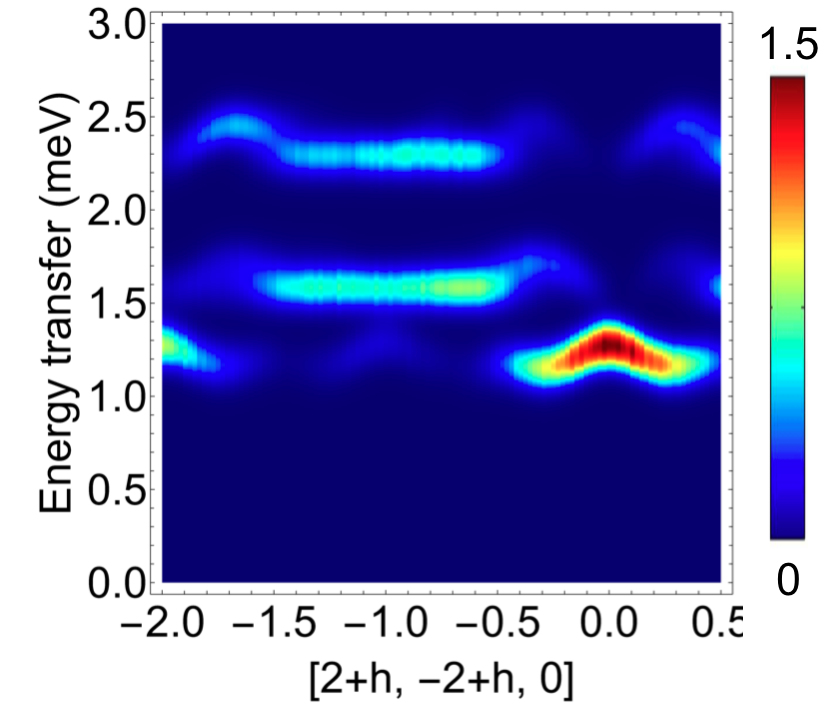}
	} 
	\\
	\subfloat[INS, E = 1.4meV, T = 90mK\label{fig:HighField_g}]{
	\includegraphics[width=0.3\linewidth]{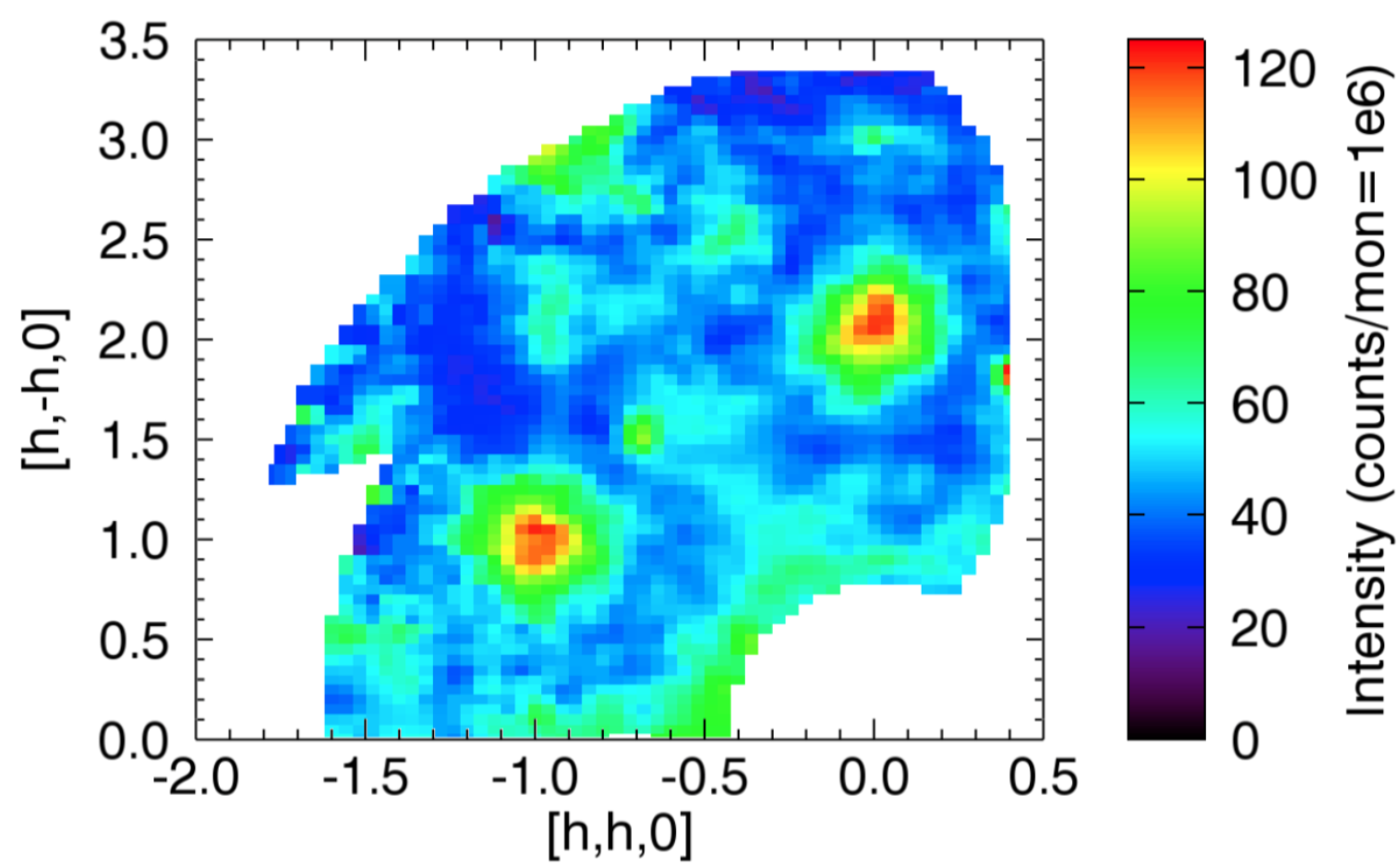}
	}
	\quad
	\subfloat[$S^{\sf LSW}_{\sf T = 0}({\bf q}, \omega = 1.4 ~ \text{meV})$ \label{fig:HighField_h}]{
	\includegraphics[width=0.28\linewidth]{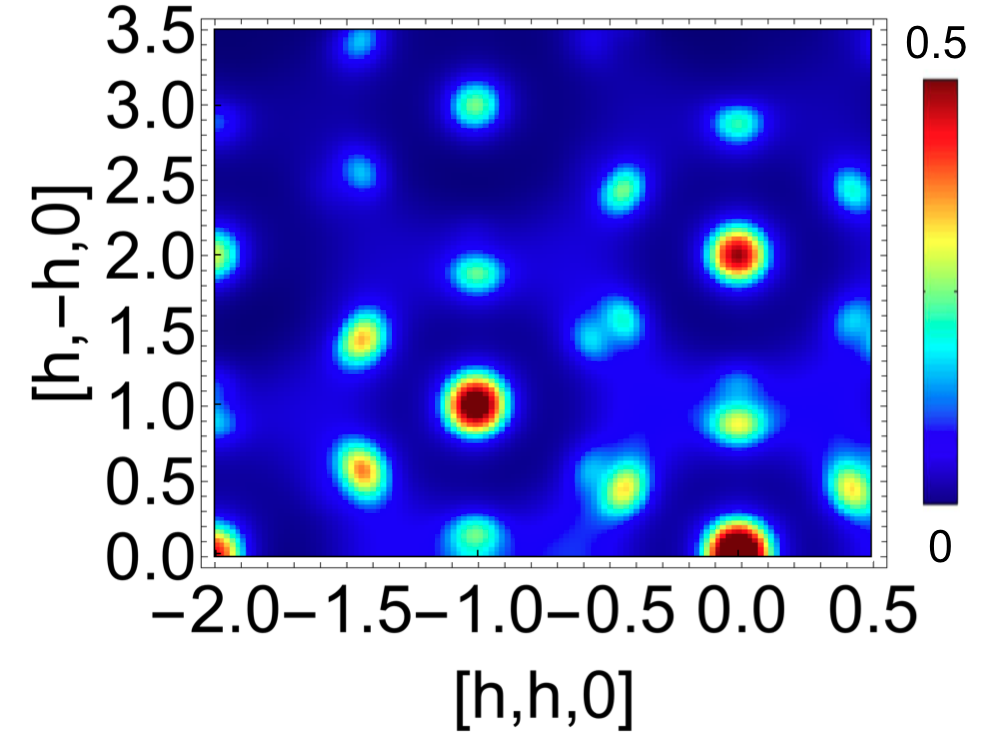}
	} 
	\quad
	\subfloat[ $\tilde{S}^{ \sf MD}_{\sf T = 10mK}({\bf q}, \omega = 1.4 ~ \text{meV})$ \label{fig:HighField_i}]{
	\includegraphics[width=0.28\linewidth]{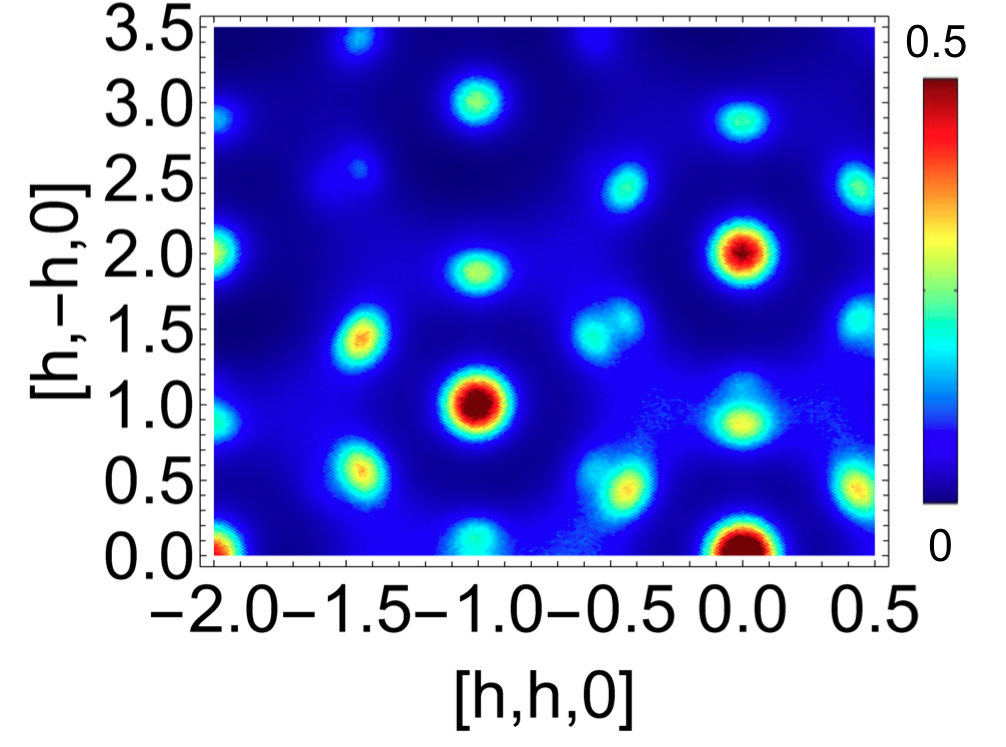}
	} 
	\caption{
	Spin dynamics of \CCO\ in the saturated state at $B = 11 $T.
	(a), (d), (g) Results for inelastic 
	neutron scattering (INS) experiments, reproduced from 
	\cite{Balz2016, Balz2017-JPCM29}, are compared 
	to (b), (e), (h) linear spin-wave (LSW) theory and (c), (f), (i) molecular dynamics 
	(MD) results of the bilayer breathing-kagome (BBK) model ${\cal H}_{\sf BBK}$
	[Eq.~(\ref{eq:H.BBK})], with parameters taken from Tab.\ref{tab:experimental.parameters}.
	(a)--(c) Gapped, dispersing spin-wave excitations along the \mbox{[h, -h, 0]} direction.
	(d)--(f) Gapped, dispersing spin-wave excitations along the \mbox{[2+h, -2 + h, 0]} direction.
	Constant energy cut at (g) -- (i) E = 1.4 meV shows bright features 
	corresponding to high intensities in the spin-wave dispersion. 
	In order to compare to INS data, results for LSW theory and MD simulations are presented 
	with a Gaussian convolution of \mbox{FWHM = 0.2 meV} and Cr$^{5+}$ form factor. 
	INS results were taken at $T = 90$mK, while LSW theory corresponds to the T = 0 quantum 
	case. 
	MD simulations have been performed at $T = 10$mK, while presented in its temperature
	corrected form $\tilde{S}(\omega, {\bf q})$ [Eq.~(\ref{eq:S.tilda})], in order to compare to
	T = 0 LSW theory. 
	}
	\label{fig:DynamicsHighField}
\end{figure*}

%%%%%%%%%%%%%%%%%%%%%%%%%%%%%%%%%%%%%

Meanwhile, for \mbox{$0.225 < B < 0.375\ \text{T}$}, where the heat--capacity 
peak splits, the maximum in $\chi^\perp_\phi(T)$ is found at the same temperature 
as the {\it lower} peak in $c(T)$ [Fig.~\ref{fig:inField_0.3}(b)], 
consistent with the existence of a lattice--nematic state at low temperatures.

%%%%%%%%%%%%%%%%%%%%%%%%%%%%%%%%%%%%%
\subsection{Competing (quasi--)ordered phases}
%%%%%%%%%%%%%%%%%%%%%%%%%%%%%%%%%%%%%
\label{sec:new.forms.of.order}

At temperatures between the two peaks in specific heat [Fig.~\ref{fig:specH_chi_B0.3}], 
we find a phase that lacks $Z_3$ order [Fig.~\ref{fig:inField_0.3}(b)], but shows 
spin correlations very similar to those observed 
in a triple--q state triangular lattice \cite{Okubo2012}; a known example
of a skyrmion lattice [Fig.~\ref{fig:inField_0.3}(c,d)].
At higher temperatures, these give way to the known correlations of the 
spiral spin liquid [Fig.~\ref{fig:inField_0.3}(e,f)]. 

%%%%%%%%%%%%%%%%%%%%%%%%%%%%%%%%%%%%%%%%
% Fig -  field evolution of spin excitations
%%%%%%%%%%%%%%%%%%%%%%%%%%%%%%%%%%%%%%%%
%
\begin{turnpage}
\begin{figure*}[h]
	\centering
	\captionsetup[subfigure]{labelformat=empty}
	\subfloat[]{
  		\includegraphics[width=0.202\textwidth]{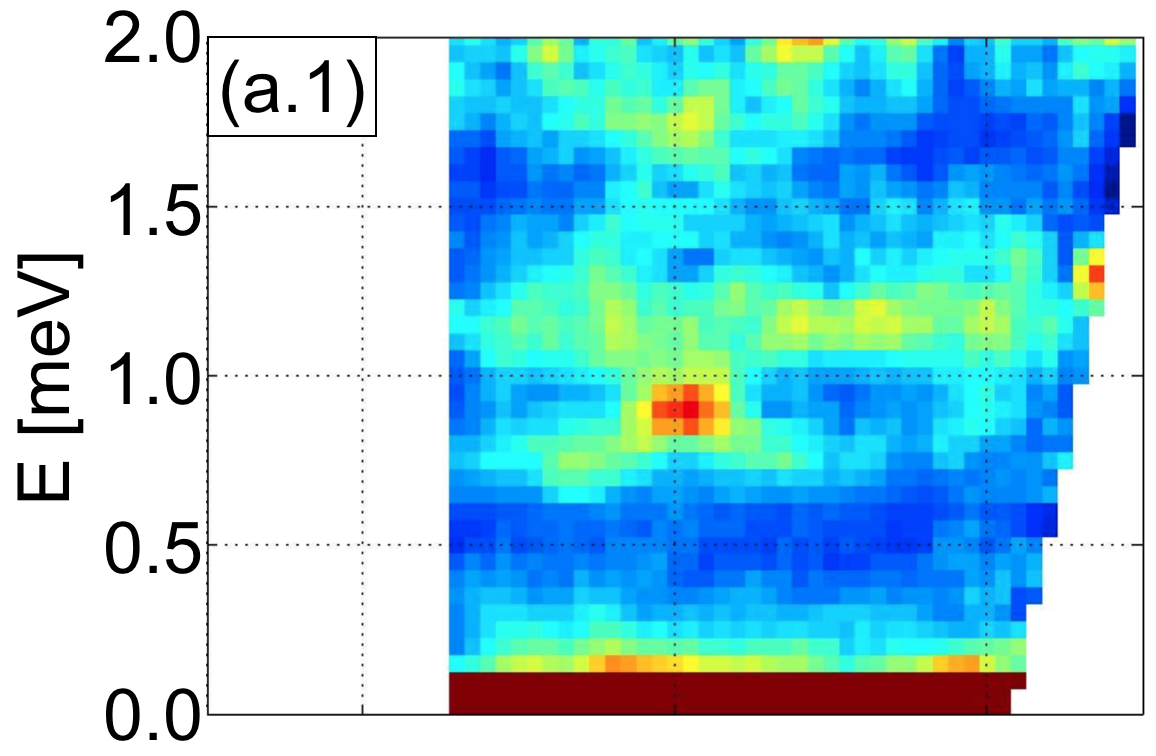}\llap{ 
		\parbox[b]{3cm}{\fcolorbox{black}{white}{$B = 7.5$T} \\\rule{0ex}{2.1cm}
		}}
	}
	\subfloat[]{
  		\includegraphics[width=0.17\textwidth]{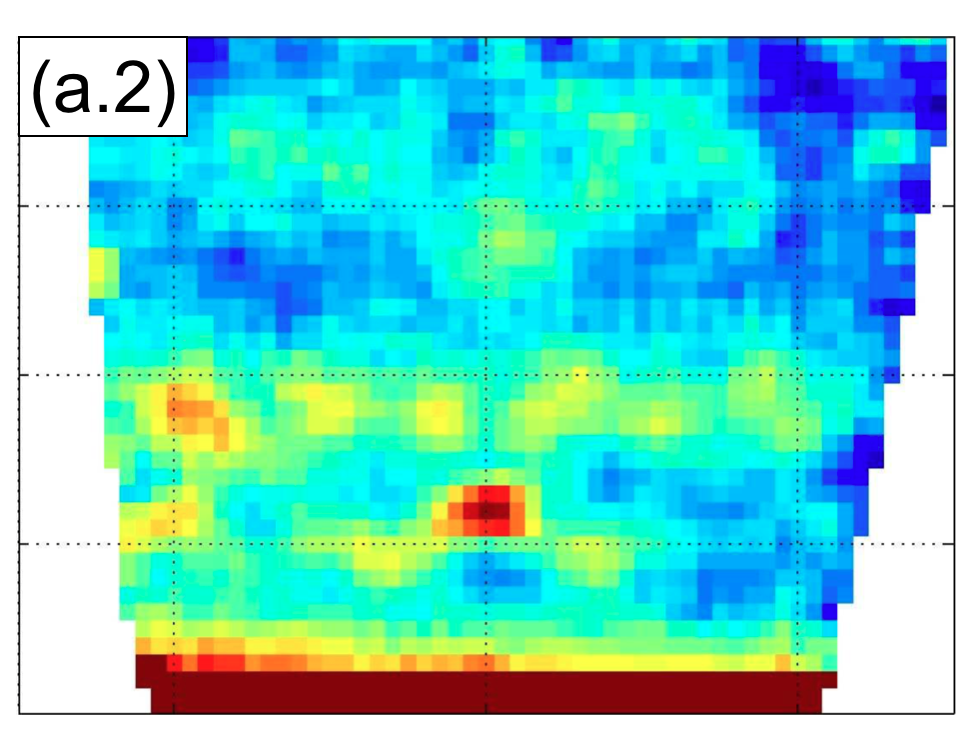}\llap{ 
		\parbox[b]{3cm}{\fcolorbox{black}{white}{$B = 5$T} \\\rule{0ex}{2.1cm}
		}}
	}
	\subfloat[]{
  		\includegraphics[width=0.17\textwidth]{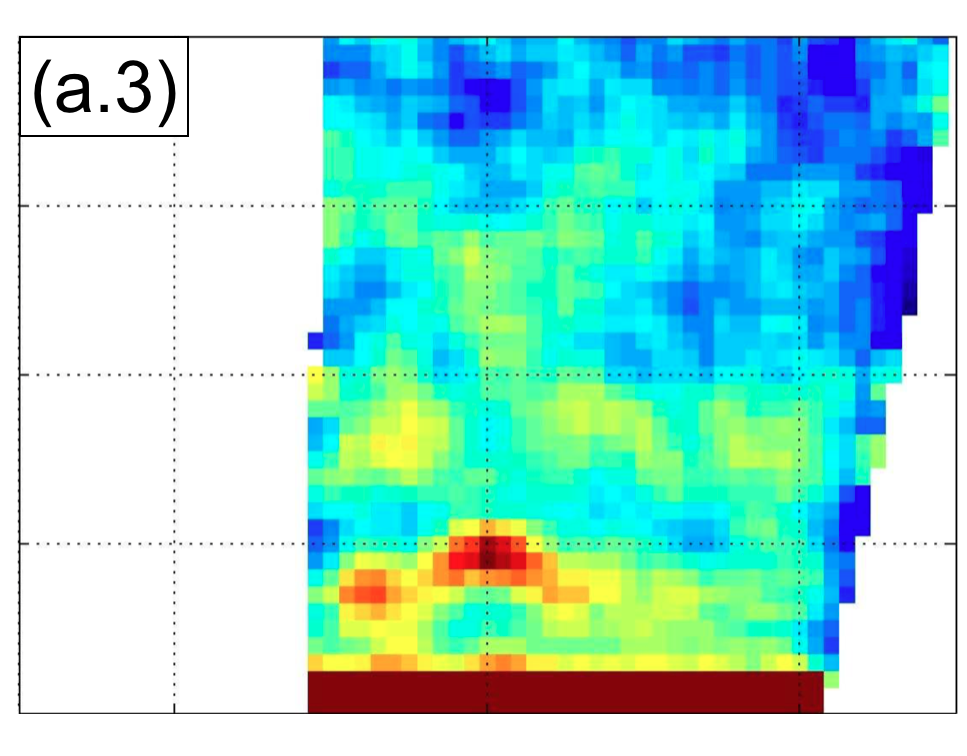}\llap{ 
		\parbox[b]{3cm}{\fcolorbox{black}{white}{$B = 4$T} \\\rule{0ex}{2.1cm}
		}}
	}
	\subfloat[]{
  		\includegraphics[width=0.17\textwidth]{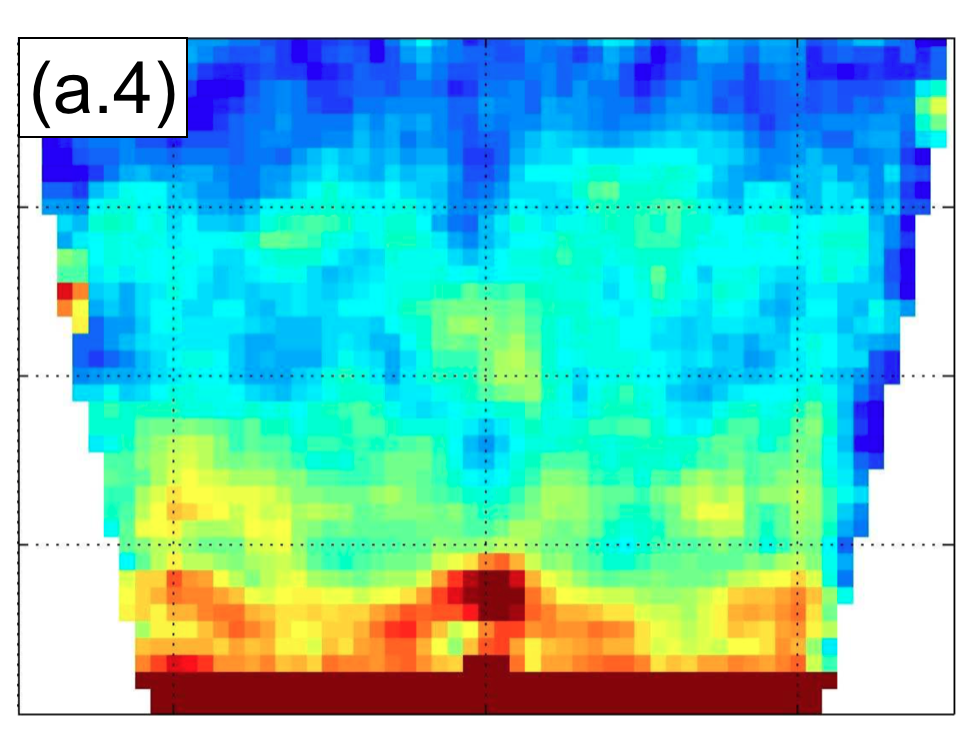}\llap{ 
		\parbox[b]{3cm}{\fcolorbox{black}{white}{$B = 3$T} \\\rule{0ex}{2.1cm}
		}}
	}
	\subfloat[]{
  		\includegraphics[width=0.17\textwidth]{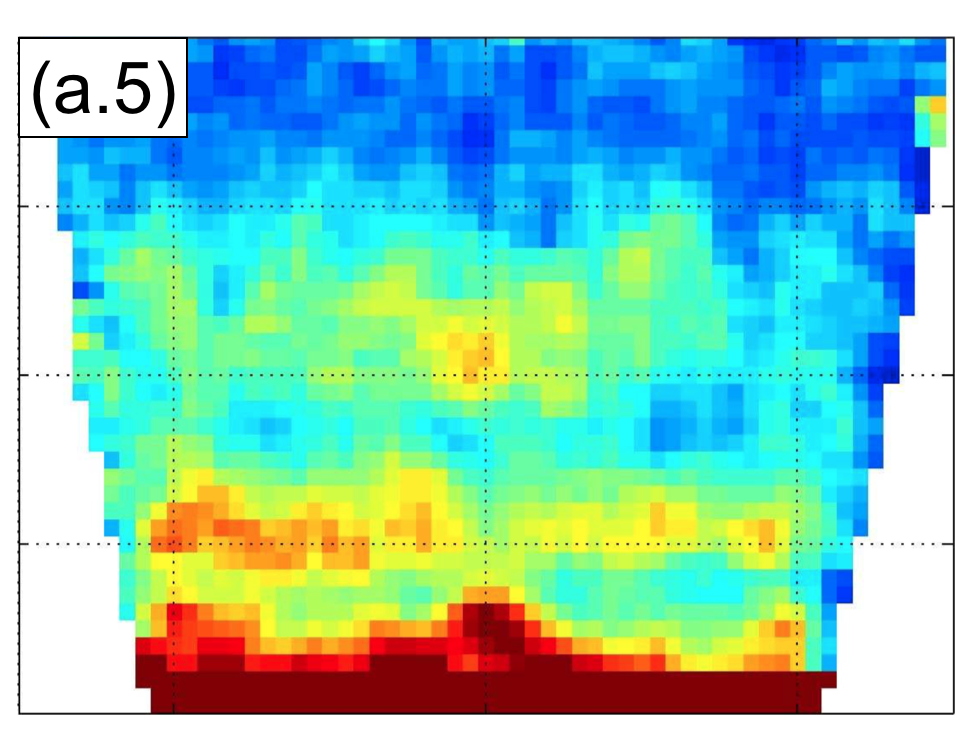}\llap{ 
		\parbox[b]{3cm}{\fcolorbox{black}{white}{$B = 2$T} \\\rule{0ex}{2.1cm}
		}}
	}
	\subfloat[]{
  		\includegraphics[width=0.17\textwidth]{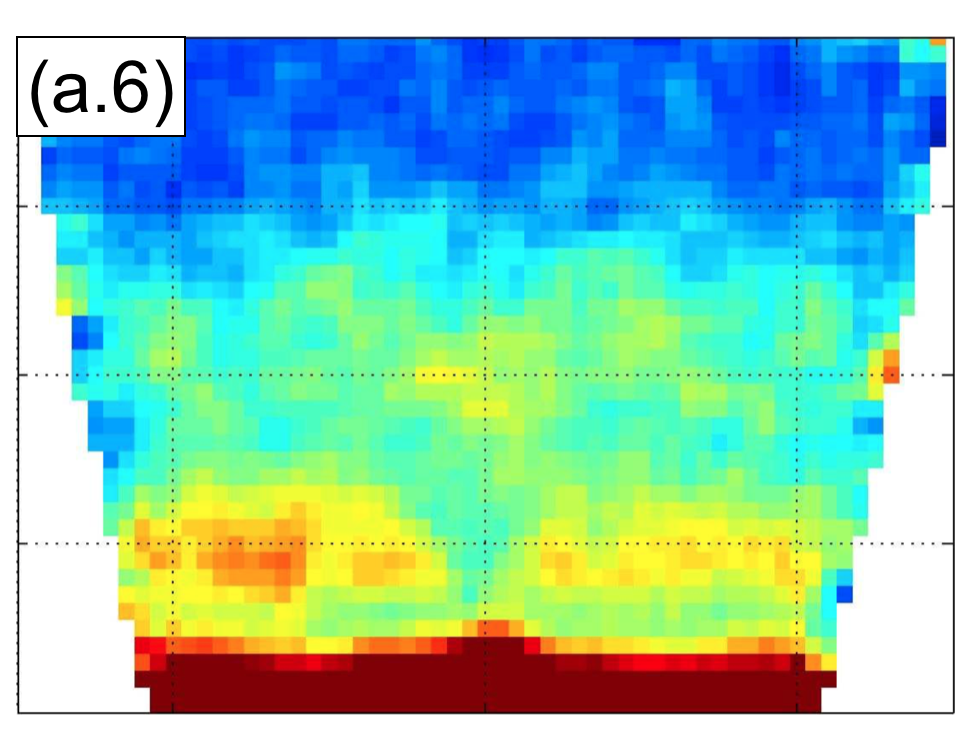}\llap{ 
		\parbox[b]{3cm}{\fcolorbox{black}{white}{$B = 1$T} \\\rule{0ex}{2.1cm}
		}}
	}
	\subfloat[]{
  		\includegraphics[width=0.198\textwidth]{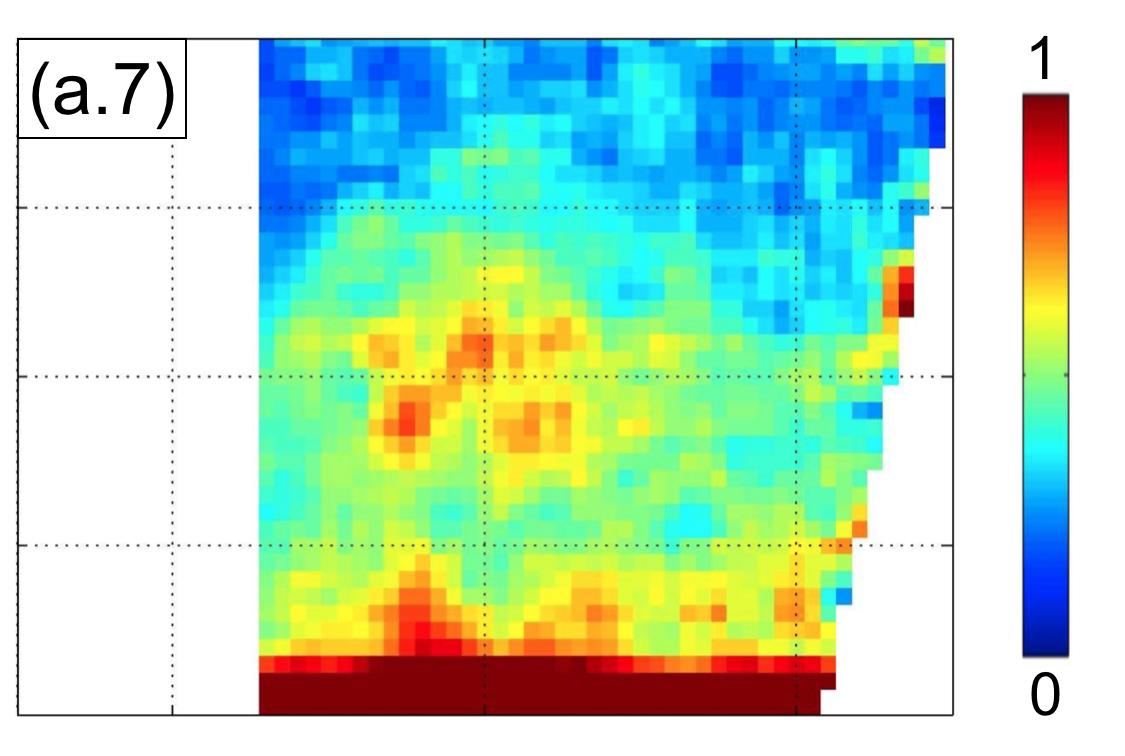}\llap{ 
		\parbox[b]{3cm}{\fcolorbox{black}{white}{$B = 0$T} \\\rule{0ex}{2.1cm}
		}}
	}\\[-6ex]
	\subfloat[]{
  		\includegraphics[width=0.202\textwidth]{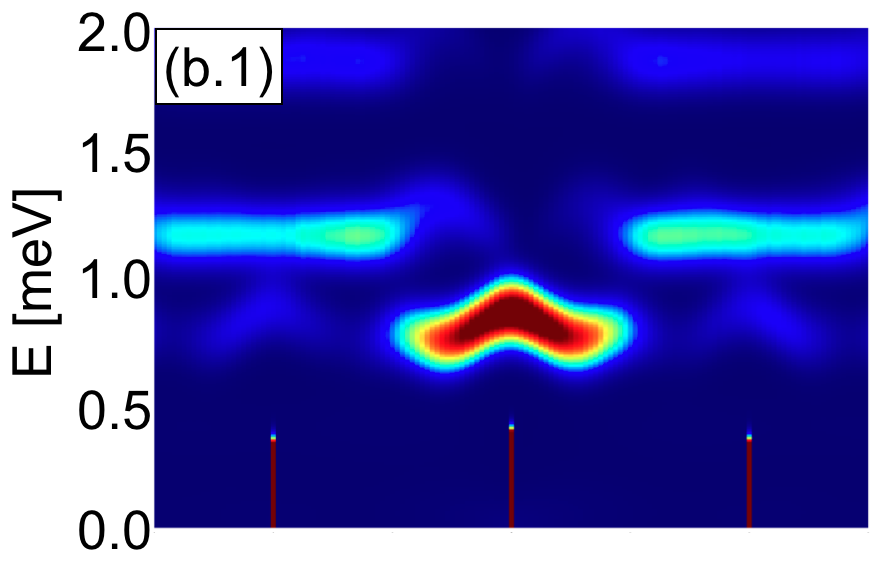}
	}
	\subfloat[]{
  		\includegraphics[width=0.17\textwidth]{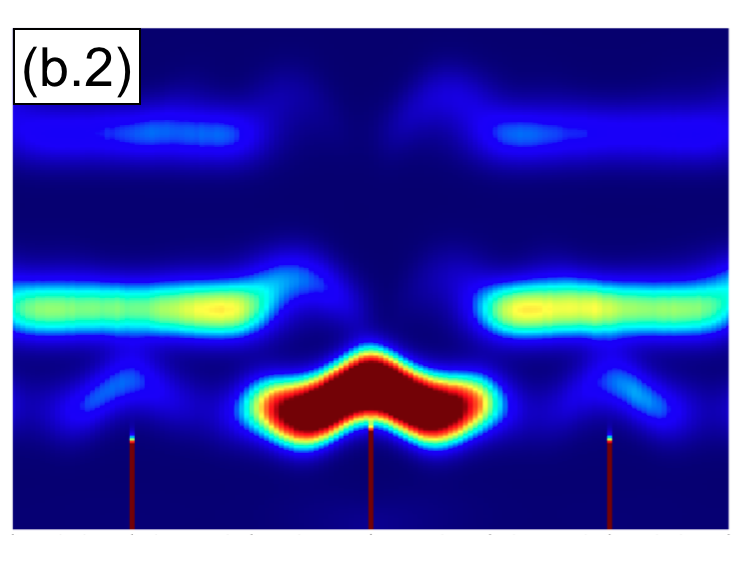}
	}
	\subfloat[]{
  		\includegraphics[width=0.17\textwidth]{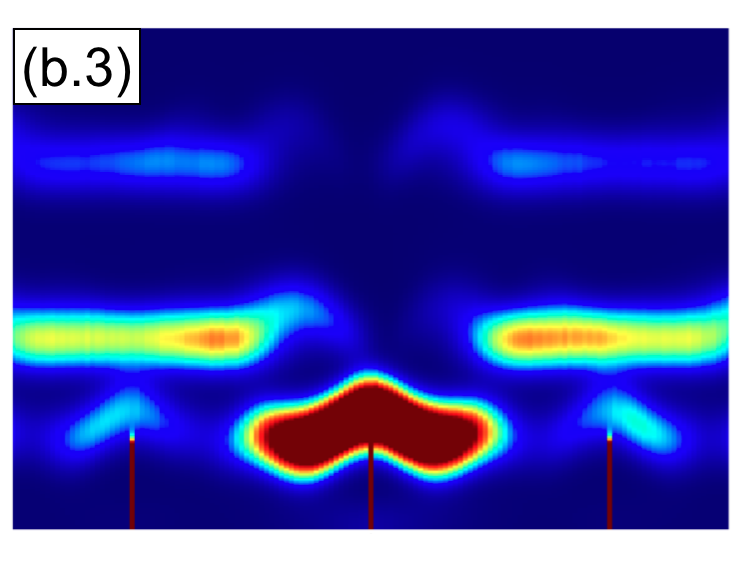}
	}
	\subfloat[]{
  		\includegraphics[width=0.17\textwidth]{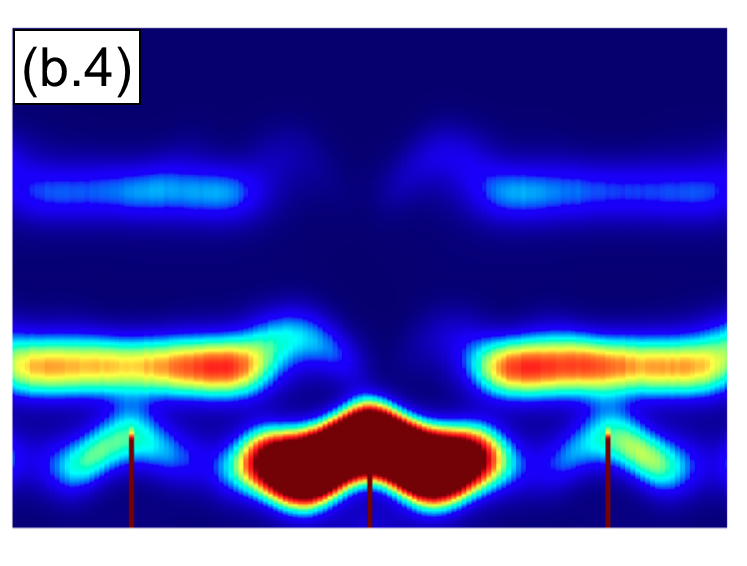}
	}
	\subfloat[]{
  		\includegraphics[width=0.17\textwidth]{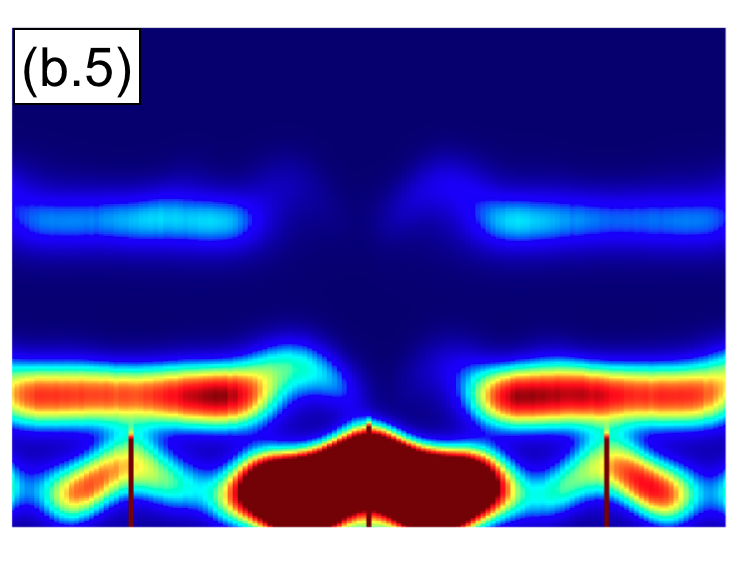}
	}
	\subfloat[]{
  		\includegraphics[width=0.17\textwidth]{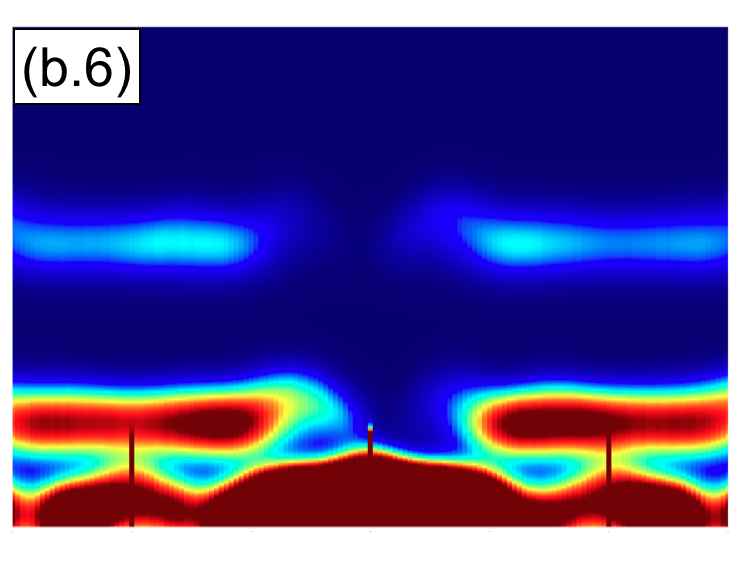}
	}
	\subfloat[]{
  		\includegraphics[width=0.193\textwidth]{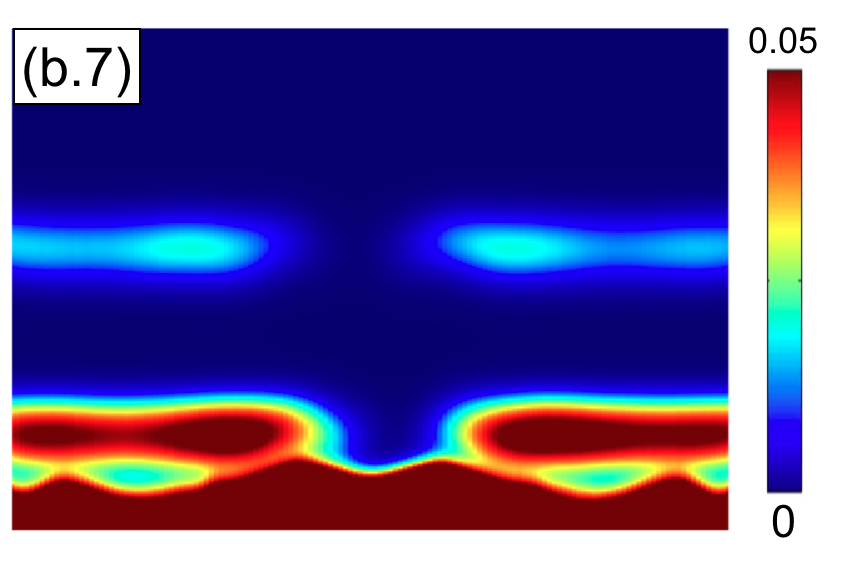}
	}\\[-6ex]
	\subfloat[]{
  		\includegraphics[width=0.202\textwidth]{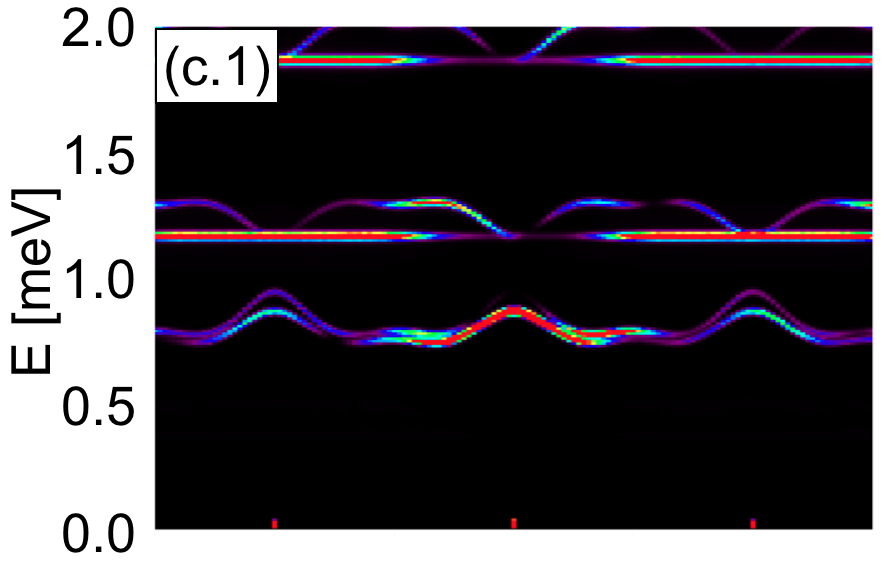}
	}
	\subfloat[]{
  		\includegraphics[width=0.17\textwidth]{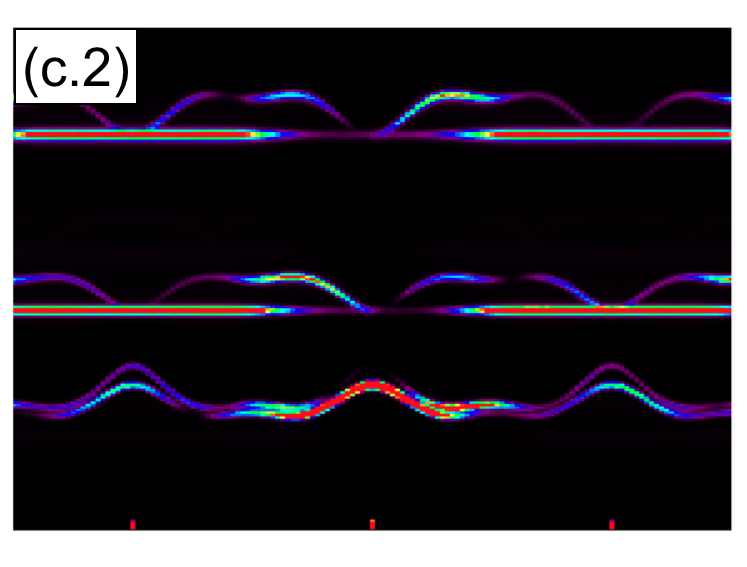}
	}
	\subfloat[]{
  		\includegraphics[width=0.17\textwidth]{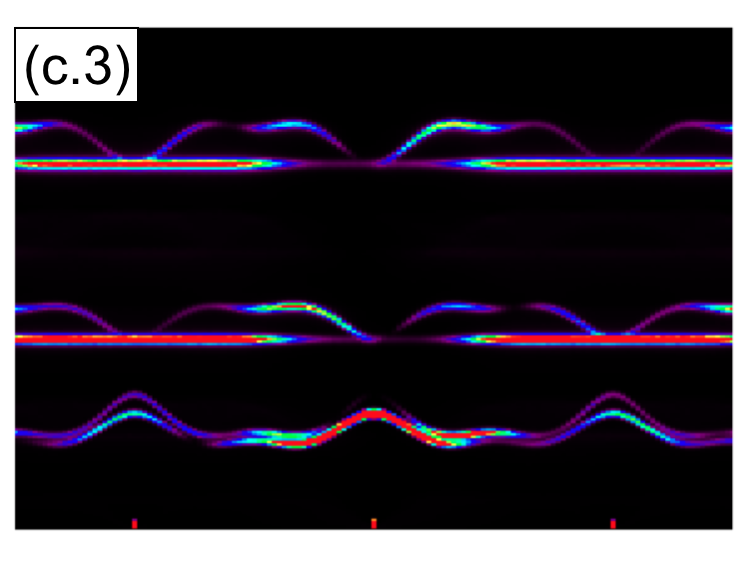}
	}
	\subfloat[]{
  		\includegraphics[width=0.17\textwidth]{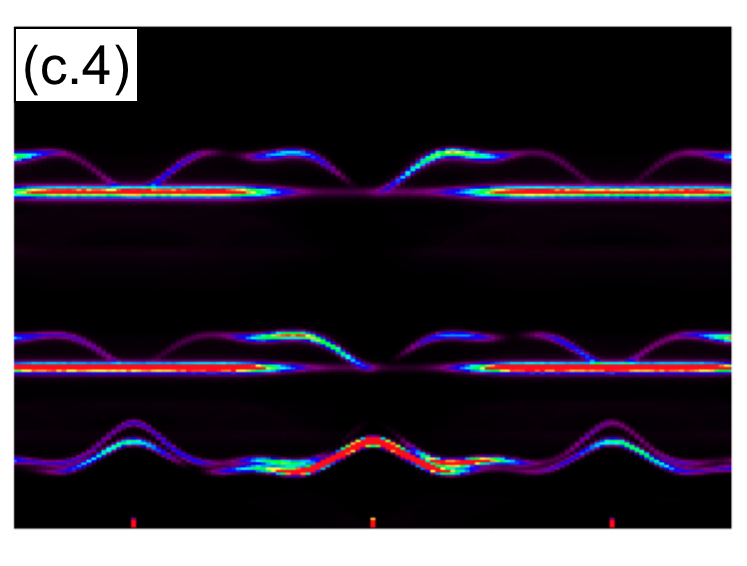}
	}
	\subfloat[]{
  		\includegraphics[width=0.17\textwidth]{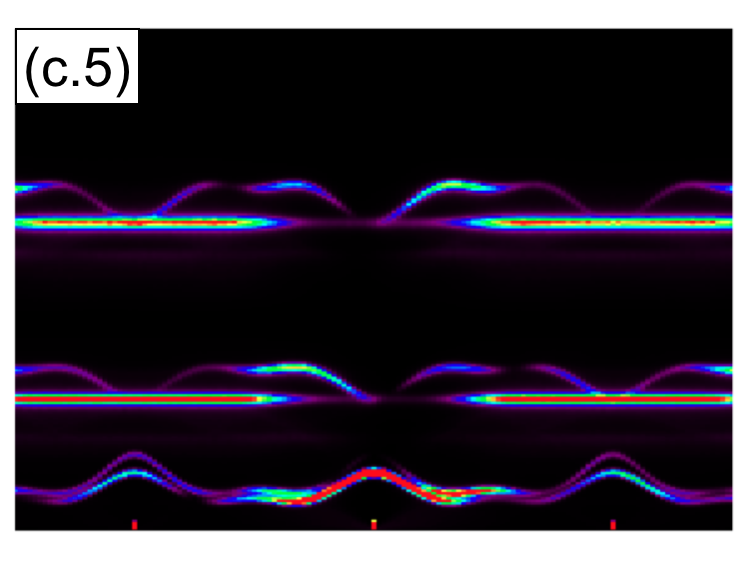}
	}
	\subfloat[]{
  		\includegraphics[width=0.17\textwidth]{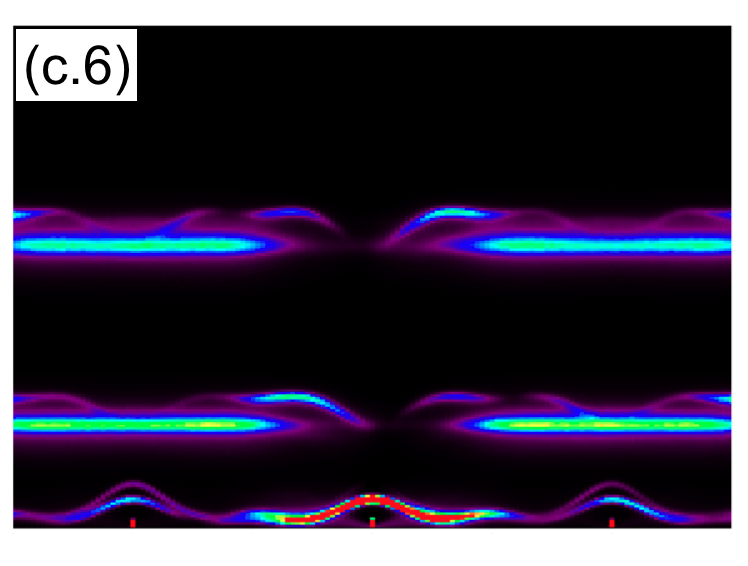}
	}
	\subfloat[]{
  		\includegraphics[width=0.193\textwidth]{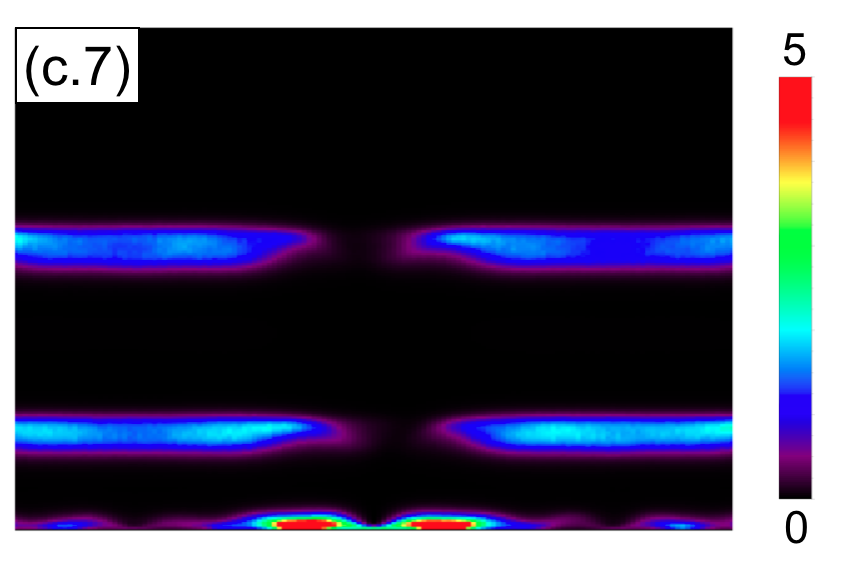}
	}\\[-6ex]
	\subfloat[]{
  		\includegraphics[width=0.202\textwidth]{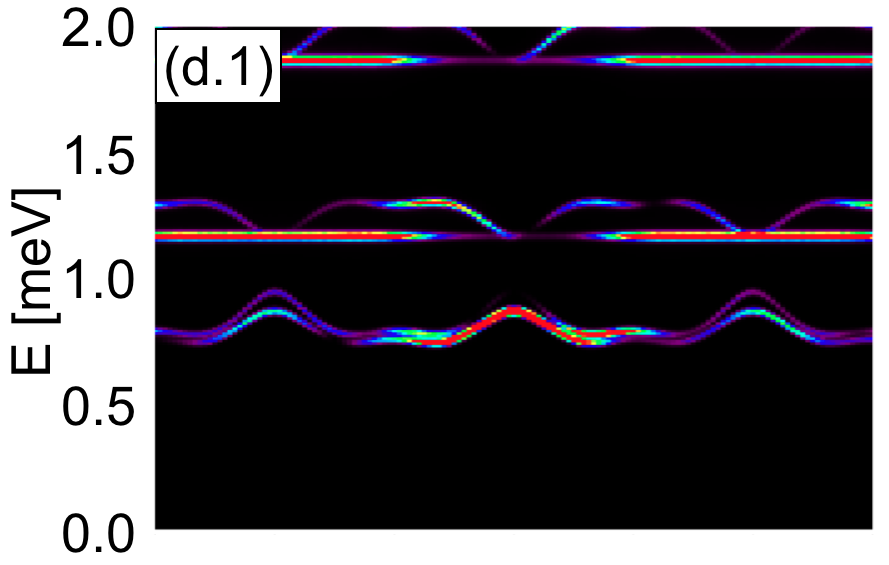}
	}
	\subfloat[]{
  		\includegraphics[width=0.17\textwidth]{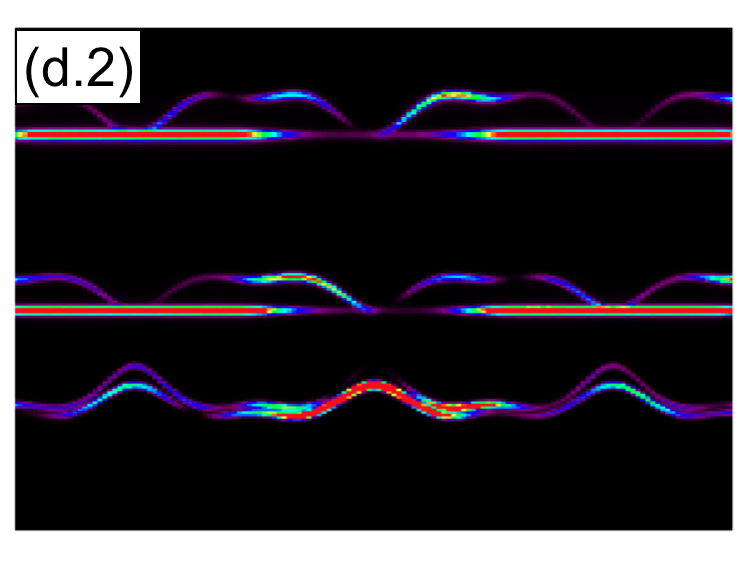}
	}
	\subfloat[]{
  		\includegraphics[width=0.17\textwidth]{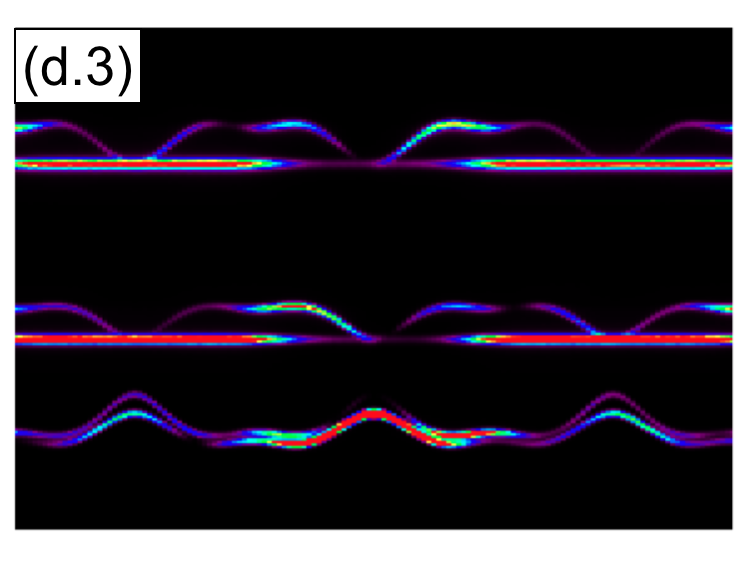}
	}
	\subfloat[]{
  		\includegraphics[width=0.17\textwidth]{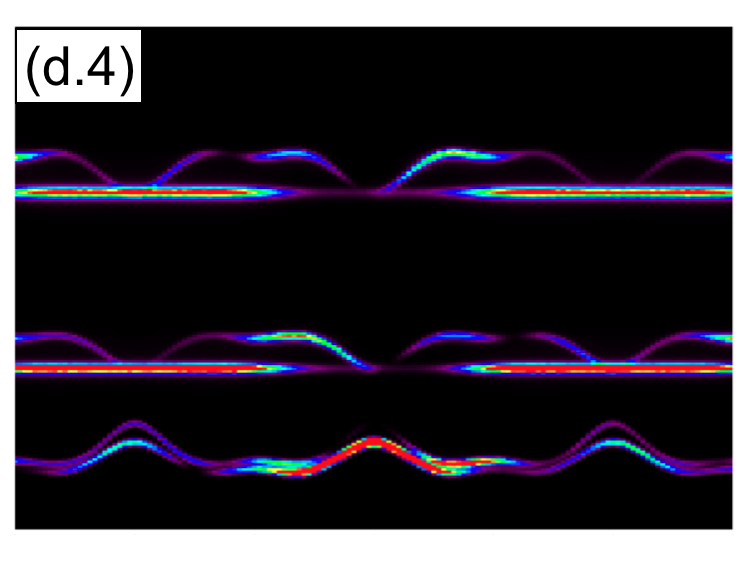}
	}
	\subfloat[]{
  		\includegraphics[width=0.17\textwidth]{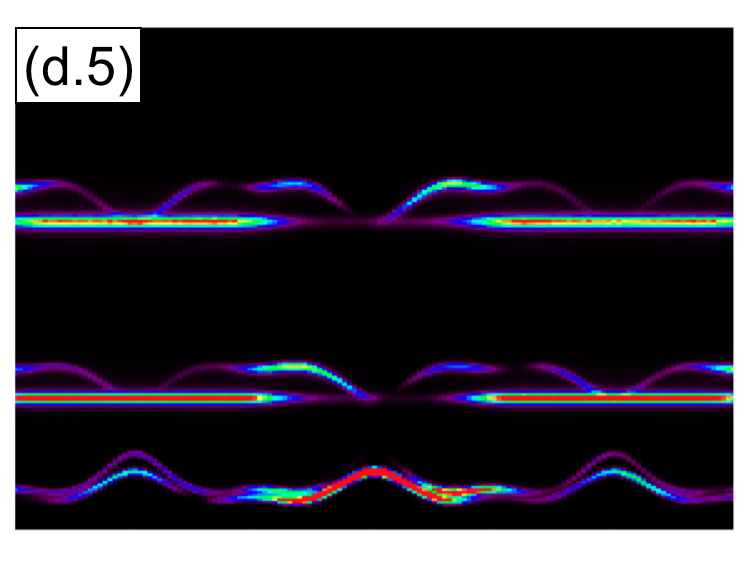}
	}
	\subfloat[]{
  		\includegraphics[width=0.17\textwidth]{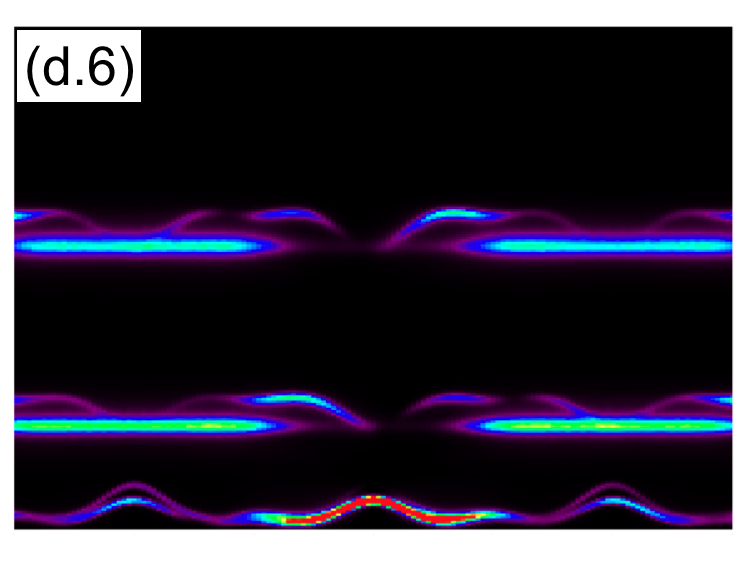}
	}
	\subfloat[]{
  		\includegraphics[width=0.193\textwidth]{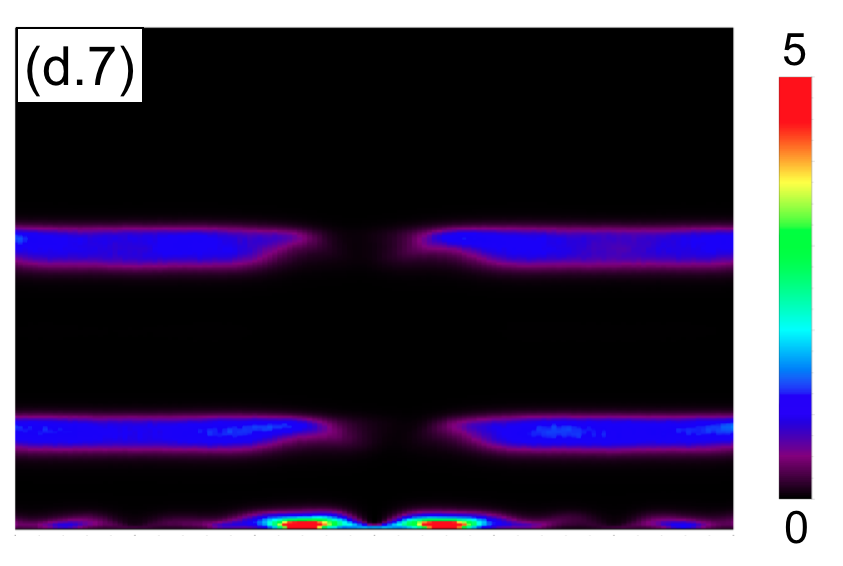}
	}\\[-6ex]
	\subfloat[]{
  		\includegraphics[width=0.202\textwidth]{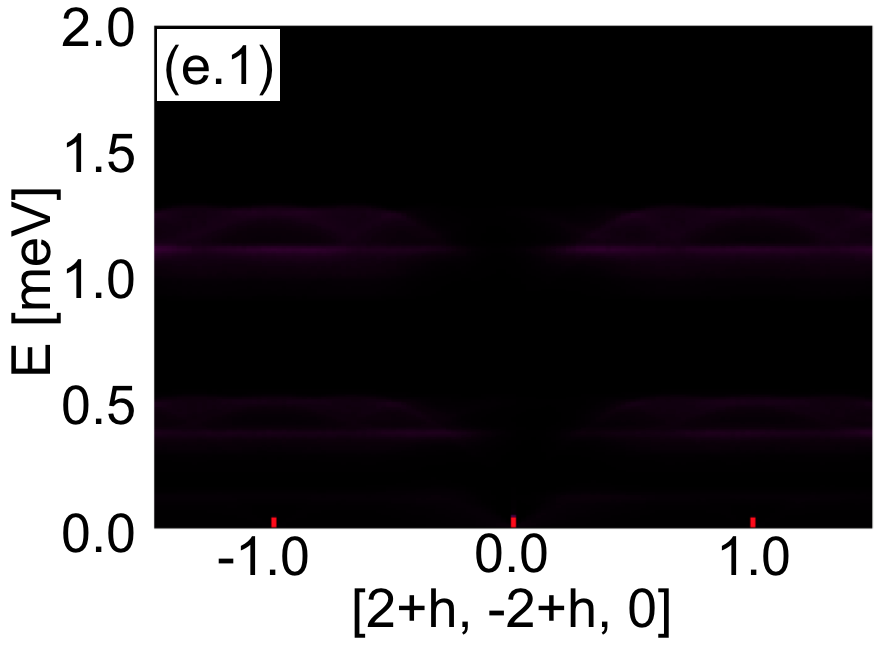}
	}
	\subfloat[]{
  		\includegraphics[width=0.17\textwidth]{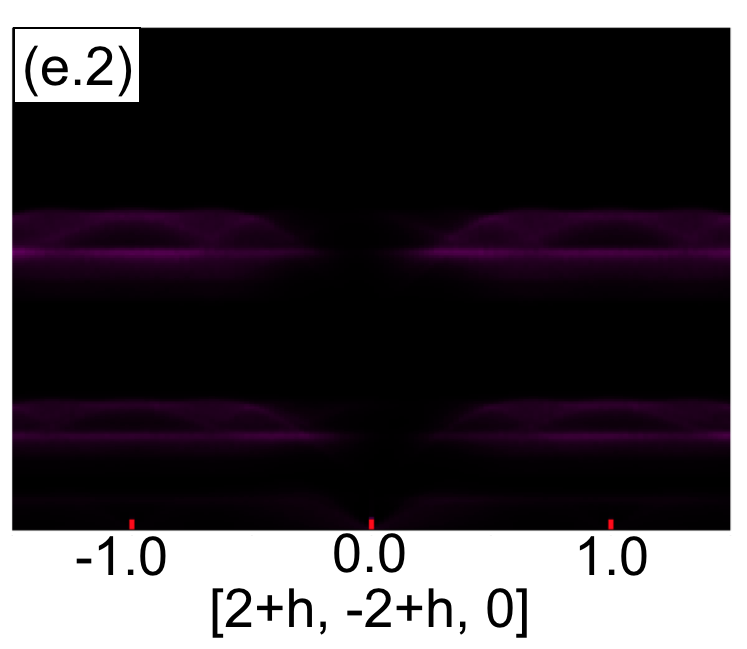}
	}
	\subfloat[]{
  		\includegraphics[width=0.17\textwidth]{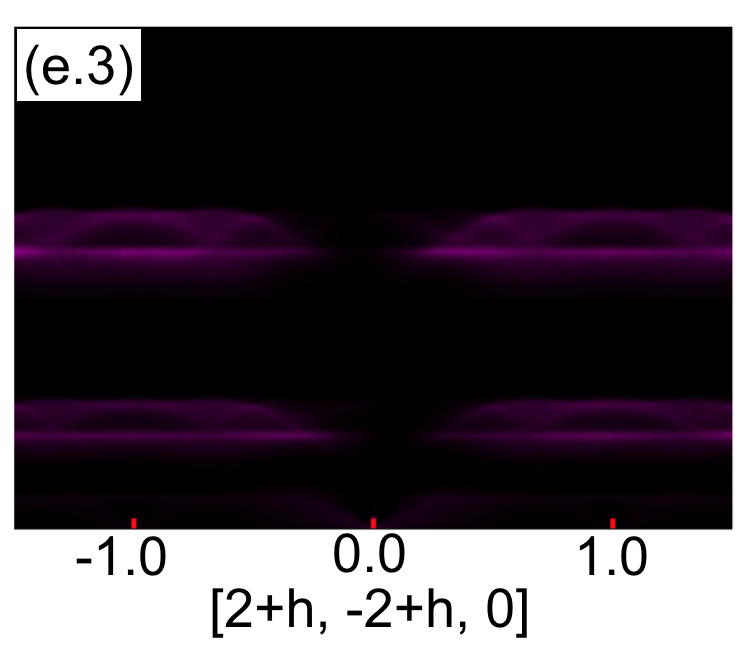}
	}
	\subfloat[]{
  		\includegraphics[width=0.17\textwidth]{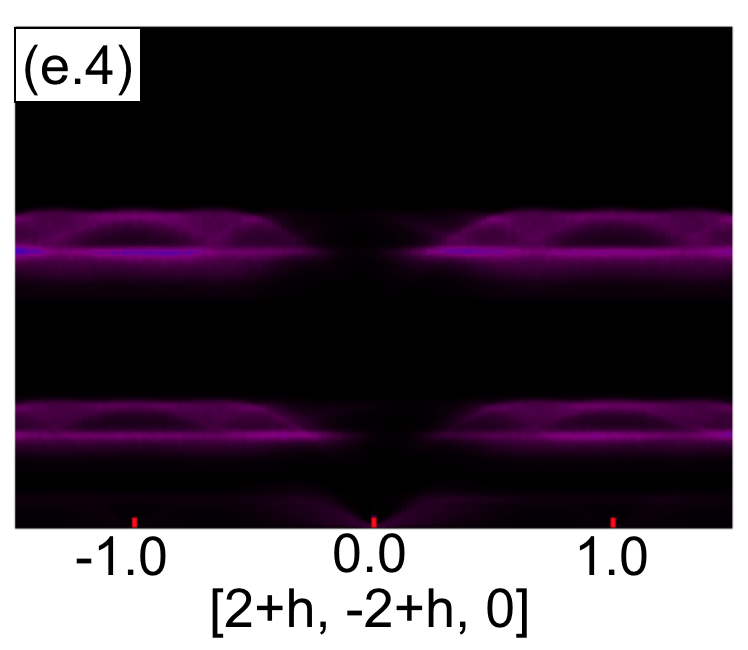}
	}
	\subfloat[]{
  		\includegraphics[width=0.17\textwidth]{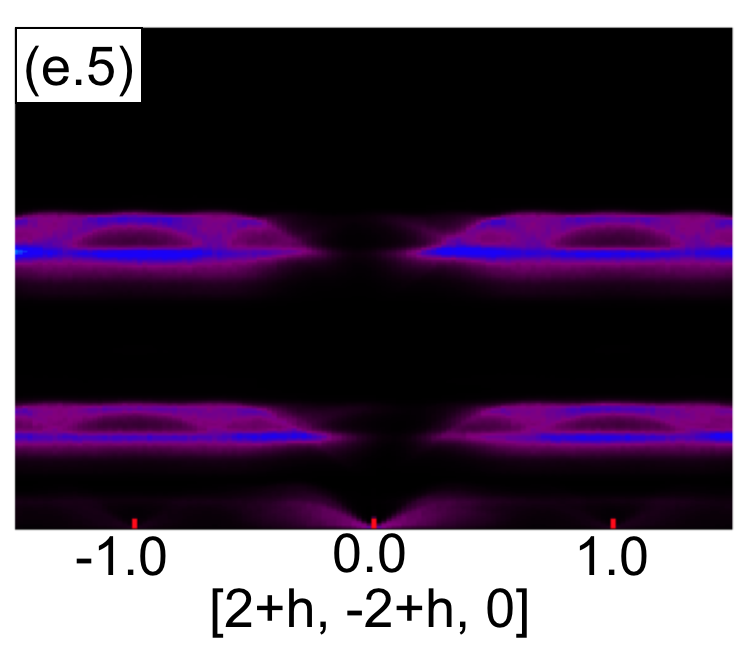}
	}
	\subfloat[]{
  		\includegraphics[width=0.17\textwidth]{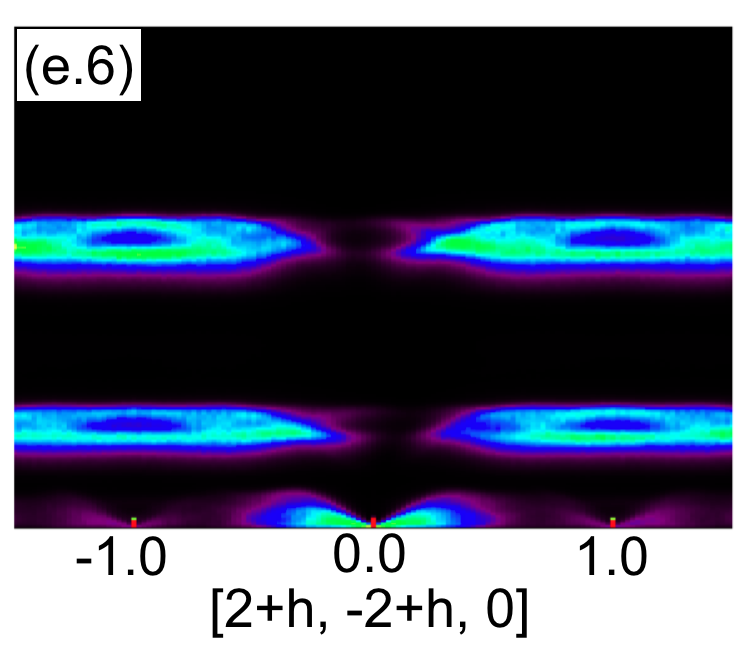}
	}
	\subfloat[]{
  		\includegraphics[width=0.193\textwidth]{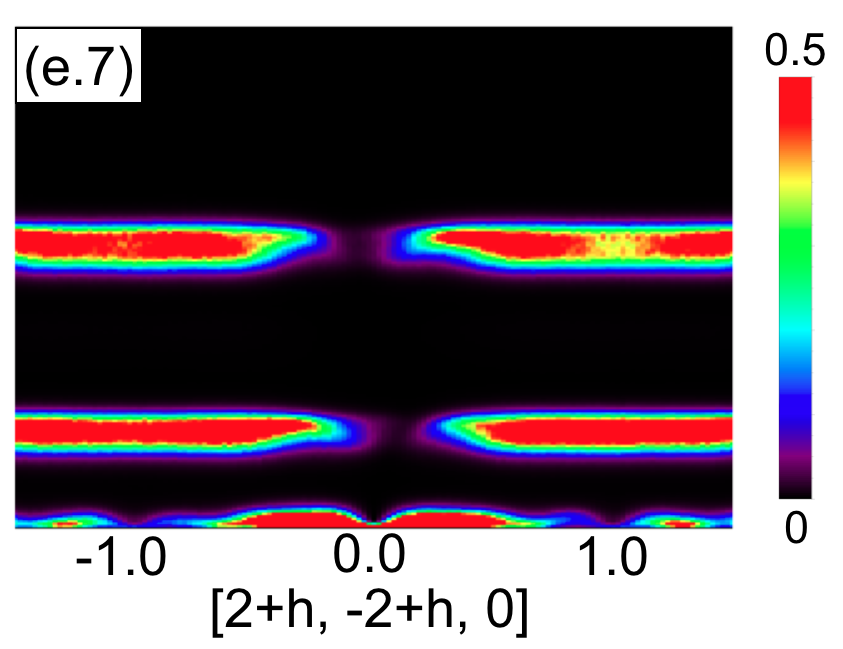}
	}
	\caption{
	Evolution of spin excitations as a function of magnetic field at $T=220$ mK, with high resolution in
	frequency space. 
	(a) Inelastic neutron scattering results for \CCO\ , reproduced from \cite{Balz2017-PRB95}, 
	for magnetic fields ranging from $B = 7.5\ \text{T}$ to $B = 0\ \text{T}$.
	(b) Predictions for inelastic scattering taken from MD simulations of 
	${\mathcal H}_{\sf BBK}$, as described in the Sec.~\ref{appendix:Numerics}.
	Results are shown along the \mbox{$[2+\text{h}, -2 + \text{h}, 0]$} direction, 
	integrated over $\pm 0.2$ r.l.u. perpendicular to the cut in reciprocal space, 
	and convoluted with a gaussian in frequency space of FWHM $\approx$ 0.2meV,
	and Cr$^{5+}$ form factor to compare with experiments in (a).
	(c) Simulation results for the first moment of the dynamical structure factor 
	$\tilde{S}({\bf q}, \omega)$ [Eq.~(\ref{eq:S.tilda})], with better energy resolution of 
	FWHM = 0.02meV.		
	(d) Contribution to $\tilde{S}({\bf q}, \omega)$ coming from transverse spin fluctuations
	$\tilde{S}^\bot({\bf q}, \omega)$ [Eq.~(\ref{eq:S.perp.tilda})].
	(e) Contribution to $\tilde{S}({\bf q}, \omega)$ coming from longitudinal spin fluctuations
	$\tilde{S}^\parallel({\bf q}, \omega)$ [Eq.~(\ref{eq:S.parallel.tilda})].
	}
	\label{fig:DynamicsEvolutionField}
\end{figure*}
\end{turnpage}

%%%%%%%%%%%%%%%%%%%%%%%%%%%%%%%%%%%%%

For higher values of magnetic field, $0.375 < B \lesssim 1\ \text{T}$, 
there is only a single sharp feature in $c(T)$, presaged by a broad 
shoulder at (slightly) higher temperature [Fig.~\ref{fig:specH_chi_B0.8}].
The sharp peak is found at higher temperatures than the anomaly in the 
$Z_3$ order--parameter susceptibility 
%$\chi_{\phi_\perp}(T)$ [Eq.~\ref{eq:phi.perp}] --- 
[Fig.~\ref{fig:inField_0.8}(b)].
For \mbox{$T \to 0$}, the peak tends in $c(T)$ to a field $B \sim 0.9\ \text{T}$ 
[Fig.~\ref{fig:c.as.function.B}], setting the outer limits of $Z_3$ order.

%%%%%%%%%%%%%%%%%%%%%%%%%%%%%%%%%%%%%

The weaker anomaly (shoulder), meanwhile, tracks to 
\mbox{$B \sim 1.1\ \text{T}$}, the critical field found in LSW theory at $T=0$
[Figs.~\ref{fig:phase.diagram.B},\ref{fig:c.as.function.B}].   
This defines a new multiple--q (II) phase, coloured pink in Fig.~\ref{fig:phase.diagram.B}, 
with correlations [Fig.~\ref{fig:inField_0.8}(c,d)] which are qualitatively 
different from those of the lattice nematic or multiple--q (I) state
found at lower field [Fig.~\ref{fig:inField_0.3}].
%
%We tentatively identify this ``pink'' phase as a second multiple--q state.
%
Unfortunately, the precise nature of the phases found at low temperature 
for this range of fields has proved extremely difficult to determine on the 
basis of simulations based on a local update, and further work would be
needed to definitively identify this state.
%
% And, since it is not directly relevant to the spin--liquid behaviour 
% of \CCO, we leave this question for future studies.

%%%%%%%%%%%%%%%%%%%%%%%%%%%%%%%%%%%%%
\subsection{Summary of results}
%%%%%%%%%%%%%%%%%%%%%%%%%%%%%%%%%%%%%
\label{sec:summary.of.thermodynamics.in.field}

Putting these results together, we arrive at the phase diagram shown 
in Fig.~\ref{fig:phase.diagram.B}.
Anomalies found in $c(T)$ and  $\chi^\perp_\phi(T)$ at low 
temperature are marked with triangular symbols.
These are taken from simulation results for a cluster of $N= 1944$ spins; 
some way from the thermodynamic limit, but large enough to offer representative results.
The single data point at $T=220\ \text{mK}$ [square symbol]  
is taken from studies of dynamics, described in Section~\ref{sec:dynamics.in.field}.

%%%%%%%%%%%%%%%%%%%%%%%%%%%%%%%%%%%%%

Five distinct phases are shown.
The phase diagram is dominated by two phases of direct relevance to 
experiments on \CCO; a correlated paramagnet in which spins on triangular 
plaquettes are ferromagnetically aligned [shaded green]; and the finite--field 
continuation of the ``spiral spin liquid'' studied in Section~\ref{sec:dynamics} 
[shaded blue];
These phases extend up to temperatures of order $T \approx 500\ \text{mK}$, 
where the three spins on FM plaquettes start to fluctuate independently, and 
the heat capacity rolls over into a paramagnetic behaviour [Fig.~\ref{fig:heat.capacity}].

%%%%%%%%%%%%%%%%%%%%%%%%%%%%%%%%%%%%%

We also identify three (quasi--)ordered phases at low temperatures; the finite--field 
continuation of the $Z_3$ lattice nematic studied in Section~\ref{sec:3.state.Potts} 
[shaded orange]; a ``multiple--q (I)'' state [shaded magenta]; and a 
 ``multiple--q (II)'' phase [shaded pink], which have both multiple--q character.   
The shaded region separating the  ``multiple--q (II)'' phase from the lattice nematic 
is not identified as a new phase, but indicates the splitting of anomalies in 
$c(T)$ and  $\chi^\perp_\phi(T)$, which is subject to strong finite--size scaling.

%%%%%%%%%%%%%%%%%%%%%%%%%%%%%%%%%%%%%

Current results leave a number of open questions about the nature of 
(quasi--)ordered phases found in the classical limit of the BBK model 
at low temperature.
However, as these do not appear to be of direct relevance to \CCO, 
and are difficult to probe with current simulation techniques, 
%s of a BBK model based on updates to a single spin, 
we leave them for future work.
Here, a good first step towards understanding these phases might be
to extend simulations of the equivalent $S=3/2$ $J_1$--$J_2$ 
honeycomb--lattice model \cite{biswas18} to finite magnetic field. 

%%%%%%%%%%%%%%%%%%%%%%%%%%%%%%%%%%%
\section{Dynamical properties of BBK model in field}		
%%%%%%%%%%%%%%%%%%%%%%%%%%%%%%%%%%%%%
\label{sec:dynamics.in.field}

Having explored the thermodynamics of the BBK model in field, 
we now turn to its dynamics, where it is possible to make explicit 
connection with the inelastic neutron scattering results of 
Balz {\it et al.} \cite{Balz2016,Balz2017-PRB95}.
We start with results for high field, where experiments were been used
to parameterise the BBK model of \CCO, before returning to the question of 
how a spin liquid emerges from a field--saturated state.

%%%%%%%%%%%%%%%%%%%%%%%%%%%%%%%%%%%
\subsection{Dynamics in high field}
%%%%%%%%%%%%%%%%%%%%%%%%%%%%%%%%%%%%%
\label{sec:dynamics.in.high.field}

In simulation as in experiment, results are most easily understood for high 
magnetic fields, where the magnetization is saturated.
Here, the fact that the BBK model is invariant under rotations about the 
magnetic field, implies that linear spin wave (LSW) theory provides an {\it exact} 
description of  the one--magnon excitations of a fully polarised state at $T=0$.
We start by exploring the properties of the six dispersing bands of spin waves 
in this case.

%%%%%%%%%%%%%%%%%%%%%%%%%%%%%%%%%%%%%

In Fig.~\ref{fig:DynamicsHighField} we show results for the dynamical 
structure factor $S({\bf q}, \omega)$ in a magnetic field of $B = 11\ \text{T}$, 
well above the saturation field $B \sim 1\ \text{T}$.
For comparison we show results taken from MD simulation ($T= 10\ \text{mK}$),
LSW theory ($T=0$), and INS data ($T= 90\ \text{mK}$).
To aid comparison, both MD and LSW results for $S({\bf q}, \omega)$ have been 
convoluted with an experimental ``resolution function'' (a Gaussian of FWHM 
$\Delta E_{\sf sim} = 0.2\ \text{meV}$), and with the atomic form factor for Cr$^{5+}$ 
[Appendix~\ref{sec:CompExp}]. 
In this limit, once thermal occupation factors have been corrected for 
[cf.~Eq.~(\ref{eq:S.tilda})], the agreement between MD and LSW results 
is essentially perfect, establishing that MD simulation also accurately 
describes these excitations.
Both LSW and MD results also offer a good account of the main features
of experiment, confirming that the parameters found by Balz {\it et al.} 
\cite{Balz2016,Balz2017-PRB95}, are a reasonable starting point 
for describing \CCO.

%%%%%%%%%%%%%%%%%%%%%%%%%%%%%%%%%%%
\subsection{Evolution of dynamics with reducing magnetic field}
%%%%%%%%%%%%%%%%%%%%%%%%%%%%%%%%%%%%%
\label{sec:field.evolution.of.dynamics}

%%%%%%%%%%%%%%%%%%%%%%%%%%%%%%%%%%%%%

We now turn to the question of how the relatively simple spin dynamics of 
the saturated state of \CCO, evolve into the complex behaviour of the 
spin liquid at zero field.  
We take as a starting point Fig.~8 of \cite{Balz2017-PRB95}, where 
INS results measured at  $T = 220\ \text{mK}$ are presented for 
magnetic fields in the range $0 \leq B \leq  7.5\ \text{T}$.
These data, for a cut through reciprocal space \mbox{$[2+\text{h}, -2 + \text{h}, 0]$},  
are reproduced in Fig~\ref{fig:DynamicsEvolutionField}(a).
They show a progressive evolution of the broad dispersing features found at high field 
[cf. Fig.~\ref{fig:DynamicsHighField}], towards the relatively diffuse scattering of 
the spin liquid at $B = 0\ \text{T}$, with results for \mbox{$E \lesssim 0.15\ \text{meV}$} 
obscured by a strong incoherent elastic background. 

%%%%%%%%%%%%%%%%%%%%%%%%%%%%%%%%%%%%%

In Fig~\ref{fig:DynamicsEvolutionField}(b), we present equivalent 
results for the dynamical structure factor of the BBK model.
These are taken from MD simulations carried out at 
$T = 220\ \text{mK}$, well above the transition temperature for the 
lattice--nematic or multiple--q states [Fig.~\ref{fig:phase.diagram.B}].
To facilitate comparison with experiment, the magnetic form factor 
of Cr$^{5+}$ has been taken into account, and results %for $S({\bf q}, \omega)$ 
have been integrated over $\pm 0.2$ r.l.u. perpendicular to the cut in 
reciprocal space.
They have also been convoluted with a Gaussian in frequency 
space of FWHM $\Delta E \approx 0.2\ \text{meV}$.
Processed in this way, simulation results provide a good account 
of experimental data for $B \geq 1\ \text{T}$, and capture many 
of the key features for $B = 0 \ \text{T}$ [cf. Fig.~\ref{fig:comparison.with.experiment}].
However, because of the information lost in convolution, it is relatively difficult 
to identify the six dispersing magnon bands of the saturated state, or to  
disentangle the different types of fluctuation within the spin liquid for 
$B \leq 1\ \text{T}$.

%%%%%%%%%%%%%%%%%%%%%%%%%%%%%%%%%%%%%

To shed more light on these questions, in Fig~\ref{fig:DynamicsEvolutionField}(c), 
we show results at the native energy--resolution of the MD simulations, 
with FWHM $\Delta E = 0.02\ \text{meV}$.
In this case, no attempt has been made to correct for the magnetic form 
factor of Cr$^{5+}$, experimental resolution, or the polarisation--dependence 
of scattering.
However, in order to compensate for the classical statistics of the MC 
simulations, we plot the temperature--corrected dynamic structure 
factor $\tilde{S}({\bf q}, \omega)$ [Eq.~(\ref{eq:S.tilda})].
At high values of field, we can distinguish six different branches 
of spin-wave excitations, in correspondence with the results of linear 
spin-wave theory \cite{Balz2017-PRB95}.
The intermediate and high--energy spin--wave branches are qualitatively similar, 
containing flat bands of localised spin-wave excitations, similar to those 
found in the Heisenberg antiferromagnet on the kagome lattice 
\cite{garanin99-PRB59, zhitomirsky08-PRB78,Taillefumier2014}.
The low--energy branches are qualitatively different, and it is the lowest of 
these that encodes the ``ring'' characteristic of the spiral spin liquid, in the 
form of a set of quasi--degenerate minimima close to the zone boundary.

%%%%%%%%%%%%%%%%%%%%%%%%%%%%%%%%%%%%%%

In Fig.~\ref{fig:DynamicsEvolutionField}(d) and 
Fig.~\ref{fig:DynamicsEvolutionField}(e), we show equivalent results, 
separated into transverse 
\begin{eqnarray}
   \tilde{S}^\perp({\bf q}, \omega) &=& \frac{1}{2} \frac{\omega}{k_B T} S^\perp({\bf q}, \omega) \; ,
   \label{eq:S.perp.tilda}
\end{eqnarray}
and longitudinal components
\begin{eqnarray}
   \tilde{S}^\parallel({\bf q}, \omega) &=& \frac{1}{2}  \frac{\omega}{k_B T} S^\parallel({\bf q}, \omega) \; ,
   \label{eq:S.parallel.tilda}
\end{eqnarray}
as defined in Appendix~\ref{appendix:Numerics} [Eqs.~(\ref{eq:S.perp.para})--(\ref{eq:S.parallel})].
 At high fields, $\tilde{S}_\perp({\bf q}, \omega)$, clearly distinguishes 
 the six branches of spin--wave excitations, while 
 $\tilde{S}_\parallel({\bf q}, \omega)$ shows spectral 
 weight at $\omega \approx 0.5\ \text{meV}$ and 
 $\omega \approx 1.2\ \text{meV}$, suggestive of bands of (highly--localised)  
 longitudinal excitations.
With reducing magnetic field, the spin--wave branches decrease in energy, 
at constant intensity, while longitudinal excitations remain fixed in energy, 
but gain in spectral weight.

%%%%%%%%%%%%%%%%%%%%%%%%%%%%%%%%%%%%%%

At $B = 1\ \text{T}$, transverse excitations show a small gap 
[Fig.~\ref{fig:DynamicsEvolutionField}(d6)], while longitudinal 
excitations [Fig.~\ref{fig:DynamicsEvolutionField}(e6)] have gained 
enough spectral weight for a zero--energy feature resembling a 
small ``volcano'' to be distinguished in the zone center.
Finally, at $B = 0\ \text{T}$, transverse [Fig.~\ref{fig:DynamicsEvolutionField}(d7)] 
and longitudinal [Fig.~\ref{fig:DynamicsEvolutionField}(e7)] excitations merge into
the three diffuse bands documented in Fig.~\ref{fig:dynamics.in.zero.field}.

%%%%%%%%%%%%%%%%%%%%%%%%%%%%%%%%%%%%%
%% Fig. X - S(q, \omega) at B = 2T -- half moon, pinch-point, and other
%%%%%%%%%%%%%%%%%%%%%%%%%%%%%%%%%%%%%

\begin{figure*}
	\centering	
	\subfloat[$\tilde{S}^{\bot}({\bf q},\omega)$ \label{fig:Sperp_2T}]{
  		\includegraphics[width=0.4\textwidth]{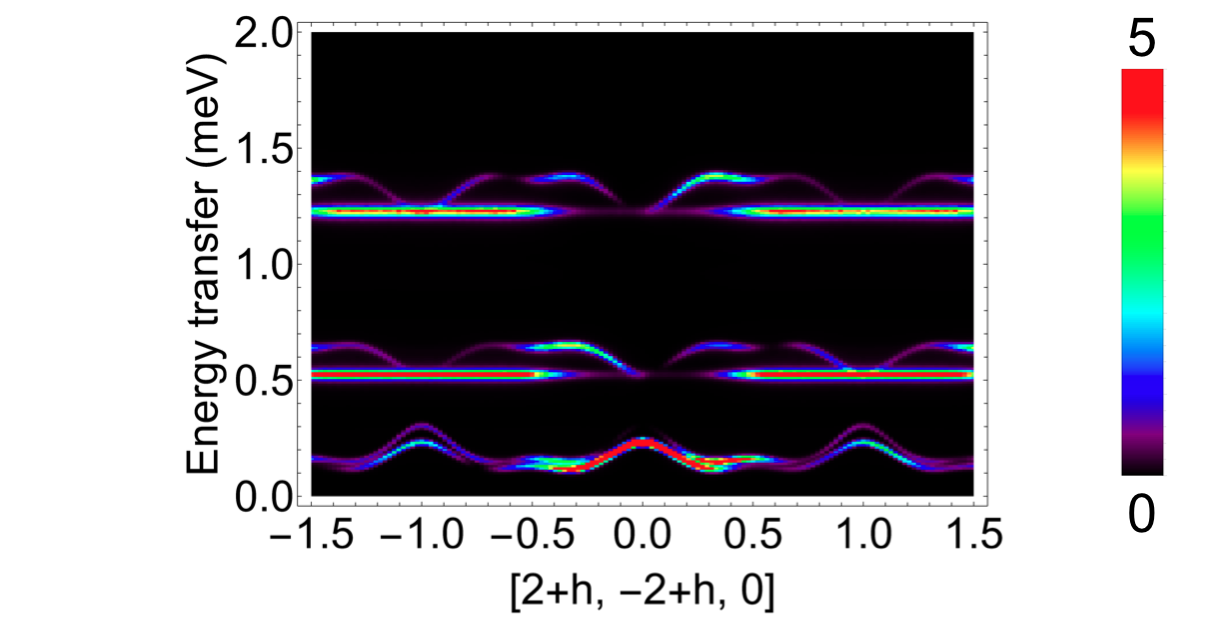}
	}
	\centering	
	\subfloat[$\tilde{S}^{\parallel}({\bf q}, \omega)$ \label{fig:Spara_2T}]{
  		\includegraphics[width=0.4\textwidth]{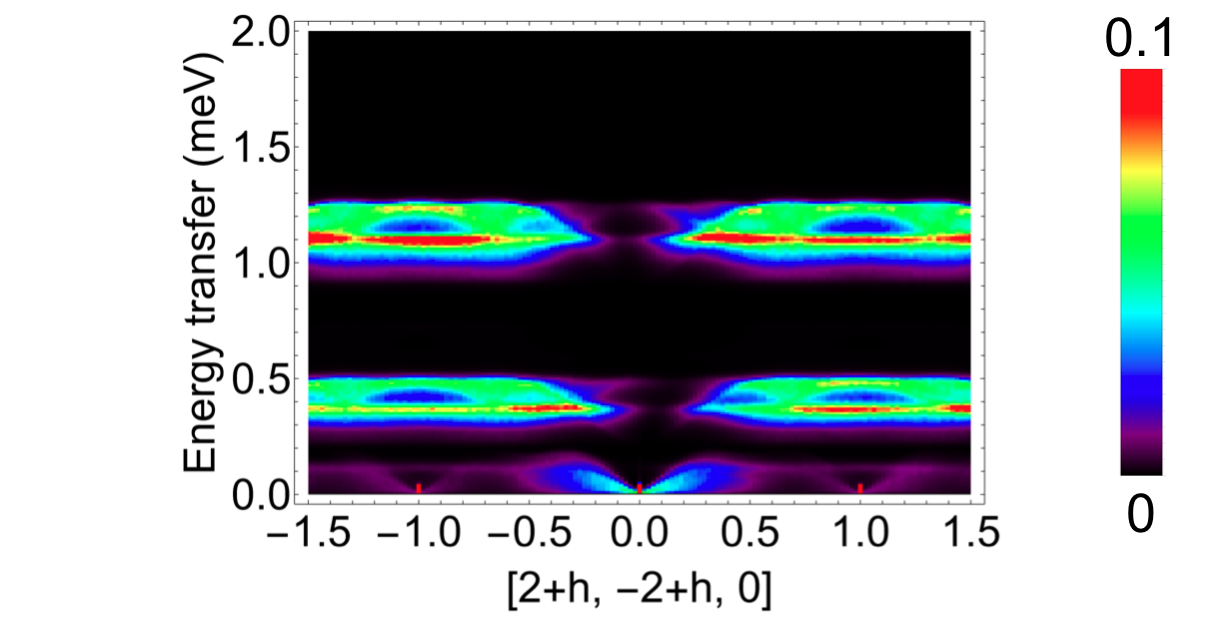}
	}\\
	\subfloat[$\tilde{S}^{\bot}({\bf q}, \omega = 0.55 \text{meV}) $ \label{fig:Sperp_2T_HalfMoon}]{
  		\includegraphics[width=0.4\textwidth]{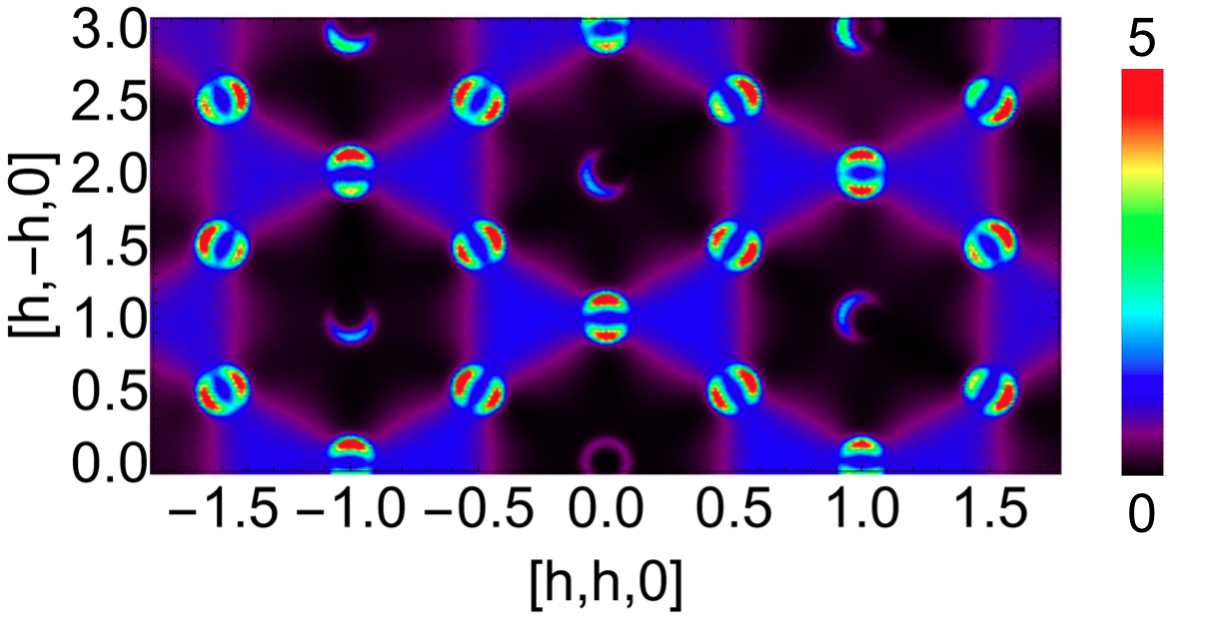}
	}
	\centering	
	\subfloat[$\tilde{S}^{\parallel}({\bf q}, \omega = 0.46 \text{meV}) $ \label{fig:Spara_2T_AboveFlatMode}]{
  		\includegraphics[width=0.4\textwidth]{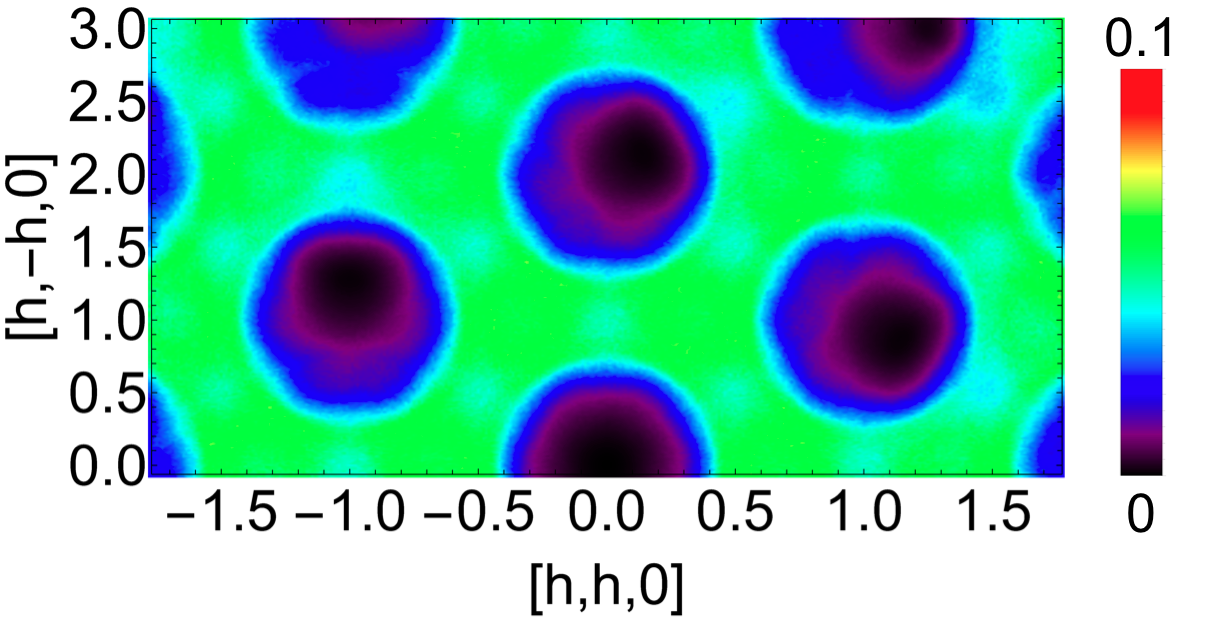}
	}\\
	\subfloat[$\tilde{S}^{\bot}({\bf q}, \omega = 0.51 \text{meV}) $ \label{fig:Sperp_2T_PinchPoint}]{
  		\includegraphics[width=0.4\textwidth]{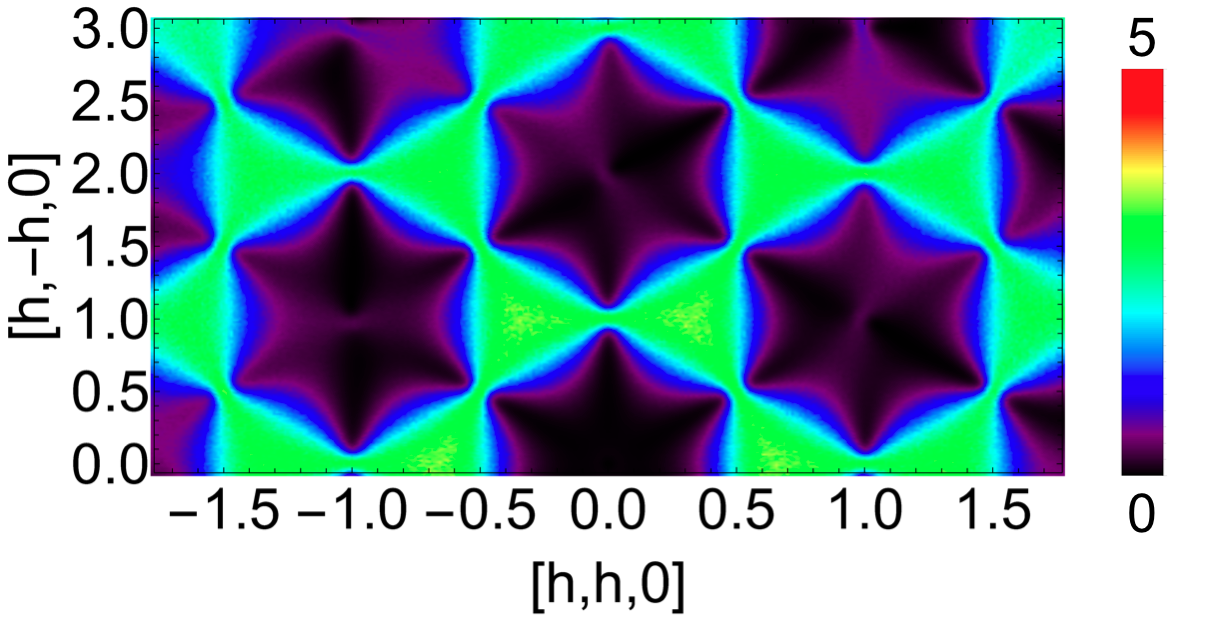}
	}
	\centering	
	\subfloat[$\tilde{S}^{\parallel}({\bf q}, \omega = 0.34 \text{meV}) $ \label{fig:Spara_2T_FlatMode}]{
  		\includegraphics[width=0.4\textwidth]{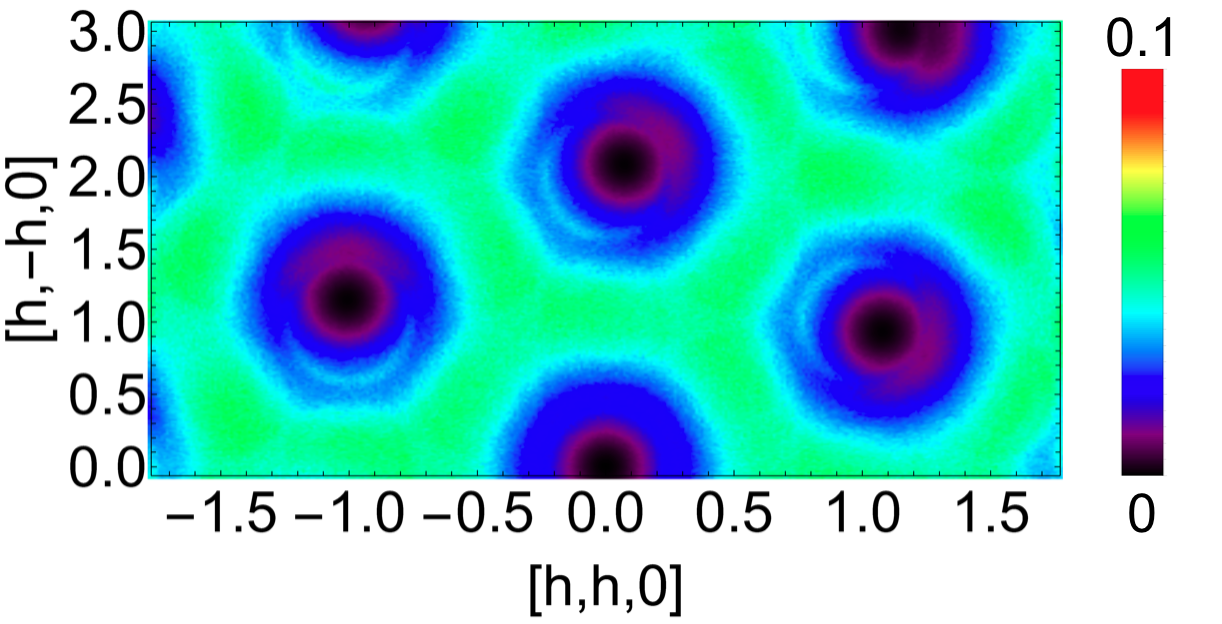}
	}\\
	\subfloat[$\tilde{S}^{\bot}({\bf q}, \omega = 0.12 \text{meV}) $ \label{fig:Sperp_2T_ring}]{
  		\includegraphics[width=0.4\textwidth]{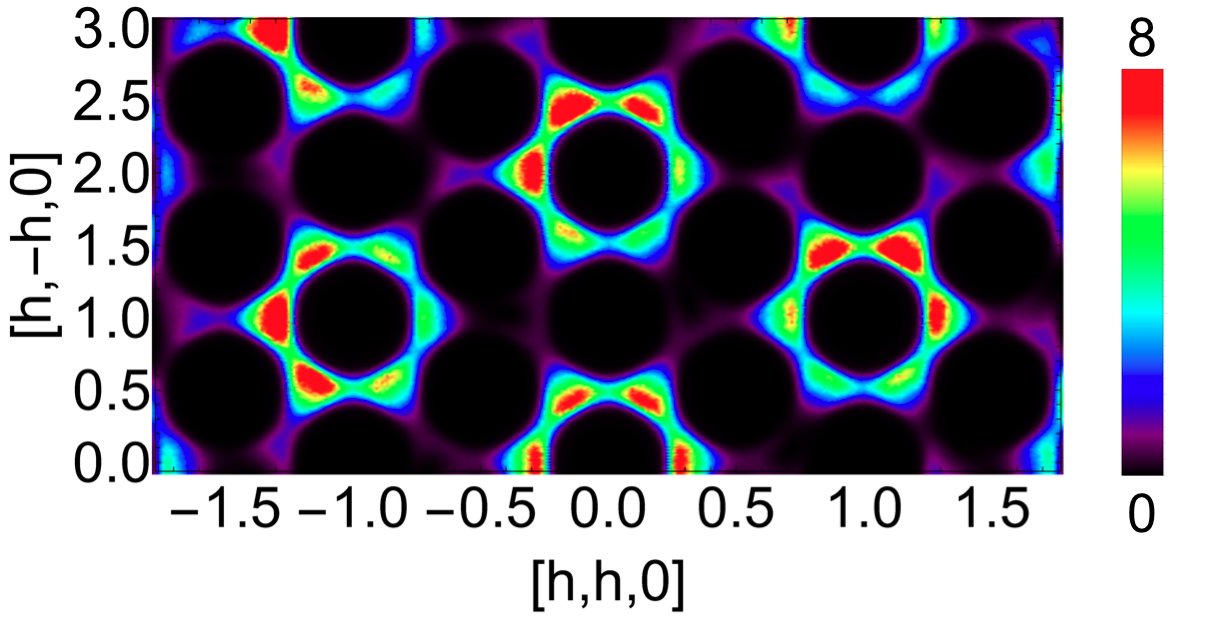}
	}
	\centering	
	\subfloat[$\tilde{S}^{\parallel}({\bf q}, \omega = 0.01 \text{meV}) $ \label{fig:Spara_2T_zeroE}]{
  		\includegraphics[width=0.4\textwidth]{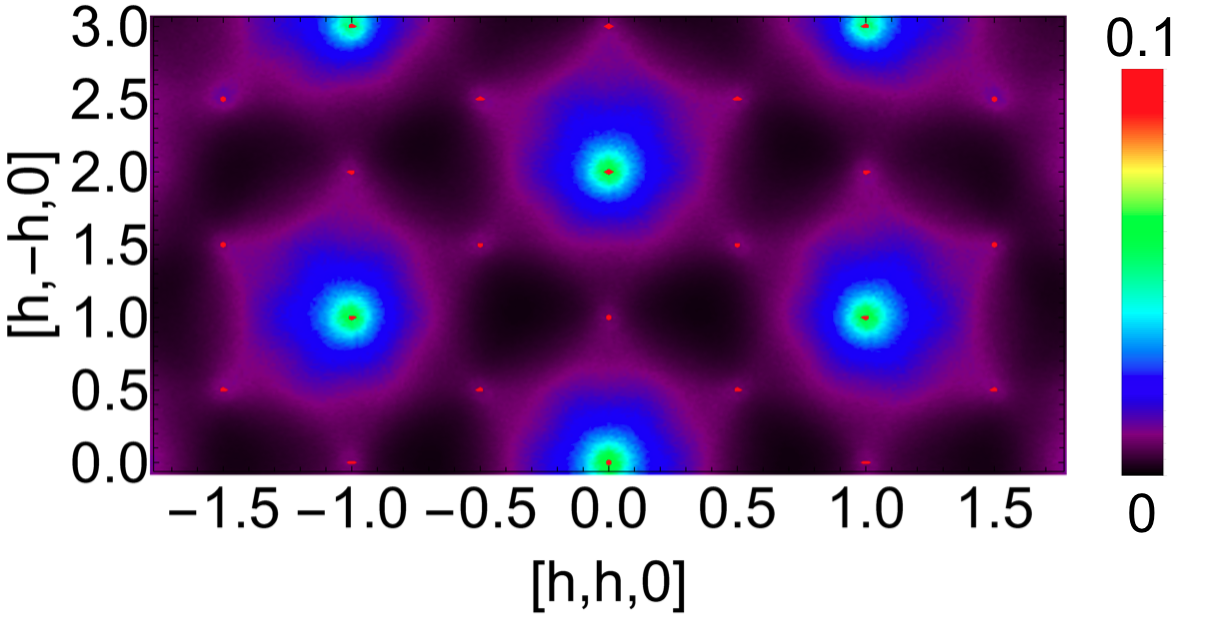}
	}\\
	\caption{
	Detail of spin dynamics at $B = 2\ \text{T}$, showing the qualitatively 
	different character of excitations in the transverse and longitudinal 
	channels.
	(a)~Temperature corrected transverse structure factor, 
	$\tilde{S}^\bot({\bf q}, \omega)$ [Eq.~(\ref{eq:S.perp.tilda})],
	showing six distinct branches of spin--wave excitations.
	(b)~Temperature corrected longitudinal structure factor,
	$\tilde{S}^\parallel({\bf q}, \omega)$ [Eq.~(\ref{eq:S.parallel.tilda})],
	showing weakly--dispersing longitudinal spin excitations 
	at three distinct energy scales.
	(c)~Transverse correlations at intermediate energy, 
	\mbox{$\omega = 0.55\ \text{meV}$}, 
	showing ``half--moon'' features overlaid on a lattice 
	of ``bow--ties'' (pinch points).  
%	(c) $S^\bot({\bf q}, \omega)$ at fixed, intermediate energy 
%	$\hbar\omega = 0.57\ \text{meV}$, showing ``half--moon'' features 
%	overlaid on a lattice of ``bow--ties'' (pinch points).  
	%
	(d)~Longitudinal  correlations at 
	\mbox{$ \omega = 0.46\ \text{meV}$}, 
	showing a network of broad scattering. 
%	(d) $S^\parallel({\bf q}, \omega)$ at fixed, intermediate energy 
%	$\hbar\omega = 0.51\ \text{meV}$, showing a network of broad 
%	scattering. 
	%
	(e)~Transverse correlations at $\omega = 0.51\ \text{meV}$, 
	showing sharp pinch--point features.
	(f)~Longitudinal  correlations at $\omega = 0.34\ \text{meV}$, 
	showing evolution of the network of broad scattering. 
	(g)~Transverse correlations at low energy, 
	$\omega = 0.12\ \text{meV}$, showing the ring--like feature
	associated with the spiral-spin liquid.
	(h)~Longitudinal correlations at 
	$\omega = 0.01\ \text{meV}$, 	showing the 
	elastic peaks associated with finite magnetisation in zone centers, 
	and accompanying broad, diffuse scattering.
	All results were obtained 
	from molecular dynamics (MD) simulation of the BBK model, Eq.~(\ref{eq:H.BBK}), 
	at $T = 220\ \text{mK}$, with parameters taken from Experiment 
	[Table~\ref{tab:experimental.parameters}].
	System sizes are of linear dimension $L = 48$ \mbox{($N = 13,824$)}, while 
	the energy resolution is FWHM = 0.02~meV, in absence of the
	Cr$^{5+}$ form factor. 
	}
	\label{fig:Eslice_SwqParaPerp}
\end{figure*}

%%%%%%%%%%%%%%%%%%%%%%%%%%%%%%%%%%%
\subsection{Longitudinal and transverse modes at $2\ \text{T}$}
%%%%%%%%%%%%%%%%%%%%%%%%%%%%%%%%%%%%%
\label{sec:dynamics.at.2T}

In Fig.~\ref{fig:Eslice_SwqParaPerp} we document the different types of 
correlation associated with transverse and longitudinal modes, at 
\mbox{$B = 2\ \text{T}$} and \mbox{$T = 220\ \text{mK}$}.
This value of field lies well within the correlated PM [Fig.~\ref{fig:phase.diagram.B}], 
and the expected six bands of transverse spin--wave excitations are 
visible in $\tilde{S}^\perp({\bf q}, \omega)$ [Fig.~\ref{fig:Sperp_2T}].
However the magnetisation is not yet saturated [Fig.~\ref{fig:M.as.function.B}], 
and significant spectral weight can also be found in the longitudinal excitations 
probed by $\tilde{S}^\parallel({\bf q}, \omega)$ [Fig.~\ref{fig:Spara_2T}].
In both cases, excitations can be divided into low--, intermediate-- and high energy bands [cf. First Animation~\cite{first.animation}], with the 
structure of intermediate-- and high--energy correlations being broadly similar.
However, while transverse excitations form sharp dispersing bands, 
longitudinal ones are much more diffuse in character.
And, while transverse excitations show an energy gap 
$\Delta \approx 0.3\ \text{meV}$, longitudinal excitations are gapless.
%
%This means that longitudinal excitations at low energies 
%cannot simply be decomposed into combinations of $T=0$ 
%spin--wave modes.

%%%%%%%%%%%%%%%%%%%%%%%%%%%%%%%%%%%%%%

When we examine the correlations at fixed energy, these differences 
become more stark.
%
% \nic{\bf [explain pinch points first, then talk about half moons]}
%
At intermediate energy, the transverse structure factor reveals 
characteristic ``half moon'' features [Fig.~\ref{fig:Sperp_2T_HalfMoon}], 
dispersing out of bow--tie like ``pinch points'' encoded on a flat band 
[Fig.~\ref{fig:Sperp_2T_PinchPoint}].
Both of these structures reflect the existence of a local 
constraint \cite{Yan2018}, a point which we return to 
%in Section~\ref{sec:half.moons} 
below.

%%%%%%%%%%%%%%%%%%%%%%%%%%%%%%%%%%%%%%

Longitudinal correlations at similar energies, meanwhile, show a broad 
web of scattering near to the zone boundary 
[Fig.~\ref{fig:Spara_2T_AboveFlatMode}, \ref{fig:Spara_2T_FlatMode}].
And at low energies, correlations look even more different, with  
$\tilde{S}^\perp({\bf q}, \omega = 0.12\ \text{meV})$ showing the same 
``ring'' of excitations near the zone boundary as the spiral spin liquid
Section~\ref{sec:thermodynamics} [Fig.~\ref{fig:Sperp_2T_ring}], 
while $\tilde{S}^\parallel({\bf q}, \omega = 0.01\ \text{meV})$ is dominated 
by a volcano-like structure near the zone center [Fig.~\ref{fig:Spara_2T_zeroE}].
Interestingly, a similar structure has been reported within the 
parton--phenomenology of Sonnenschein {\it et al.} \cite{Sonnenschein2019}, 
where gapless excitations arise from low--energy 
particle--hole pairs spanning the Fermi surface, 
and is also seen in exact--diagonalization studies \cite{shimokawa-in-prep}.

%%%%%%%%%%%%%%%%%%%%%%%%%%%%%%%%%%%
\subsection{Characterisation of pinch points and half moons}
%%%%%%%%%%%%%%%%%%%%%%%%%%%%%%%%%%%%%
\label{sec:half.moons}

The combination of pinch--point and half--moon features observed in 
the spin--wave bands of the field--saturated state [Fig.~\ref{fig:Eslice_SwqParaPerp}] 
is a ubiquitous feature of frustrated magnets realising (or proximate to) 
a classical Coulombic spin liquid 
\cite{Robert2008,Guitteny2013,Fennell2014,Taillefumier2014,Petit2016,Rau2016,Udagawa2016,benton16-PRB94,Mizoguchi2017,Lhotel2017,Yan2018,Mizoguchi2018}, 
and is well--characterised in the case of the Kagome--lattice antiferromagnet \cite{Robert2008,Taillefumier2014,Yan2018}.
While they might at first sight appear different, pinch points and half moons 
have a common origin, stemming from the local constraint associated 
with the (proximate) spin liquid \cite{Henley2010}.
In what follows, we outline how it is possible to construct a theory of these 
features by generalising the semi--classical analysis given in \cite{Yan2018} 
to the BBK lattice.

\begin{figure}[t]
\centering  
\includegraphics[width=0.95\columnwidth]{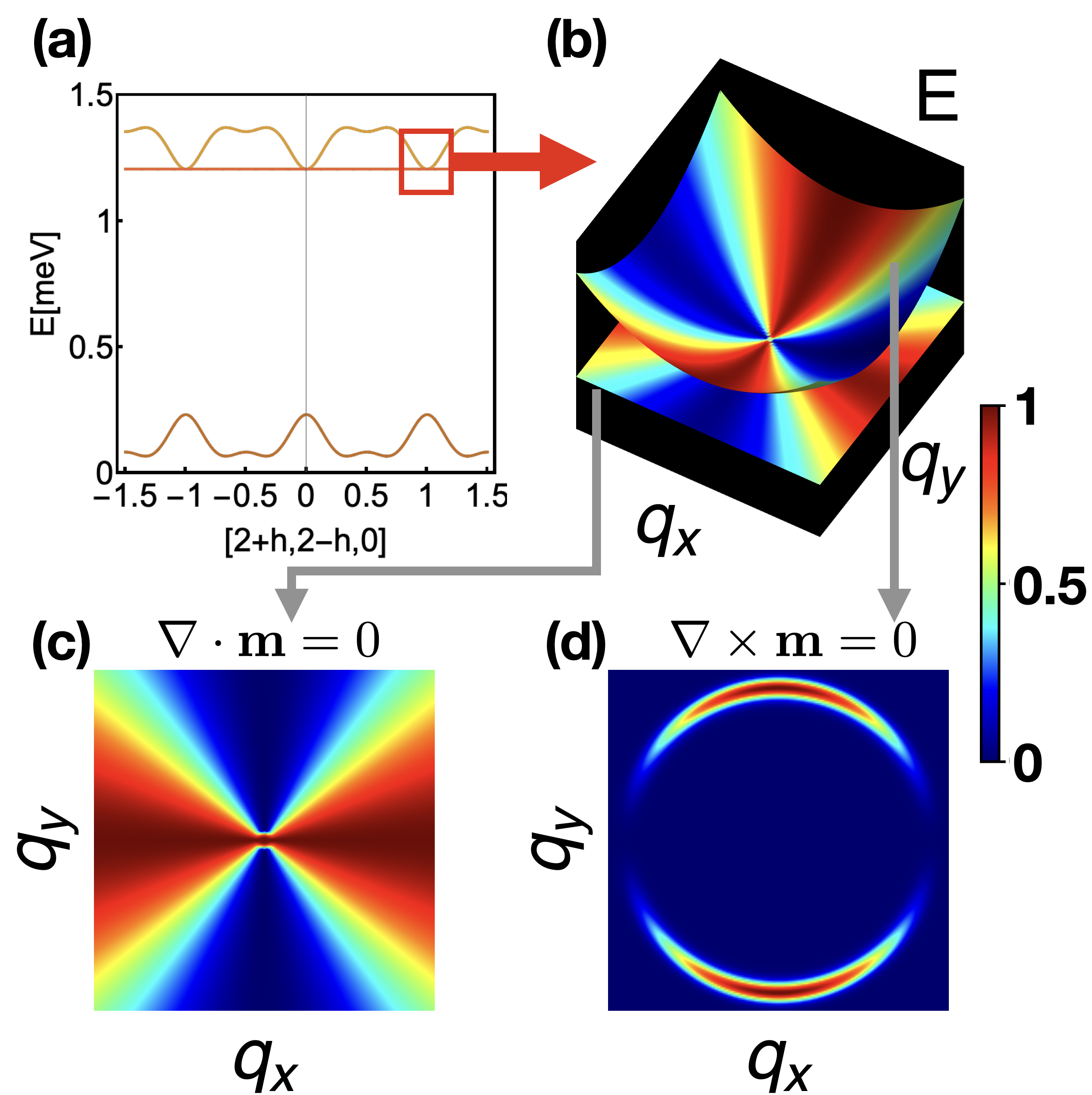}
\caption{ 
	Schematic illustration of the pinch points and half moons coming 
	from a single breathing--Kagome layer in applied magnetic field.
	(a) Spin--wave dispersion of field--saturated state.
	The upper pair of bands correspond to excitations satisfying either 
	a zero--divergence or a zero--curl condition.
	(b) Detail of dispersion in zone center, where the upper pair of bands touch.
	The contribution made by each band to the equal--time structure factor
	$S({\bf q})$ is shown in colour.
	(c) Correlations at fixed energy in the flat band, showing the pinch point 
	in the dynamical structure factor $S({\bf q},\omega)$, coming from the 
	divergence--free condition.
	(d) Correlations at fixed energy in the neighbouring dispersing band, 
	showing the pair of half moons in $S({\bf q},\omega)$, coming from the 
	curl--free condition.
	Results are shown for the upper layer of the breathing bilayer Kagome (BBK) 
	model of \CCO\ [Fig.~\ref{fig:BBK.model}, Eq.~(\ref{eq:H.BBK})], with parameters taken 
	from experiment [Table~\ref{tab:experimental.parameters}], and 
	magnetic field $B=2\ \text{T}$.
}
\label{fig:BK.pinch.point}
\end{figure}

%%%%%%%%%%

Before considering the BBK model of \CCO, it is helpful to consider the simpler
example of a Kagome lattice antiferromagnet whose magnetisation has been 
saturated by magnetic field, as studied in \cite{Yan2018}.  
Here we specialise to a single breathing--Kagome (BK) layer, 
with parameters $J_{22}$ and $J_{31}$ taken from the BBK model of \CCO\ 
[Table~\ref{tab:experimental.parameters}].
The primitive unit cell for this BK lattice is a triangle, 
containing three sites, and the field--saturated state therefore supports three bands 
of transverse spin--wave excitations, shown in Fig.~\ref{fig:BK.pinch.point}(a).
We focus on the two upper bands of excitations, one flat, and one dispersing, 
which touch in zone centers  [Fig.~\ref{fig:BK.pinch.point}(b)].
What is interesting about these two magnon bands is that the eigenvectors
associated with the flat band have a divergence--free character, while the 
eigenvectors associated with the dispersing band have a curl--free character \cite{Yan2018}.
And within a (semi--)classical evolution of spin configurations, these two types of 
excitations entirely decouple from one another \cite{benton16,Yan2018}.

When it comes to evaluation of equal--time structure factors $S({\bf q})$
[Fig.~\ref{fig:BK.pinch.point}(b)], both zero--divergence and zero--curl 
excitations exhibit ``pinch points'', singular features resembling a ``bow tie'' \cite{Henley2005}.
In the case of the zero--divergence excitations, these 
pinch points are visible in the dynamical structure factor $S({\bf q},\omega)$ 
when $\omega$ is tuned to the energy of the flat band 
[Fig.~\ref{fig:BK.pinch.point}(c)].
Meanwhile, since the pinch points of the zero--curl excitations are 
inscribed on a dispersing band, they manifest as ``half moons''  
in $S({\bf q},\omega)$ [Fig.~\ref{fig:BK.pinch.point}(d)].

\begin{figure}[t]
	\centering  
	\includegraphics[width=0.95\columnwidth]{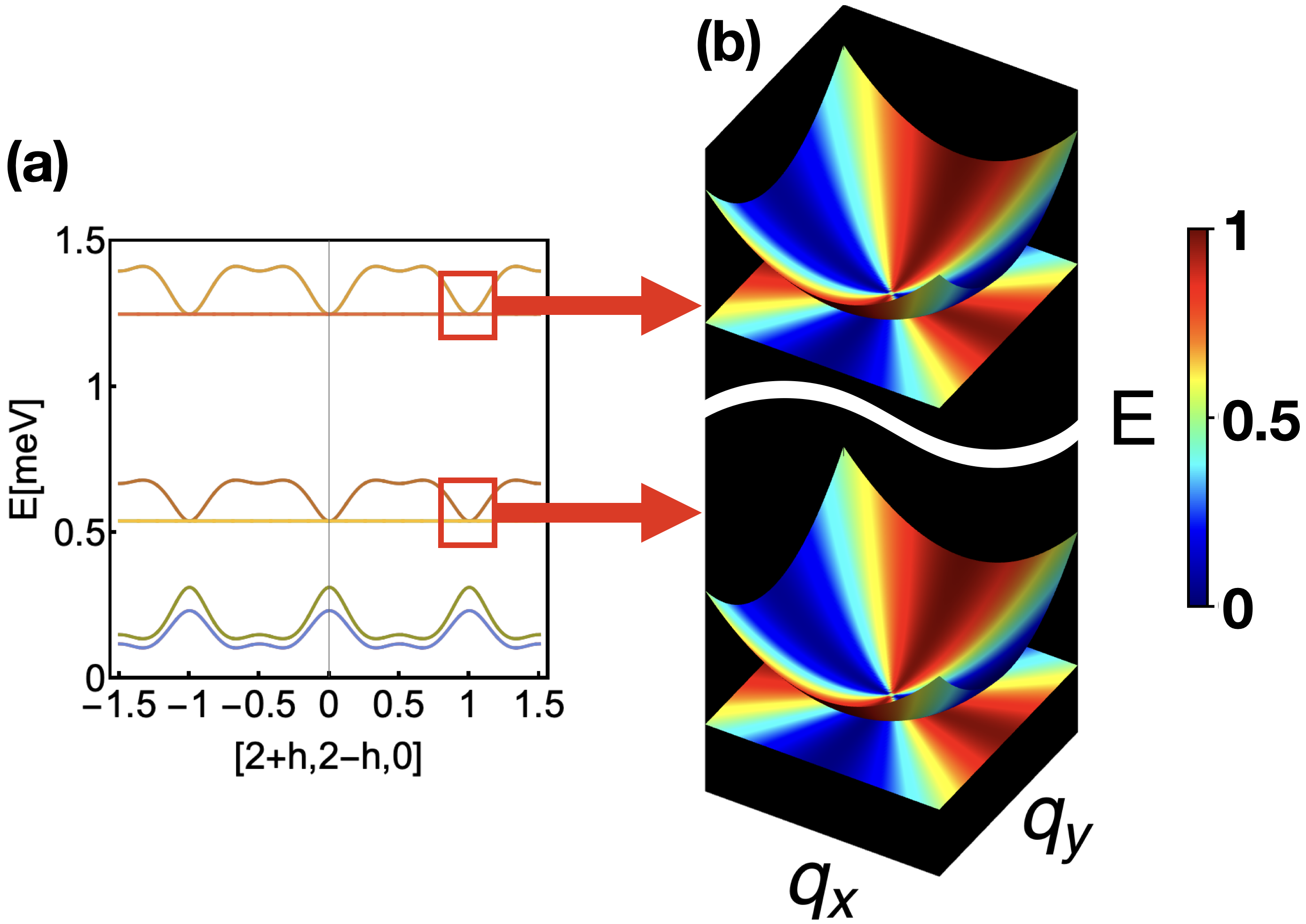}
	\caption{ Schematic illustration of the relationship between flat and dispersing bands and pinch points
		in the bilayer---breathing Kagome (BBK) model of \CCO.
		(a) Spin--wave dispersion of field--saturated state, showing pairs of bands 
		at low, intermediate and high energy.
		The two dispersive bands at intermediate and high energy 
		correspond to curl--free excitations on each layer, mixed by interlayer interactions 
		[Eq.~\eqref{eq:EoM.curl.free.a},\eqref{eq:EoM.curl.free.b}].
		Meanwhile, the adjoining flat bands correspond to the divergence--free 
		excitations on each layer, mixed by interlayer interactions 
		[Eq.~\eqref{eq:EoM.div.free.a},\eqref{eq:EoM.div.free.b}].
		(b) Detail of the upper pairs of bands near their zone--center touching points.
		Pinch points are imprinted on all four bands as a consequence of the curl-- and 
		divergence--free conditions  [Eq.~\eqref{eq:curl.free},\eqref{eq:divergence.free}].
		These are visible as either pinch points or half moons 
		in the dynamical structure factor $S({\bf q},\omega)$, as illustrated 
		in Fig.~\ref{fig:BK.pinch.point}.
		Results are shown for the BBK model Eq.~(\ref{eq:H.BBK}), with parameters taken 
		from experiment [Table~\ref{tab:experimental.parameters}], and magnetic field $B=2\ \text{T}$.
	}
	\label{fig:BBK.pinch.point}
\end{figure}

We now return to the (weakly) localised bands of excitations found at intermediate 
and high energies in the BBK model of \CCO.
The properties of these bands are not very different from the Kagome lattice model
considered in \cite{Yan2018}, with the obvious caveat that there are now two 
copies of each type of excitation, and so four bands in total.
The fact that these ``duplicate'' bands are not degenerate 
reflects the fact that exchange interactions in the two Kagome layers of \CCO\ are 
not identical [Table~\ref{tab:experimental.parameters}], and that interlayer coupling 
mixes the bands associated with each layer.
However, the key physics -- decoupling of the curl--free  
and the divergence--free excitations -- remains valid, and gives rise to 
pinch--point and half--moon features in $S({\bf q},\omega)$, 
[Fig.~\ref{fig:BBK.pinch.point}].

%%%%%%%%%%%%%%%%%%%%%%%%%%%%%%%%%%%%%

Now let us 
develop the mathematics of this picture.
We start by constructing the long-wavelength fields that
describe the transverse excitations of a polarised state,
and define the vector field.
We explicitly consider a single bilayer of a breathing Kagome 
lattice [Fig.~\ref{fig:BBK.model}], as two (breathing) Kagome lattices, 
for each of which the primitive unit cell is a triangle, labelling triangles in ``top'' and 
``bottom'' layers of the lattice "t" and "b" respectively.
We then introduce fields describing transverse spin excitations
which transform with the $\mathsf{E}$ (vector) and $\mathbf{A}$ (scalar) 
irreps of the point group of a triangle
\begin{eqnarray}
	{\bf m}_{\mathsf{t/b}}  = \sum\limits_{i=1}^3 S^-_{\mathsf{t/b,} i} {\bf u}_i 	
	\; 	, \qquad
	\phi_{\mathsf{t/b}}   =  \sum_{i=1}^3 S^-_{\mathsf{t/b,} i} 		,
\label{eq:fields}	
\end{eqnarray}
where
\begin{eqnarray}
S^-_i = S_i^x - i S_i^y
\end{eqnarray}
are spin lowering operators, and 
\begin{equation}	
	{\bf u}_1=(0,1),\ {\bf u}_2=(-\sqrt{3}/2,1/2),\ {\bf u}_3=(\sqrt{3}/2,1/2) \; ,
\label{eqn:u-vec}
\end{equation}
are (unit) vectors pointing from the center towards the corners 
of each triangle.
${\bf m}_{\mathsf{t/b}} $ are the vector fields and $\phi_{\mathsf{t/b}}$ the scalar fields of interest.
%,
%and are illustrated in Fig.~\ref{fig:BBK.pinch.point}a,b.}

%%%%%%%%%%%%%%%%%%%%%%%%%%%%%%%%%%%%%

By evaluating the commutation relations between ${\bf m}_{\mathsf{t/b}}$  and the 
BBK Hamiltonian [Eq.~(\ref{eq:H.BBK})], we obtain the equations of motion (EoM) for ${\bf m}_{\mathsf{t/b}} $.
Their diagonalization exhibits the dispersion relations for the bands involved, 
and also the corresponding eigenstates.
In the long wavelength limit, they take the form
\begin{eqnarray}
-i\frac{\partial}{\partial_t} {\bf m}_{\mathsf{t}} =&   &
- \frac{1}{4}a_0^2 J_{\mathsf{t}+} {\nabla} ({\nabla} \cdot {\bf m}_{\mathsf{t}}) \nonumber\\
&& +
\left( H^z-3 J_{\mathsf{t}+} - \frac{1}{2} J_\text{inter} \right)  {\bf m}_{\mathsf{t}} \nonumber \\
& &+ J_\text{inter}{\bf m}_{\mathsf{b}}	\; ,	
\label{eq:EoM.mt}\\
-i\frac{\partial}{\partial_t} {\bf m}_{\mathsf{b}} =& & 
- \frac{1}{4}a_0^2 J_{\mathsf{b}+} {\nabla} ({\nabla} \cdot {\bf m}_{\mathsf{b}}) \nonumber \\
&&+
\left( H^z-3 J_{\mathsf{b}+} - \frac{1}{2} J_\text{inter} \right) {\bf m}_{\mathsf{b}} \nonumber \\
&&+ J_\text{inter}{\bf m}_{\mathsf{t}} \; , 
\label{eq:EoM.mb}
\end{eqnarray}
where we assumed a perfect bilayer Kagome lattice with lattice constant $a_0$ for simplicity 
and
\begin{eqnarray}
J_{\mathsf{t}+} &=& \frac{1}{2} \left(J_{31} + J_{22} \right) \; ,\nonumber \\
J_{\mathsf{b}+} &=& \frac{1}{2} \left(J_{32} + J_{21} \right) \; , \\
J_\text{inter}  &=& J_0 \; .\nonumber
\end{eqnarray}
We have further simplified the equations of motion by neglecting terms
coupling the fields ${\bf m}_{\mathsf{t/b}}$ to the 
fields $\phi_{\mathsf{t/b}}$ in Eq.~(\ref{eq:EoM.mt}) and Eq.~(\ref{eq:EoM.mb}).
They can be included at the cost of a more complicated description of the 
problem, but do not play any significant role in the long wavelength limit.
% on account 
%of the large energy gap between the bands associated with ${\bf m}_{\mathsf{t/b}}$ 
%and those associated with $\phi_{\mathsf{t/b}}$.

%%%%%%%%%%%%%%%%%%%%%%%%%%%%%%%%%%%

Note that the first terms on the right hand side of the EoM
	is curl-free. 
This means the EoM can be partially decoupled 
by introducing a Helmholtz--Hodge decomposition of the vector field 	 
\begin{eqnarray}
 {\bf m}_{\mathsf{t/b}} = {\bf m}_{\mathsf{t/b,DF}} + {\bf m}_{\mathsf{t/b,CF}}
\end{eqnarray}
in terms of divergence--free components
\begin{eqnarray}
	&&\nabla \cdot {\bf m}_{\mathsf{t/b,DF}} = 0 	
	\label{eq:divergence.free}	
\end{eqnarray}
and curl--free components
\begin{eqnarray}
	&&\nabla \times {\bf m}_{\mathsf{t/b,CF}} 
	\equiv
	\left( -\frac{\partial}{\partial_y},\frac{\partial}{\partial_x} \right) \cdot {\bf m}_{\mathsf{t/b,CF}} = 0			.
		\label{eq:curl.free}	
\end{eqnarray}

%\han{
%	The two copies of curl-free vector fields are illustrated in  Fig.~\ref{fig:BBK.pinch.point}(a),
%	and the 
%	divergence-free vector fields are illustrated in 
% Fig.~\ref{fig:BBK.pinch.point}(b).
%}
%
This leads to the decoupled EoM for
divergence--free components 	
\begin{eqnarray}
	-i\frac{\partial}{\partial_t} {\bf m}_{\mathsf{t,DF}} =& & 
	\left( H^z-3 J_{\mathsf{t}+} - \frac{1}{2} J_\text{inter} \right)  {\bf m}_{\mathsf{t,DF}} \nonumber \\
	& &+ J_\text{inter}{\bf m}_{\mathsf{b,DF}}
	\label{eq:EoM.div.free.a}  \; ,\\
	-i\frac{\partial}{\partial_t} {\bf m}_{\mathsf{b,DF}} =& & 
	\left( H^z-3 J_{\mathsf{b}+} - \frac{1}{2} J_\text{inter} \right) {\bf m}_{\mathsf{b,DF}} \nonumber \\
	&&+ J_\text{inter}{\bf m}_{\mathsf{t,DF}} \; .
	\label{eq:EoM.div.free.b}  
\end{eqnarray}	
And the 
decoupled EoM for curl--free components  are 
%of $ {\bf m}_{\mathsf{t/b}} $
%
\begin{eqnarray}	
	-i\frac{\partial}{\partial_t} {\bf m}_{\mathsf{t,CF}} =&   &
	- \frac{1}{4}a_0^2 J_{\mathsf{t,CF}+} {\nabla} ({\nabla} \cdot {\bf m}_{\mathsf{t,CF}}) \nonumber \\
	&& +
	\left( H^z-3 J_{\mathsf{t}+} - \frac{1}{2} J_\text{inter} \right)  {\bf m}_{\mathsf{t,CF}} \nonumber \\
	& &+ J_\text{inter}{\bf m}_{\mathsf{b,CF}}	\; ,  \label{eq:EoM.curl.free.a} 	\\
	-i\frac{\partial}{\partial_t} {\bf m}_{\mathsf{b,CF}} =& & 
	- \frac{1}{4}a_0^2 J_{\mathsf{b}+} {\nabla} ({\nabla} \cdot {\bf m}_{\mathsf{b,CF}}) \nonumber \\
	&& +
	\left( H^z-3 J_{\mathsf{b}+} - \frac{1}{2} J_\text{inter} \right) {\bf m}_{\mathsf{b,CF}} \nonumber \\
	&&+ J_\text{inter}{\bf m}_{\mathsf{t,CF}} \; .
	\label{eq:EoM.curl.free.b}  
\end{eqnarray}% 
%
 
%%%%%%%%%%%%%%%%%%%%%%%%%%%%%%%%%%%%%%%%%

Let us look at the EoMs for the divergence--free components ${\bf m}_{\mathsf{b/t,DF}}$ first.
We will find they are responsible for the flat bands with pinch points.
The EoM for them,
Eqs.~(\ref{eq:EoM.div.free.a}) and (\ref{eq:EoM.div.free.b}),
couple the divergence-free components between themselves,
but they decouple from all the other degrees of freedom.
In the absence of spatial derivatives on the right hand side of EoM, they 
describe a pair of flat bands, split by a gap of  
\begin{eqnarray}	
	\Delta_\mathsf{DF} = 2\sqrt{9(J_{\mathsf{b}+} - J_{\mathsf{t}+})^2 + 4  J_\text{inter}^2} \approx 0.7\ \text{meV} \; ,
\label{eq:Delta.DF}
\end{eqnarray}	
as found in MD simulation [Fig.~\ref{fig:Sperp_2T}], and  illustrated in Fig.~\ref{fig:BBK.pinch.point}a.
Furthermore, the divergence--free condition, Eq.~(\ref{eq:divergence.free}), 
implies that correlations of these fields show an (algebraic) singularity \nic{with the form of a ``pinch point''}
\begin{equation}
	  q^\alpha{m}_{\mathsf{DF}}^\alpha=0  \rightarrow 
	\langle  {m}_{\mathsf{DF}}^\alpha ({\bf q}) {m}_{\mathsf{DF}}^\beta ({\bf -q}) \rangle
	\sim \delta^{\alpha\beta} - \frac{q^\alpha q^\beta}{{\bf q}^2} \; .
	\label{eq:pinch.point}
\end{equation}
This leads to a corresponding singularity in the dynamical spin structure factor
\begin{eqnarray}
	S^\perp_\lambda ({\bf q}, \omega) 
		&\propto& (1-{q^\lambda_\alpha q^\lambda_\beta}/{{\bf q}^2})
       		 \delta(\omega - \omega^\lambda ({\bf q})) \; , 
%		 && {\bf q}^\text{curl} =  {\bf q}  \; , \; {\bf q}^\text{div} = \tilde{\bf q} \; ,
\label{eq:S.q.div.free}
\end{eqnarray}
where $\omega^\lambda ({\bf q})$ is a constant  energy of the associated (flat) band.
This is the origin of the bow--tie like pinch point feature seen in MD simulation results 
[compare Fig.~\ref{fig:Sperp_2T_PinchPoint} with  Fig.~\ref{fig:BK.pinch.point}c ].

%%%%%%%%%%%%%%%%%%%%%%%%%%%%%%%%%%%%%%%%%

Very similar considerations apply to the EoM for curl--free components of 
${\bf m}$, Eqs.~(\ref{eq:EoM.curl.free.a}), (\ref{eq:EoM.curl.free.b}).
They are responsible for the dispersive bands with pinch points [Fig.~\ref{fig:BBK.pinch.point}b].
In zone centers, the two bands derived from these EoM are 
degenerate with the two flat bands derived from divergence--free 
components of ${\bf m}$.
However the presence of a term with spatial derivative, 
$ {\nabla} ({\nabla} \cdot {\bf m}_{\mathsf{CF}})$, 
leads to non-flat dispersion of the bands. 
The gap between the flat and dispersive band opens quadratically
in zone centers [Fig.~\ref{fig:Sperp_2T} and Fig.~\ref{fig:BBK.pinch.point}a].

%%%%%%%%%%%%%%%%%%%%%%%%%%%%%%%%%%%%%%%%%

The curl--free condition is also encoded
in the correlation function as a pinch point, 
but this time rotated by $\pi/2$ with respect 
to those of the divergence--free bands
\begin{equation}
	\langle  {m}_{\mathsf{CF}}^\alpha ({\bf q}) {m}_{\mathsf{CF}}^\beta ({\bf -q}) \rangle
	\sim \delta^{\alpha\beta} - \frac{\tilde{q}^i \tilde{q}^j}{{\bf q}^2}
		\quad
	{\bf \tilde{p}} = (-p_y,p_x)
\end{equation}
Once again, the associated singularity is imprinted on the dynamical 
spin structure factor
\begin{eqnarray}
	S^\perp_\lambda ({\bf q}, \omega) 
		&\propto& (1 - \frac{\tilde{q}^i \tilde{q}^j}{{\bf q}^2})
       		 \delta(\omega - \omega^\lambda ({\bf q})) \; , 
\label{eq:S.q.curl.free}
\end{eqnarray}
where $\omega^\lambda ({\bf q})$ is the energy of the associated (dispersing) 
bands.
However, because the bands associated with zero--curl states 
have a finite dispersion, this singularity appears not as a pinch--point 
but as a pair of half moons in cross-sections at constant energy 
[compare Fig.~\ref{fig:Sperp_2T_HalfMoon} with  Fig.~\ref{fig:BK.pinch.point}d].

Comparing quantitative results for the BBK model  [Fig.~\ref{fig:BBK.pinch.point}]
with those for a single BK layer [Fig.~\ref{fig:BK.pinch.point}], we see that interlayer 
coupling has relatively little effect on bands carrying pinch points and half moons.
This reflects the fact that the interlayer coupling $J_0 \approx -0.08\ \text{meV}$ 
is small compared with the splitting of the bands $\Delta_\mathsf{DF} \approx 0.7\ \text{meV}$ 
[Eq.~(\ref{eq:Delta.DF})].
This should be contrasted with the situation at low energies, 
where interlayer coupling plays a crucial role in determining the 
effective honeycomb lattice model, Eq.~(\ref{eq:H.HC}), and thereby the 
spiral spin liquid ground state.

To summarise, the four bands of transverse spin excitations 
found in MD simulation of the BBK model at intermediate and high energy 
[cf. Section~\ref{sec:dynamics.in.high.field}, 
\ref{sec:field.evolution.of.dynamics}, \ref{sec:dynamics.at.2T}] can be understood  
as two sets of flat bands, satisfying a zero--divergence condition, 
and two sets of (weakly) dispersing bands, satisfying a zero--curl condition.
Each of these bands support specific features in dynamical structure 
factors --- pinch points and half moons --- which are characteristic of the 
corresponding local constraint.
Elsewhere, equivalent features at finite energy have been characterised as a
``dynamical spin liquid'' \cite{Petit2016, benton16, lhotel18}.
And it is a special feature of the BBK model of \CCO\ that the dynamical spin 
liquid found at intermediate and higher energies ultimately 
coexists with a completely different from of the spin liquid at low energy.
It is the way in which this spiral spin liquid emerges as the ground state, 
as magnetic field is reduced, that we turn to below.

%%%%%%%%%%%%%%%%%%%%%%%%%%%%%%%%%%%
\subsection{Closing of gap to transverse excitations}
%%%%%%%%%%%%%%%%%%%%%%%%%%%%%%%%%%%%%
\label{sec:gap.closing}

Since the correlations of the spiral spin liquid [Fig.~\ref{fig:thermodyn}c] 
are encoded in the lowest lying transverse spin excitations of the high--field 
paramagnet [Fig.~\ref{fig:Sperp_2T_ring}], we can identify the onset of the 
spin liquid, with the closing of the gap to these excitations.
In Fig.~\ref{fig:BandGapEvolutionField} we show the gap $\Delta(B)$ 
to the lowest lying excitations in $S^\perp({\bf q}, \omega)$ evolves as 
a function of magnetic field, as found in MD simulations carried out at 
\mbox{$T = 220\ \text{mK}$}.
At higher values of field, where the magnetisation is (approximately) 
saturated, $\Delta(B)$ tracks the results of linear spin wave theory at 
$T=0$ [red line in Fig.~\ref{fig:BandGapEvolutionField}(a)], 
which would extrapolate to the gap closing at $B = 1.0\ \text{T}$.
However for fields $B \lesssim 2\ \text{T}$, $\Delta(B)$ starts to deviate from the 
spin wave prediction [inset to Fig.~\ref{fig:BandGapEvolutionField}(a)], 
finally closing for $B \gtrapprox 0.7\ \text{T}$.
In Fig.~\ref{fig:BandGapEvolutionField}(b) and (c), we show results for 
$S^\perp({\bf q}, \omega)$ at $B=0.5\ \text{T}$, just below, and at 
$B=1.0\ \text{T}$, a little above, the closing of the gap.
The corresponding estimates of the field at which the gap closes, 
form the basis of the phase boundary between correlated paramagnet 
and spiral spin liquid shown in Fig.~\ref{fig:phase.diagram.B}.

%%%%%%%%%%%%%%%%%%%%%%%%%%%%%%%%%%%
\subsection{Summary of results for dynamics in field}
%%%%%%%%%%%%%%%%%%%%%%%%%%%%%%%%%%%%%
\label{sec:summary.dynamics.in.field}

In summary, the dynamics of the BBK model of \CCO\ in field reveals 
a number of interesting features.
Firstly, dynamics in longitudinal and transverse channels are very different, 
with transverse dynamics showing a gap at high fields, while longitudinal 
dynamics remain gapless at all fields [Fig.~\ref{fig:DynamicsEvolutionField}].

%%%%%%%%%%%%%%%%%%%%%%%%%%%%%%%%%%%%%

\begin{figure}[t]
	\centering
	\captionsetup[subfigure]{labelformat=empty}
	\begin{minipage}[t]{0.99\columnwidth}
		\centering
		\subfloat[\label{fig:Gap_B}]{
  		\includegraphics[width=0.96\columnwidth]{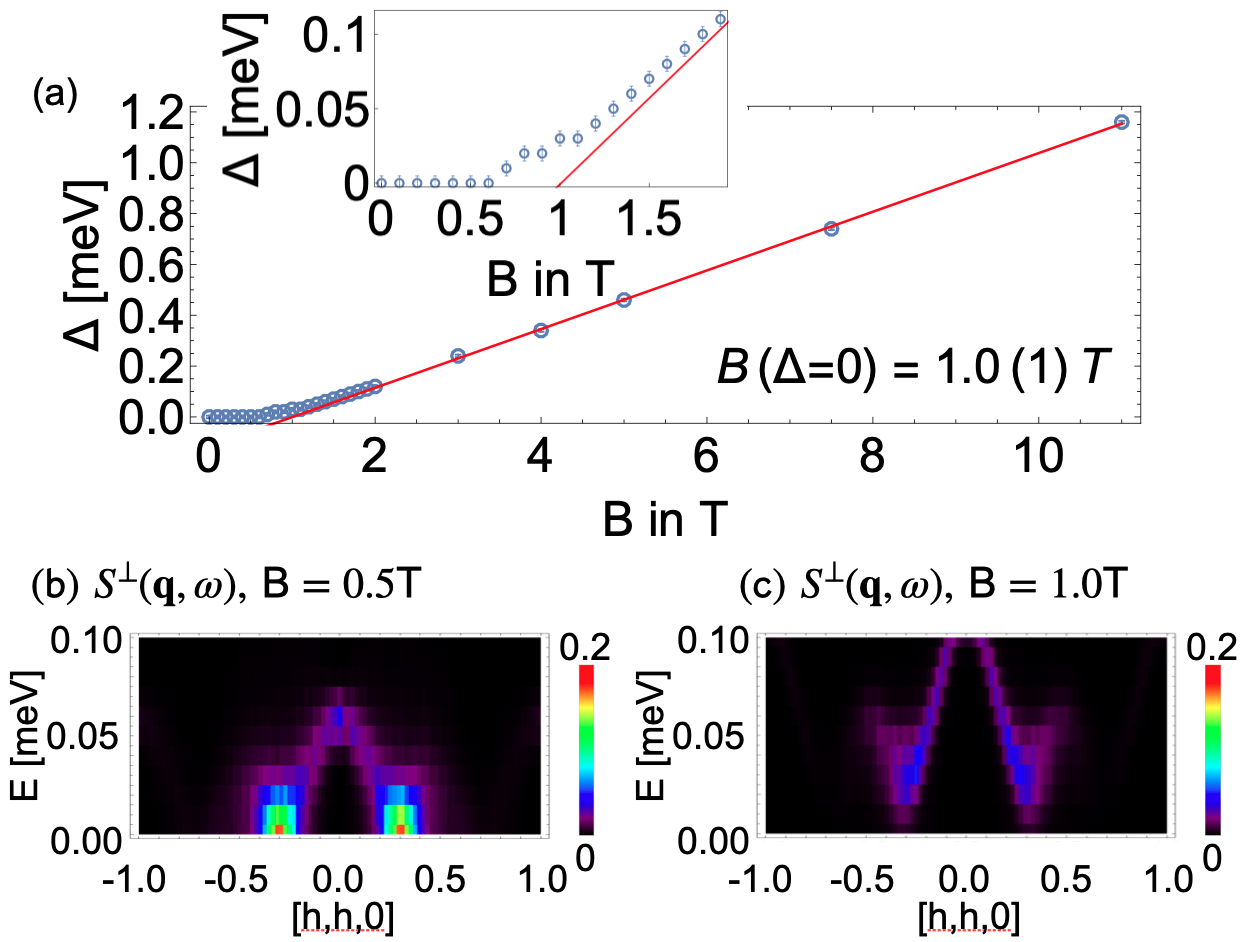}
		} 
	\end{minipage}
	\caption{
	Evolution of gap $\Delta$ to low--lying transverse 
	spin excitations in applied magnetic field.
	(a)~Results for $\Delta(B)$, as found in molecular 
	dynamics (MD) simulation at  $T = 220\ \text{mK}$ 
	[cf. Fig.~\ref{fig:DynamicsEvolutionField}]. 	
	Red line: linear behaviour found for spin--wave excitations 
	about the field--saturated state at $T=0$.
	%
	%\mbox{(fit: $B_{\sf fit} = -0.116 + 0.115 $ T)}.
	%
	Inset: Detail of $\Delta$ for $B \leq 2T$.
	(b)~Detail of MD results for transverse structure factor 
	$S_\perp({\bf q}, \omega)$ [Eq.~(\ref{eq:S.perp})] at $B=0.5\ \text{T}$, 
	showing gapless spin excitations.
	(c)~Equivalent results for $B=1.0\ \text{T}$, showing 
	gap to transverse spin excitations.
	Simulations were carried out %for $T = 220\ \text{mK}$, 
	for a cluster of linear dimension L = 48 \mbox{(N = 13,824)}, 
	using the %bilayer breathing Kagome (BBK) 
	BBK model Eq.~(\ref{eq:H.BBK}), with parameters taken from 
	experiment [Table~\ref{tab:experimental.parameters}].
	Error bars in (a) and pixel size in (b), (c) represent the finite-energy 
	resolution of MD simulations, $\delta E = 0.01\ \text{meV}$.
	}
	\label{fig:BandGapEvolutionField}
\end{figure}

Secondly, the lowest--lying transverse spin excitations in high field 
show the same ``ring''--like (quasi--)degeneracy as the classical spin  
liquid in zero field [cf Fig.~\ref{fig:Swq.1.meV} with 
Fig.~\ref{fig:Sperp_2T_ring}].
We therefore associate the closing of the gap to these  excitations, 
at a value of magnetic field which depends on temperature 
with the onset of the low--field spin--liquid state [Fig.~\ref{fig:BandGapEvolutionField}].

%%%%%%%%%%%%%%%%%%%%%%%%%%%%%%%%%%%%%

Thirdly, transverse dynamics at high field also reveal flat 
bands at intermediate energy, which carry pinch--point correlations,
resembling the ``bow--tie'' patterns observed in the spin liquid 
[Fig.~\ref{fig:Sperp_2T_PinchPoint}].
These are accompanied by half--moon motifs, inscribed on a 
dispersing band which intersects the flat band at zone centers 
[Fig.~\ref{fig:Sperp_2T_HalfMoon}].

%%%%%%%%%%%%%%%%%%%%%%%%%%%%%%%%%%%%%

We return to each of these features where we discuss the implication
for experiments on \CCO, below.

%%%%%%%%%%%%%%%%%%%%%%%%%%%%%%%%%%%%%
%  Fig. X - evolution of spin-wave gap with field  
%%%%%%%%%%%%%%%%%%%%%%%%%%%%%%%%%%%%%

%%%%%%%%%%%%%%%%%%%%%%%%%%%%%%%%%%%%%
\section{Application to $\text{Ca}_{10}\text{Cr}_7\text{O}_{28}$}
%%%%%%%%%%%%%%%%%%%%%%%%%%%%%%%%%%%%%
\label{sec:cco.v.bbk}

%%%%%%%%%%%%%%%%%%%%%%%%%%%%%%%%%%%%%
% connecting the finite--energy and finite--temperature properties 
% of \CCO\ with the spin--1/2 BBK model, Eq.~(\ref{eq:H.BBK})
%%%%%%%%%%%%%%%%%%%%%%%%%%%%%%%%%%%%%

In Section~\ref{section:open.questions.cco} we identified four 
open questions about the BBK model of \CCO.
We now return to these, addressing each in turn, before 
discussing some of the features of \CCO which still remain to be 
understood.

%%%%%%%%%%%%%%%%%%%%%%%%%%%%%%%%%%%%%

The first challenge was to connect the finite--energy and 
finite--temperature properties of \CCO\ with the \mbox{spin--1/2}  
BBK model, Eq.~(\ref{eq:H.BBK}).
When it comes to finite--temperature properties, in the absence of magnetic 
field, our main results are summarised in the phase diagram,  
%found in Monte Carlo (MC) simulation, 
Fig.~\ref{fig:thermodyn}(a), and the associated 
predictions for the equal--time structure factor, %$S({\bf q})$ 
Fig.~\ref{fig:thermodyn}(c,d) and Fig.~\ref{fig:Sq}.
We find that experimental parameters [Table~\ref{tab:experimental.parameters}], 
place \CCO\ within a ``spiral spin liquid'' regime.
This is characterised by ring--like 
correlations in $S({\bf q})$, and occurs for temperatures ranging from a 
lattice--nematic ordering transition at $T \approx 70\ \text{mK}$, to a crossover 
into a high--temperature paramagnet, occurring for $T \sim 500-1000\ \text{mK}$.
This is consistent with experiment, where a spin liquid, characterised by 
ring--like structures in $S({\bf q})$, occurs in the same range of 
temperatures \cite{Balz2016,Balz2017-PRB95}.
And the low--temperature lattice--nematic [Section~\ref{sec:3.state.Potts}], while not 
observed in \CCO, is consistent with earlier simulations of an effective 
honeycomb lattice model \cite{biswas18}.
Moreover, our conclusions about spin liquids at finite temperature prove 
to be robust for a wide range of parameters, and so are relatively insensitive to 
the uncertainty in estimates of exchange coupling taken from 
experiment \cite{Balz2016,Balz2017-PRB95}.

%%%%%%%%%%%%%%%%%%%%%%%%%%%%%%%%%%%%%

Molecular Dynamics (MD) simulations give further insight into the nature of 
this spin liquid, through its finite--energy properties.
Results for the dynamical structure factor $S({\bf q},\omega)$ 
are summarised in Fig.~\ref{fig:dynamics.in.zero.field}, in the 
comparison with experiment, Fig.~\ref{fig:comparison.with.experiment}, 
and in the two animations \cite{first.animation,second.animation}.
A key feature is the separation of dynamics into three distinct time--scales; 
a long timescale associated with the slow, collective excitations of groups 
of three spins on ferromagnetically--correlated plaquettes, and intermediate 
and short timescales associated with qualitatively different excitations.
Correlations at low energies echo the spin--liquid, with ring--like 
features in $S({\bf q},\omega)$, while dynamics at higher energies 
show a diffuse web of scattering, with bow--tie like features visible in 
a subset of zone centers.
Ring and bow--tie features would usually be associated with qualitatively 
different spin--liquid states, and in this sense the BBK model %of \CCO\ 
appears to support different spin liquids, on different timescales, at the same time.

%%%%%%%%%%%%%%%%%%%%%%%%%%%%%%%%%%%%%

Experiment on \CCO\ also shows qualitatively different  
correlations on different time scales, with hints of rings at low 
energy, and bow--tie like features at intermediate and high 
energies \cite{Balz2016,Balz2017-PRB95}.
Where experiment and (classical) simulation differ is in the extent to 
which higher--energy excitations form a continuum, with MD results 
showing weakly--dispersing excitations at relatively well--defined energies.

%%%%%%%%%%%%%%%%%%%%%%%%%%%%%%%%%%%%%
% extending the analysis of this model to finite magnetic field 
%%%%%%%%%%%%%%%%%%%%%%%%%%%%%%%%%%%%%

The second challenge was to extend the analysis of the BBK model 
to finite values of magnetic field.
Here, key results are summarised in the phase diagrams Fig.~\ref{fig:schematic.phase.diagram.B} 
and Fig.~\ref{fig:phase.diagram.B}, and the associated predictions 
for spin dynamics, Fig.~\ref{fig:DynamicsHighField}, Fig.~\ref{fig:DynamicsEvolutionField}
and Fig.~\ref{fig:Eslice_SwqParaPerp}.
We find that spin--liquid correlations persist up to a temperature--dependent field 
\mbox{$B \sim 1\ \text{T}$}, but with dramatically different correlations at different 
energy scales, and in the transverse and longitudinal channels.
At higher fields, classical simulations find a highly--correlated paramagnet, 
in which transverse spin excitations are gapped, but longitudinal excitations 
remain gapless.
Meanwhile, at low temperatures, the lattice nematic gives way to a series of 
complex forms of order.

%%%%%%%%%%%%%%%%%%%%%%%%%%%%%%%%%%%%%

Once again, the phenomenology of the BBK model in field has 
many similarities with experiments on \CCO. 
Fits to the spin wave excitations of the saturated paramagnet have 
already been documented \cite{Balz2016,Balz2017-PRB95}.
And the zero--field magnetic susceptibility of \CCO\ \cite{Balz2017-PRB95,Kshetrimayum2020}, 
is also very similar to that found in the BBK model at low temperatures, 
with the caveat that the classical statistics of our MC simulations lead to a much 
stronger temperature dependence than is seen in experiment [Fig.~\ref{fig:M.as.function.B}].
Measurements of \CCO\ at low temperatures also exhibit a qualitative 
change in behaviour of $C/T$ at $B \sim 1\ \text{T}$ \cite{Balz2017-PRB95}, 
consistent with the opening of a gap found in MD simulations.
No sign has yet been seen, however, of the complex competing 
orders found in classical MC simulations at the lowest 
temperatures.

%%%%%%%%%%%%%%%%%%%%%%%%%%%%%%%%%%%%%

% identifying the mechanism driving its low--temperature spin liquid state.

The third challenge was to identify the mechanism driving the 
low temperature spin liquid state.
Here simulations of the BBK model offer a number of new insights.
Firstly, they show a clear separation of dynamics within the spin liquid,  
between slow collective rotations of FM polarised triangular plaquettes, 
and fast excitations of individual spins [Second Animation \cite{second.animation}].
This confirms the validity of modelling the low--energy dynamics of \CCO\ 
in terms of spin--3/2 moments on a honeycomb lattice, a construction which 
has played an important role in existing classical \cite{biswas18} 
and quantum \cite{Sonnenschein2019} theories of \CCO.
Secondly, the correlations found at low energy are consistent with 
a spiral spin liquid, while those at higher energies resemble those  
of a Kagome anitferromagnet 
[Fig.~\ref{fig:comparison.with.experiment}, Fig.~\ref{fig:dynamics.in.zero.field}].
And thirdly, the onset of spiral spin liquid behaviour can be  
identified with the closing of the gap to transverse excitations at 
$B \sim 1\ \text{T}$ [Fig.~\ref{fig:BandGapEvolutionField}].
A key feature of this transition is the (quasi--)degeneracy of the 
``ring'' of excitations at the bottom of the spin--wave spectrum, 
all of which participate in the resulting spin liquid.
While the equivalent quantum theory remains an open problem, 
this scenario has much in common with the formation of a chiral 
spin liquid through the condensation of hard--core bosons 
in a system with a continuous, ``moat''--like degeneracy \cite{sedrakyan15-PRL114}.

%%%%%%%%%%%%%%%%%%%%%%%%%%%%%%%%%%%%%

% identifying interesting properties of the BBK model which may, as yet, 
% be obscure in experimental data for \CCO

The fourth challenge was to identify interesting properties of the BBK model 
which may, as yet,  be obscure in experimental data for \CCO.  
Here, MD simulations reveal a number of features of the dynamics 
of the BBK model which it would be rewarding to look for in experiment.
In particular, they suggest that the application of a magnetic field could offer
new insights into the spin liquid, by separating longitudinal and transverse 
dynamics with dramatically different character 
[Fig.~\ref{fig:DynamicsEvolutionField}, Fig.~\ref{fig:Eslice_SwqParaPerp}].
This is something which could be probed using polarised neutron scattering.
Our results also suggest that \CCO\ is a suitable system for investigating 
the physics of pinch points and half moons \cite{Yan2018}, which play a 
prominent role in the excitations at finite energy.
These could most easily be studied in the high--field, saturated 
paramagnetic state, cf. Fig.~\ref{fig:Sperp_2T_HalfMoon}.

%%%%%%%%%%%%%%%%%%%%%%%%%%%%%%%%%%%%%

Taken together, these results represent significant progress in 
understanding \CCO.
None the less, there are a number of features in experiment which 
remain to be understood.
One is the presence of spectral weight at finite energies for 
wave vector ${\bf q} =0$ in the absence of magnetic field \cite{Balz2017-PRB95}, 
something which is forbidden for a spin--rotationally invariant model like 
${\cal H}_{\sf BBK}$~[Eq.~(\ref{eq:H.BBK})].
A likely explanation is the presence of anisotropic exchange interactions, 
including Dzyaloshinskii--Moriya (DM) terms. 
These are allowed by the symmetry of the lattice, and would endow the 
magnon bands of the field--saturated state with a topological character 
\cite{Judit-unpub}.
They would also contribute to the finite lifetime of  
excitations \cite{Chernyshev2015}, which could help to explain their 
relatively broad character in experiment \cite{Balz2017-PRB95}.

%%%%%%%%%%%%%%%%%%%%%%%%%%%%%%%%%%%%%

Another area where further investigation is merited, is the question of how the  
thermodynamics and spin dynamics of the BBK model 
of \CCO\ change once quantum statistics, and entanglement, are taken 
into account.
Exact diagonalisation (ED), and thermal pure quantum state (TPQ)  
studies of the BBK model, will form the subject of a second  
Article \cite{shimokawa-in-prep}.
However we can already make some comparison with published 
results from other quantum approaches 
\cite{Balz2016,Kshetrimayum2020,Sonnenschein2019}, 
which suggest that the role of quantum and thermal fluctuations 
in this problem may not be very different.

%%%%%%%%%%%%%%%%%%%%%%%%%%%%%%%%%%%%%

The biggest difference appear to arise in the zero--field ground state, 
where PFFRG calculations finds static correlations (characteristic ``ring'' in 
$S({\bf q}, \omega=0)$) consistent with a spin liquid \cite{Balz2016}.
Meanwhile, tensor--network calculations report a vanishing average magnetic 
moment per site, coupled to a finite magnetic susceptibility, also 
consistent with a QSL \cite{Kshetrimayum2020}.
In contrast, our MC simulations show a concentration of weight in $S({\bf q})$
at the discrete set of wave vectors associated with a lattice--nematic state 
[Section~\ref{sec:thermodynamics}] and, by their nature, always exhibit a 
finite moment on each site.
However, the ordered states found in simulation only occur at 
very low temperatures.
And given that the model is two--dimensional, and spin excitations 
are (quasi--)degenerate on line--like loci, its is highly likely that 
quantum fluctuations would eliminate the ordered 
moments at the level of individual sites at zero temperature, 
cf. \cite{Smerald2010}, just as thermal fluctuations do at higher temperatures.
Whether the $Z_3$ symmetry--breaking would survive 
in the absence of an ordered moment is a separate, and interesting, 
question \cite{Mulder2010}.

%%%%%%%%%%%%%%%%%%%%%%%%%%%%%%%%%%%%%

Even at a classical level, there are intriguing similarities 
in the thermodynamics, with tensor--network calculations of magnetization 
suggestive of phase transitions at $B \approx 0.3\ \text{T}$, $B \approx 0.8\ \text{T}$ 
and $B \approx 1.0\ \text{T}$ \cite{Kshetrimayum2020}, scales similar 
to the transitions between different ordered states found in MC simulation 
at low temperature [Section~\ref{sec:thermodynamics.in.field}].
And when it comes to dynamics, it is encouraging to note the extent 
to which MD simulations reproduce features seen in 
%both exact diagonalization (ED) calculations \cite{shimokawa-in-prep}, and in
the spinon phenomenology of Sonnenschein {\it et al.} \cite{Sonnenschein2019}.   
In particular the ``volcano'' feature observed in the structure factor
associated with longitudinal fluctuations $S_\parallel({\bf q},\omega)$ 
[Section~\ref{sec:dynamics.in.field}] is very reminiscent of the low--energy 
cone of particle--hole excitations about the Fermi surface.

%%%%%%%%%%%%%%%%%%%%%%%%%%%%%%%%%%%%%

Needless to say, (semi--)classical simulations, by themselves, cannot 
be used to argue for the existence of spinons in \CCO. 
A method with access to quantum entanglement is required \cite{shimokawa-in-prep}. 
However, the congurence of experiment, simulation of the BBK model, and parton 
approaches does suggest that, phenomenologically, a spinon picture may not be 
so wide of the mark.
And, as a general comment on the present state of theory of \CCO, it is interesting 
to note that the picture advanced by Sonnenschein {\it et al.} \cite{Sonnenschein2019} 
is essentially that of a heavy Fermion superconductor \cite{hewson-book,solyom-book, fulde-book}, 
but with spinons, rather than electrons, as heavy quasi--particles.
% with an effective mass of $XXX$ 

%%%%%%%%%%%%%%%%%%%%%%%%%%%%%%%%%%%%%

In both heavy Fermion superconductors, and the parton phenomenology of 
Sonnenschein {\it et al.}~\cite{Sonnenschein2019}, low--energy properties 
are dictated by the existence of a Fermi--surface, with heavy quasiparticles that 
become unstable against pairing at low temperatures.
An immediate implication is that the specific heat coefficent $\gamma$, and 
magnetic susceptibility $\chi$ are strongly enhanced, 
as observed in  \CCO\ [Table~\ref{tab:heavy.fermions}].
However an important difference between the parton theory, and conventional 
(heavy) Fermi liquids, is that long range interactions between neutral spinons are 
not screened in the same way as interactions between electrons.
In (2+1)D this inexorably drives $U(1)$ QSL towards a strong--coupling fixed point \cite{Lee1989,Lee1992,Polchinski1994,Nayak1994,Altschuler1994,Kim1994,Senthil2004,Montrunich2005},  
as exemplified by the $Z_2$ QSL ground state conjectured by Sonnenschien {\it et al.}~\cite{Sonnenschein2019}.
And it follows that, if \CCO\ really does have (emergent) heavy Fermi 
quasiparticles, these should show signs of strong correlation.

%%%%%%%%%%%%%%%%%%%%%%%%%%%%%%%%%%%%%

A common cross--check on the degree of correlation within a heavy Fermion system 
%
%A useful way of characterising the correlations % within a heavy--Fermion material, 
is to calculate the Wilson ratio $R_W$, a dimensionless number formed from 
the (para)magnetic susceptibility $\chi_0$, and the linear coefficient of specific 
heat, $\gamma$ \cite{solyom-book,fulde-book}.
By construction the Wilson ratio of a free electron gas $R_W =1$, while for highly—correlated 
electron bands, it takes on higher values.
Substituting experimental parameters for \CCO\ [Table~\ref{tab:heavy.fermions}],  
we find %a Wilson ratio 
\begin{eqnarray}
R_W=\frac{\pi^2}{\mu_0} \frac{\chi_0}{\mu_\text{eff}^2}\frac{k_B^2}{\gamma} 
        \approx 16.2 \; , 
\label{eq:wilson.ratio} 
\end{eqnarray}
where details of the estimate are given in Appendix~\ref{sec:wilson.ratio}.
This value is substantially greater than the $R_W \approx 4$ observed for the 
heavy Fermion system CeCu$_6$ \cite{Ott1987,Amato1987}, and 
would place \CCO\ in a strongly--interacting Fermi--liquid regime.
%

%%%%%%%%%%%%%%%%%%%%%%%%%%%%%%%%%%%%%

We conclude the spinon phenomonlology of \CCO\  \cite{Sonnenschein2019} 
remains an interesting conjecture, that does not suffer from any obvious 
contradiction with experiment, and is therefore worthy of further investigation.
And in this context it could be interesting to investigate other properties  
usually controlled by the Fermi surface in metals, such as (thermal) 
transport, and NMR $1/T_1$ relaxation rates.
All of this, however, lies outside the scope of the present article.

%%%%%%%%%%%%%%%%%%%%%%%%%%%%%%%%%%%%%
\section{Conclusions}
%%%%%%%%%%%%%%%%%%%%%%%%%%%%%%%%%%%%%
\label{sec:conclusions}

\CCO\ is a remarkable magnet, in which \mbox{spin-1/2} Cr$^{5+}$ ions form 
a bilayer breathing-kagome (BBK) lattice with complex, 
competing exchange interactions \cite{Balz2017-JPCM29}.  
A combination of heat-capacity, magnetization, $\mu$SR, neutron-scattering, 
and AC susceptibility experiments reveal \CCO\ to be a gapless quantum 
spin liquid (QSL), showing no sign of magnetic order down to $19\ \text{mK}$ 
\cite{Balz2016,Balz2017-PRB95}.
This spin liquid is charcaterised by spin fluctuations which 
show qualitatively different character on different timescales.

%%%%%%%%%%%%%%%%%%%%%%%%%%%%%%%%%%%%%

To better understand the nature and origin of the spin liquid in \CCO, we have 
carried out large-scale semi-classical molecular-dynamics (MD) simulations of the 
minimal model of \CCO, a Heisenberg model on the BBK lattice, with 
parameters taken from experiment \cite{Balz2016,Balz2017-PRB95}.
%
%As %the results in Fig.~\ref{fig:spin.dynamics.0T}, and 
%the Animations in the Supplementary Information 
%demonstrate, 
These simulations reveal a state where spins continue to fluctuate at 
very low temperatures, but the character of these 
fluctuations depends strongly on the timescale on which the dynamics 
are resolved, as shown in the Animations \cite{first.animation,second.animation}.   
This persists up to a (temperature--dependent) critical field 
$B_c \sim 1\ \text{T}$  [Fig.~\ref{fig:phase.diagram.B}], for temperatures 
ranging from an ordering temperature \mbox{$T \sim 70\ \text{mK}$}, 
to a crossover into a high--temperature paramagnet for \mbox{$T \sim 500\ \text{mK}$} 
[Fig.~\ref{fig:schematic.phase.diagram.B}].

%%%%%%%%%%%%%%%%%%%%%%%%%%%%%%%%%%%%%

Within this state, we identify fluctuations at low energies with 
a ``spiral spin liquid'', characterised by a ring of scattering in ${\bf q}$-space, 
and formed when the gap to a (quasi-)degenerate set of excitations 
closes at $B \lesssim 1\ \text{T}$ [cf. Fig.~\ref{fig:BandGapEvolutionField}].
This spiral spin liquid can be described by an effective 
spin-3/2 Heisenberg model on a honeycomb lattice, formed
by three spin-1/2 moments on the triangular plaquettes of 
the BBK lattice in \CCO\ [cf. Fig.~\ref{fig:BBK.model}, Fig.~\ref{fig:HC.model}].
The FM correlations of spins within these plaquettes 
are evident in the collective motion resolved at low energy 
in MD simulation (cf. Second Animation~\cite{second.animation}). 

%%%%%%%%%%%%%%%%%%%%%%%%%%%%%%%%%%%%%

Meanwhile, fluctuations at higher energy inherit their character from 
the kagome-lattice antiferromagnet, and for \mbox{$B \gtrsim 1\ \text{T}$}, 
are characterised by sharp pinch--points in scattering 
Fig.~\ref{fig:Sperp_2T_PinchPoint}.
These pinch points are encoded in spin fluctuations perpendicular 
to the applied magnetic field and can be resolved in MD simulations 
as collective rotations of antiferromagnetically correlated spins on 
shorter timescales (cf. Second Animation \cite{second.animation}).
In the limit $B \to 0\ \text{T}$,  pinch points ultimately merge with excitations 
in the longitudinal channel to give rise to the broader ``bow--tie'' features 
observed in inelastic neutron scattering at higher energy %for $B = 0\ \text{T}$ 
[cf. Fig.~\ref{fig:comparison.with.experiment}].

%%%%%%%%%%%%%%%%%%%%%%%%%%%%%%%%%%%%%

These simulations capture many of the features of \CCO; correctly reproducing 
the value of the critical field, $B \lesssim 1\ \text{T}$ \cite{Balz2016,Balz2017-PRB95}; 
providing insight into the different structures seen in inelastic neutron scattering 
\cite{Balz2016,Balz2017-PRB95}; 
and resolving the origin of the ring features found in Pseudo-Fermion Functional 
Renormalisation group (PFFRG) calculations \cite{Balz2016}.
To the best of our knowledge, they also provide the first theoretical example of 
a system which behaves like different types of spin liquid on different timescales.

%%%%%%%%%%%%%%%%%%%%%%%%%%%%%%%%%%%%%

Given this disparity of behaviour, it is tempting to ask just how many spin 
liquids there are in \CCO?
Since a quantum system should have one, unique, ground state, 
at low temperature the answer to this 
%rhetorical
%somewhat provocative 
question must, ultimately, be: ``{\it one}''.
None the less, the success of semi--classical simulations in describing 
experiment %on \CCO\ 
suggests that this one ground state must incorporate 
two different types of correlations; one described by effective spin-3/2 moments 
on a honeycomb lattice; and one corresponding to antiferromagnetic fluctuations of 
individual spin-1/2 moments on a bilayer breathing-kagome lattice.
Unraveling the properties of this single, massively-entangled QSL,    
represents an exciting challenge for theory and experiment alike.
And in a second paper, we will 
return to this in the context of quantum simulations 
of the BBK model of \CCO\  \cite{shimokawa-in-prep}.

%%%%%%%%%%%%%%%%%%%%%%%%%%%%%%%%%%%%%
\begin{acknowledgments}
%%%%%%%%%%%%%%%%%%%%%%%%%%%%%%%%%%%%%

The authors are pleased to acknowledge helpful conversations with 
Owen Benton, Ludovic Jaubert, Jonas Sonnenschein and Mathieu Taillefumier, 
and are indebted to Christian Balz and Bella Lake for extended discussions 
and sharing information about experiments on \CCO.
We are grateful for the help and support provided by Pavel Puchenkov
from the Scientific Computing and Data Analysis section of Research 
Support Division at OIST. 
This work was supported by the Theory of Quantum Matter Unit, OIST.
Numerical calculations we carried out using HPC Facilities provided by OIST.

%%%%%%%%%%%%%%%%%%%%%%%%%%%%%%%%%%%%%
\end{acknowledgments}
%%%%%%%%%%%%%%%%%%%%%%%%%%%%%%%%%%%%%

%%%%%%%%%%%%%%%%%%%%%%%%%%%%%%%%%%%%%
\appendix				
%%%%%%%%%%%%%%%%%%%%%%%%%%%%%%%%%%%%%%
\label{sec:appendix}

%%%%%%%%%%%%%%%%%%%%%%%%%%%%%%%%%%%
\section{Numerical Methods}		
%%%%%%%%%%%%%%%%%%%%%%%%%%%%%%%%%%%
\label{appendix:Numerics}

%%%%%%%%%%%%%%%%%%%%%%%%%%%%%%%%%%%%
\subsection{Classical Monte Carlo}     
%%%%%%%%%%%%%%%%%%%%%%%%%%%%%%%%%%%%
\label{sec:MC}

All of the results presented in this Article are based on spin configurations drawn from 
classical Monte Carlo simulations of ${\cal H}_\text{\sf BBK}$ [Eq.~(\ref{eq:H.BBK})].
Monte Carlo simulations  were performed by using a local heat--bath 
algorithm \cite{Olive1986, Miyatake1986}, in combination with parallel 
tempering \cite{Swendsen1986, Earl2005}, and over--relaxation 
techniques \cite{Creutz1987}. 
Here, we chose the heat-bath algorithm, which automatically adjusts the 
solid angle for updated spins at their given temperatures. 
In this way this method is rejection free, and therefore outperforming the 
conventional single-spin flip Metropolis algorithm \cite{Metropolis1953} 
at very low-temperatures.

%%%%%%%%%%%%%%%%%%%%%%%%%%%%%%%%%%%%

Within simulations, a single MC step consists of $N$ local heat--bath 
updates on randomly chosen sites, and two over--relaxation steps, each 
comprising a \mbox{$\pi$--rotation} of all the spins in the lattice 
about their local exchange fields.
Simulations were performed in parallel for replicas at a range of different 
temperatures, with replica--exchange 
initiated by the parallel tempering algorithm every $10^2$ MC steps.
Results for thermodynamic quantities were averaged over $10^6$ statistically independent  
samples, after initial $10^6$ MC steps for simulated annealing and $10^6$ MC steps for 
thermalisation.

%except for the phase diagram, shown in Fig.~\ref{fig:model.and.phase.diagram}, 
%which was calculated for a cluster of \mbox{$N = 1944$} sites.

%%%%%%%%%%%%%%%%%%%%%%%%%%%%%%%%%%%%%
\subsection{Semi-Classical Molecular Dynamics}			\label{sec:MD}
%%%%%%%%%%%%%%%%%%%%%%%%%%%%%%%%%%%%%

To interpret the INS data for \CCO\ \cite{Balz2016,Balz2017-PRB95}, we rely on 
Molecular Dynamics (MD) simulations.
These are based on the numerical integration of the %(semi--)classical, 
Heisenberg equations of motion 
%
%
%%%%%%%%%%%%%%%%%%%%%%%%%%%%%%%%%%%%%
\begin{eqnarray}
	\frac{d {\bf S}_i}{d t} 	= \frac{\mathrm{i}}{\hbar} \big[ {\cal H}_{\sf BBK}, {\bf S}_i \big]   
					= \bigg( \sum_j J_{ij} {\bf S}_j - B^z \hat{\mathbf{z}}\bigg) \times {\bf S}_i		\; ,
\label{eq:E.of.M2}
\end{eqnarray}
%%%%%%%%%%%%%%%%%%%%%%%%%%%%%%%%%%%%%
%
where $j$ accounts for all nearest--neighbouring sites of $i$ and $J_{ij}$ is given in 
Table~\ref{tab:experimental.parameters}.

%%%%%%%%%%%%%%%%%%%%%%%%%%%%%%%%%%%%%

Spin configurations for MD simulation were taken from the thermal 
ensemble generated by classical MC simulations of ${\cal H}_{\sf BBK}$ 
at \mbox{$T=220\ \text{mK}$}, for parameters taken from experiment 
[cf. Table~\ref{tab:experimental.parameters}].
Numerical integration of Eq.~(\ref{eq:E.of.M2}) was then carried out 
using a 4$^{th}$ order Runge-Kutta algorithm, as described in 
\cite{NumericalRecipes2007, OrdinaryDiffEquations1}.   
Simulations were performed for $N_t = 600$ time steps, with a 
time--increment $\delta t$ of 
\begin{equation}
	\delta t = \frac{t_{\text{max}}}{N_{t}} = \frac{2 \pi}{\omega_{\text{max}} }
	\label{eq:detlaT}
\end{equation}
setting a maximaum resolvable frequency of \mbox{$\omega_{\text{max}} = 6\ \text{meV}$}.

%%%%%%%%%%%%%%%%%%%%%%%%%%%%%%%%%%%%%

The dynamical structure factor 
\begin{equation}
	S({\bf q}, \omega) = \frac{1}{\sqrt{N_t} N}
					\sum_{i, j}^N  \mathrm{e}^{ \mathrm{i} {\bf q} ({\bf r}_{i} - {\bf r}_j ) } 
					\sum_n^{N_t}  \mathrm{e}^{ \mathrm{i}  \omega \ n \delta t } \: 
					\langle {\bf S}_{i} (0) \cdot {\bf S}_{j} (t) \rangle   \; ,	
\label{eq:S.q.omega}
\end{equation}
was calculated using Fast Fourier Transform (FFT) \cite{FFTW05}, and averaged 
over spin dynamics obtained from $500$ independent initial spin configurations.
Where simulations were carried out in applied magnetic field, this was resolved 
in to contributions coming from transverse and longitudinal fluctuations, i.e.  
\begin{eqnarray}
    S({\bf q},\omega) = S^\bot({\bf q}, \omega) + S^\parallel({\bf q}, \omega) 
   \label{eq:S.perp.para}
\end{eqnarray}
where
\begin{eqnarray}
	S^\bot({\bf q}, \omega)  &=& \frac{1}{\sqrt{N_t} }
		\sum_n^{N_t}  \mathrm{e}^{ \mathrm{i}  \omega \ n \delta t } \: 
		\langle {\bf S}^\bot_{\bf q}(t) \cdot {\bf S}^\bot_{\bf -q}(0) \rangle    \; ,  \\ 
	{\bf S}^\bot_i &=& (S^x_i, S^y_i)   \; ,	
	\label{eq:S.perp2} 
\end{eqnarray}
and
\begin{eqnarray}
	S^\parallel({\bf q}, \omega)  &=& \frac{1}{\sqrt{N_t} }
		\sum_n^{N_t}  \mathrm{e}^{ \mathrm{i}  \omega \ n \delta t } \: 
		\langle S^z_{\bf q}(t) \cdot S^z_{\bf -q}(0) \rangle    \; .	
	\label{eq:S.parallel}
\end{eqnarray}

%%%%%%%%%%%%%%%%%%%%%%%%%%%%%%%%%%%%%%

To avoid numerical artifacts (Gibbs phenomenon \cite{MathimaticalPhyics}), coming from 
discontinuities of the finite time-window at \mbox{$t = 0$} and \mbox{$t = t_{\text{max}}$}, the time 
sequence of spin configurations has been multiplied by a Gaussian 
envelop prior to Fourier transform, imposing a maximally possible Gaussian energy resolution of 
\mbox{FWHM = 0.02\ meV}, on the numerically-obtained $S({\bf q}, \omega)$.

%%%%%%%%%%%%%%%%%%%%%%%%%%%%%%%%%%%%%
\subsection{Correcting for classical statistics}		
%%%%%%%%%%%%%%%%%%%%%%%%%%%%%%%%%%%%%
\label{sec:MD.corrected}

The MD simulations described above inherit the classical statistics of 
the classical MC simulations from which spin configurations are drawn.
At low temperatures, where spins are treated as classical $O(3)$ vectors, 
it is possible to decompose the excitations about a given ground state 
into $2N$ individual harmonic modes.
(This approach can also be generalised to the ensemble of ground 
states found in classical spin liquids \cite{Shannon2010}).
Each of these modes functions like a (classical) harmonic oscillator, with 
amplitude 
\begin{eqnarray}
\langle x^2 \rangle \sim \frac{T}{\omega}
\label{eq:classical.statistics}
\end{eqnarray}
and at low temperatures, the spin excitations found in 
classical MC simulation follows this distribution of amplitudes.

%%%%%%%%%%%%%%%%%%%%%%%%%%%%%%%%%%%%%

MD simulations evolve spin configurations in time according to the Heisenberg 
equation of motion [Eq.~(\ref{eq:E.of.M2})], with eigenenergies equivalent to 
those found in linear spin wave (LSW) approximation.
However, while the harmonic excitations of a LSW theory 
are quantised as Bosons, with amplitude determined by a Bose factor,  
MD simulations inherit the classical statistics of MC simulations which, 
at low temperatures, are described by Eq.~(\ref{eq:classical.statistics}).
It follows that MD simulations function in a ``mixed ensemble'', 
with classical statistics, but semi-classical dynamics. 
And to obtain a valid prediction for semi--classical dynamics in the limit 
$T \to 0$, that can be compared directly with LSW theory at $T=0$, 
it is therefore necessary to  ``divide out'' the classical statistical factor, 
Eq.~(\ref{eq:classical.statistics}).

%%%%%%%%%%%%%%%%%%%%%%%%%%%%%%%%%%%%%

A careful analysis of the dynamical structure factor $S_{\sf MD} ({\bf q}, \omega)$
found in MD simulation of a spin--1/2 moment, taking into account the mixed 
ensemble, leads to the result quoted as Eq.~(\ref{eq:S.tilda}) of main text
\begin{eqnarray}
   \tilde{S}({\bf q}, \omega)  
  %=  S^{\sf quantum}_{\sf T = 0}({\bf q}, \omega)  
   					 = \frac{1}{2} \frac{\omega}{k_B T} S_{\sf MD} 
   ({\bf q}, \omega) \; ,
\end{eqnarray}
where $ \tilde{S}({\bf q}, \omega)$ is the corresponding 
prediction for quantum (semi--classical) dynamics in the limit $T \to 0$.
Details of this calculation will be reported elsewhere \cite{remund-in-preparation}.

%%%%%%%%%%%%%%%%%%%%%%%%%%%%%%%%%%%%%%
\section{Real-time animation}
%%%%%%%%%%%%%%%%%%%%%%%%%%%%%%%%%%%%%
\label{appendix:animation}

Here we provide technical details of the animations of MD simulation results 
discussed in Section~\ref{sec:first.animation} and Section~\ref{sec:second.animation}.
Both animations were prepared using the open-source software package 
``Blender'' \cite{blender}.

%%%%%%%%%%%%%%%%%%%%%%%%%%%%%%%%%%%%%%
\subsection{First Animation}
%%%%%%%%%%%%%%%%%%%%%%%%%%%%%%%%%%%%%

In the First Animation \cite{first.animation}, we show a ``fly through'' of spins on the breathing 
bilayer-kagome (BBK) lattice,  for a cluster of $N = 5400$ sites, at a temperature of 
$T = 220 \text{mK}$.
The total number of time steps in the simulation was  $N_t = 6000$, and in order 
to obtain a smooth rotation of spins in the animation, the time step for 
each frame has been set to $\delta t \approx 0.1\ \hbar \text{meV}^{-1}$ 
[cf.~Eq.(\ref{eq:detlaT})], which is of the order of a femto second.
To emphasize the dynamics on different timescales, spins have been 
color-coded according to their speed of rotation, with red indicating
fast rotation, and green denoting slow rotation.

%%%%%%%%%%%%%%%%%%%%%%%%%%%%%%%%%%%%%%
\subsection{Second Animation}
%%%%%%%%%%%%%%%%%%%%%%%%%%%%%%%%%%%%%

In the Second Animation \cite{second.animation} we show results taken from a single 
MD simulation of ${\mathcal H}_{\sf BBK}$, equivalent to that shown in the First Animation. 
However in this case, the time sequence for the spin dynamics %coming from MD simulation
has been separated into slow, intermediate and fast components, to emphasise the  
dynamics on different timescales.
This is accomplished by first performing a fast Fourier transform (FFT) on the time
sequence of each spin, then filtering the resulting signal in frequency space, 
using a digital analogue of a ``band-pass'' filter.
The frequencies used for this band-pass filter are equivalent to energies of
$0.750$--$1.500\ \text{meV}$ (fast fluctuations); 
$0.225$--$0.750\ \text{meV}$ (intermediate fluctuations); 
and $0.00$--$0.225\ \text{meV}$ (slow fluctuations). 
After filtering in frequency, a second FFT %back to time %from frequency to time 
is used to reconstruct separate time sequences for slow, intermediate 
and fast fluctuations.
The final result for each of these %three time sequences 
is presented in the three panels of the Second Animation.

%%%%%%%%%%%%%%%%%%%%%%%%%%%%%%%%%%%%%

In the second part of the Second Animation, the speed of playback for three time sequences has 
been adjusted, so as to match the characteristic speed of the relevant fluctuations.
To accommodate this, a much longer sequence of $N_t = 130000$ time steps has been 
generated from MD simulation, using a much shorter time increment of 
\mbox{$\delta t \approx  0.035\ \hbar^{-1} \text{meV}^{-1}$}.
The adjustment of playback speed has been accomplished within ``Blender'' \cite{blender}, 
by adjusting the number of frames included for each time sequence. 
Viewed in this way, the very different dynamics on different timescales is 
self-evident.

%%%%%%%%%%%%%%%%%%%%%%%%%%%%%%%%%%%%%
\section{Comparison with Experiment} 		
%%%%%%%%%%%%%%%%%%%%%%%%%%%%%%%%%%%%%
\label{sec:CompExp}

Predictions for inelastic neutron scattering are plotted as 
\begin{eqnarray}
	\frac{d^2 \sigma}{d\Omega d E_f} 
					&\propto& I({\bf q}, \omega)
\end{eqnarray}
where we calculate 					
\begin{eqnarray}
	 I({\bf q}, \omega) &=& \mathcal{F}({\bf q})^2  \sum_{\alpha, \beta}  
					\bigg( \delta_{\alpha \beta} - \frac{q_{\alpha}q_{\beta}}{{\bf q}^2} \bigg)
					S^{\alpha \beta}({\bf q}, \omega)   \; .
\end{eqnarray}
Here $\mathcal{F}({\bf q})$ is the atomic form factor appropriate to a Cr$^{5+}$ ion, 
and following \cite{NeutronDataBooklet}, we write 
\begin{equation}
	\mathcal{F}({\bf q}) 
	=   \langle j_0 ({\bf q}) \rangle 
	+ \bigg(1 - \frac{2}{g} \bigg) \langle j_2 ({\bf q}) \rangle	\; .
\end{equation}
We consider gyromagnetic ratio $g = 2$, implying that 
$\langle j_2 ({\bf q}) \rangle$ plays no role.
The remaining function, $\langle j_0 ({\bf q}) \rangle $ can be parameterised as
%
%%%%%%%%%%%%%%%%%%%%%%%%%%%%%%%%%%%%%
\begin{equation}
	\langle j_0 ({\bf q}) \rangle	
	= 	A e^{- a ( |{\bf q} | / 4 \pi)^2} + B e^{- b ( | {\bf q} | / 4 \pi)^2} + C e^{- c (| {\bf q} | / 4 \pi)^2} + D	\; ,
	\label{eq:FormFactor_j0}
\end{equation}
%%%%%%%%%%%%%%%%%%%%%%%%%%%%%%%%%%%%%
%
where, be consistent with earlier work \cite{balz-private}, 
coefficients are taken to be 
%
%%%%%%%%%%%%%%%%%%%%%%%%%%%%%%%%%%%%%
\begin{align}
	A &= -0.2602	\; ,	\  B = 0.33655	\; ,	\  C = 0.90596	\; , \ 	D = 0.0159	\\
	a &= 0.03958	\; ,	\  b = 15.24915	\; ,	\   c = 3.2568	\; .
\end{align}
%%%%%%%%%%%%%%%%%%%%%%%%%%%%%%%%%%%%%
%
For comparison with experiment, following [\onlinecite{Balz2016}]
and [\onlinecite{Balz2017-PRB95}], 
MD results for $S({\bf q}, \omega)$ have further been convoluted in energy with 
a Gaussian of \mbox{FWHM = 0.2\ meV}.
%%%%%%%%%%%%%%%%%%%%%%%%%%%%%%%%%%%%%
 
%%%%%%%%%%%%%%%%%%%%%%%%%%%%%%%%%%%%%%
\section{Estimate of Wilson Ratio}
%%%%%%%%%%%%%%%%%%%%%%%%%%%%%%%%%%%%%
\label{sec:wilson.ratio}

The Wilson ratio is a dimensionless ratio of the (para)magnetic 
susceptibility of a metal, $\chi_0$, to the linear coefficient 
of its specific heat, $\gamma$, with parameters chosen such that the 
ratio takes on the value one in a free electron gas.
It is defined to be \cite{solyom-book,fulde-book}
\begin{equation}
	R_W=\frac{\pi^2}{\mu_0} \frac{\chi_0}{\mu_\text{eff}^2}\frac{k_B^2}{\gamma}  \; ,
\end{equation}
where the effective magnetic moment 
\begin{equation}
	\mu_\text{eff} = \sqrt{j(j+1)}g\mu_B \;.
\end{equation}

%%%%%%%%%%%%%%%%%%%%%%%%%%%%%%

For comparison with experiment on \CCO, we take 
\begin{align}
j=1/2\ \; \text{and} \; g=2 \; .
\end{align} 
The remaining parameters, in cgs units, are given by
\begin{align}
	k_B &= 1.381 \times 10^{-16}\ \frac{\text{erg}}{\text{K}} \;,\\
	\mu_B &= 9.274 \times 10^{-21}\ \frac{\text{erg}}{\text{G}} \; ,\\
	\mu_0 & = 1\ \frac{\text{G}}{\text{Oe}} \; .
\end{align}
Taking values of $\chi_0$ and $\gamma$ from experiment [Table~\ref{tab:heavy.fermions}], 
\begin{align}	
	\chi_0 
	=  3\  \frac{\text{emu}}{\text{mol Oe }} 
	= 3\  \frac{\text{erg}}{\text{mol Oe G}} \;,
\end{align}
and
\begin{align}	
	\gamma 
	= 1.35 \times 10^{4}\ \frac{\text{mJ}}{\text{mol }\text{K}^2} 
	=  1.35 \times 10^{8}\ \frac{\text{erg}}{\text{mol }\text{K}^2} \;,
\end{align}
we obtain an estimate of the Wilson Ratio of \CCO\ 
\begin{eqnarray}
	R_W \approx 16.2 \; .
\end{eqnarray}

%%%%%%%%%%%%%%%%%%%%%%%%%%%%%%

For completeness, we note that taking $j=1/2$ and $g=2$ leads to two 
other commonly quoted expressions for the Wilson ratio 
\begin{equation}
	R_W=\frac{4\pi^2}{3\mu_0} \frac{\chi_0}{g^2\mu_B^2}\frac{k_B^2}{\gamma}
	=\frac{\pi^2}{3\mu_0} \frac{\chi_0}{\mu_B^2}\frac{k_B^2}{\gamma} \;.
\end{equation}
It should also be noted that in some literature the units are chosen 
such that $\mu_0$ can be omitted.
%
%%
%\begin{equation}
%	\mu_0 = 1 \frac{\text{G}}{\text{Oe}} \;,
%\end{equation}
%
 
%%%%%%%%%%%%%%%%%%%%%%%%%%%%%%%%%%%%%
\bibliography{paper}
%%%%%%%%%%%%%%%%%%%%%%%%%%%%%%%%%%%%%

%%%%%%%%%%%%%%%%%%%%%%%%%%%%%%%%%%%%%
\end{document}